\documentclass[prb,twocolumn]{revtex4-2}
\usepackage{latexsym}
\usepackage{graphicx}

\usepackage{pict2e}
\usepackage{amsmath}
\usepackage{amssymb}
\usepackage{subfigure}
\usepackage{array}
\usepackage{verbatim}
\usepackage{amsmath}
\usepackage{color}
\usepackage{xcolor}
\usepackage{soul}
\usepackage{tabu}
\usepackage{multirow}
\usepackage{lipsum}
\usepackage[normalem]{ulem}
\begin{document}
\title{Gradient-expansion of the inhomogeneous electron-gas revisited}
\author{Mario Benites, Angel Rosado and Efstratios Manousakis}
  \address{
    Department  of  Physics, Florida State University}
  \begin{abstract}
    In the present work, we revisit the problem of the inhomogeneous electron gas under the influence of a weak external potential, which allows us to calculate the gradient corrections to the density functional within linear response, an approach known as the gradient expansion approximation. To obtain the exchange ($b_x$) and correlation ($b_{c}$)
    contributions to the coefficient $b_{xc}$, i.e., to the prefactor of the $q^2$ term of the proper-polarization function, we revisited all the previous calculations and expose misconceptions which led to    incorrect conclusions. We used
various ways to apply a necessary regularization to the singular Coulomb interaction potential. We found that the separate exchange ($b_x$) and correlation ($b_c$) contributions to the coefficient $b_{xc}$
    have regularization-scheme dependent values even though the regulator is set to zero at the end of the calculation. This implies that it is impossible to define such a separation meaningfully.
    On the contrary, we found that when the regulator is set to zero at the end of the calculation, the combination $b_{xc}$  is regularization-scheme independent and, thus, has a unique value.  We conclude that it is incorrect to separate those two terms when constructing a generalized-gradient-approximation (GGA) contribution to the density functional. This appears to be a common approach
   in most popular GGA functionals, where various
constraints are applied to each contribution separately.

    \end{abstract}
\maketitle
\section{introduction}
The problem of the interacting electron gas has been a focus of interest for
nearly a century\cite{PhysRev.46.1002,PhysRev.82.625,PhysRev.85.338,PhysRev.92.609,PhysRev.92.626,PhysRev.106.364,PhysRev.109.741,PhysRev.139.A796,PhysRev.136.B864,Ma-Brueckner,Geldart-Taylor1970,Geldart-Rasolt,Sham1971,PhysRevB.21.5469_Langreth_Perdew,PhysRevLett.45.566,Gross1981,Kleinman1984,Kleinman-Antoniewicz,Kleinman-Lee1988,Engel-Vosko1990,PhysRevB.54.17402,Svendsen1995,Vignale,benites2024}
and it is covered by classic many-body theory books\cite{Fetter,Mahan,Pines,Gorkov}
 as it allows a study of the electronic ground-state of
materials within the density functional theory (DFT). Using DFT one solves the
one-body Schr\"odinger-like Kohn-Sham equations\cite{PhysRev.140.A1133}, which
treat the many-body problem as an equivalent system of $N$ non-interacting electrons in the presence of an effective external field. In addition to
the external field generated by the ions, this field contains the
Hartree-term and the ``exchange-correlation" potential $V_{xc}$.
The latter is a functional of the spatially dependent local density field $n(\vec{r})$ produced by the presence of all of the electrons of the system in the interacting ground-state.  

The universal character of the functional $V_{xc}$\cite{PhysRev.140.A1133},
i.e., that it is material independent, allows the determination of its contributions using many-body theory of the electron gas system. A first approximation to the $V_{xc}$ functional
 is to consider the homogeneous electron gas; this allows us to obtain the so-called local part of the functional $V_{xc}$ and
such a simplification is known as the local density approximation (LDA).

Starting from the pioneering work of Ma and Brueckner\cite{Ma-Brueckner} (MB), the gradient-expansion approximation (GEA)\cite{PhysRevLett.77.3865,Geldart-Rasolt,PhysRevLett.100.136406-PBEsol} was introduced as an extension to the LDA exchange-correlation energy functional. The GEA adds a correction to the LDA exchange-correlation energy which is, in addition to the density, a functional of the gradient of the density;
this term is obtained from the response of the electron gas to a weak external spatially varying perturbation. In GEA, in the weak perturbation limit, where the density is  slowly varying, and in
the large-density limit, the problem translates into finding a coefficient of the square of the density gradient term. This is done by calculating
the static proper-polarization function $\Pi^*(q,\omega=0)$ and extracting the terms of
order $q^2$.

However, when trying to describe the electronic structure of real materials, the dimensionless quantity
$s=\nabla n(\vec r)/(2 k_F n(\vec r))$ (where $k_F$ is the Fermi wavevector), in certain regions of a given material,
is not necessarily small. In order to capture the contribution of such
non-perturbative effects in the functional,  a generalized-gradient approximation (GGA) has been attempted\cite{PhysRevLett.77.3865}. Within the GGA one introduces an ad hoc form of the
functional of the density gradient which should be forced to obey known constraints to yield the results of the GEA in the regime of its validity.

A controversy seems to exist in previous attempts to calculate the exchange $b_x$ and correlation $b_c$ contributions to the coefficient $b_{xc}=b_x+b_c$ of the terms of order $q^2$ in $\Pi^*(q,0)$
(which are related to the coefficient of $s^2$) in the high density and low $q$ limit.

MB\cite{Ma-Brueckner} were the first to derive the leading contribution to $b_c$ within the
random-phase approximation (RPA). Geldart and Taylor (GT)\cite {Geldart-Taylor1970} were the first to report a value for $b_x$. However, a year later, Sham\cite{Sham1971} found a different value for $b_x$ using a Yukawa-like potential $e^2 e^{-\lambda r}/r$ to regulate the Coulomb interaction with the regulator $\lambda$ kept finite while carrying out the integrations and set to zero at the end of the calculation. 

Later, MB calculations were verified by Geldart and Rasolt\cite{Geldart-Rasolt} by introducing an infrared (IR) cutoff at one of the integrals contributing to $b_c$.
Langreth and Perdew\cite{PhysRevB.21.5469_Langreth_Perdew} (LP) performed a calculation
of the full exchange-correlation coefficient $b_{xc}$ with a
wavevector decomposition method; they found that their value of $b_{xc}$
equals the sum of the value obtained by MB for $b_c$ plus the value obtained by Sham\cite{Sham1971} for $b_x$.

Antoniewicz and Kleinman\cite{Kleinman-Antoniewicz} (AK), recalculated $b_x$ and their results agree with those of GT, and then within the same decade, Kleinman-Lee\cite{Kleinman-Lee1988} (KL) obtained the same result. After this, Kleinman and Tamura\cite{Kleinman-Tamura} (KT) recalculated $b_c$, and their result disagrees with the value found by MB. Furthermore, Engel and Vosko\cite{Engel-Vosko1990} (EV) reported an analytical calculation of $b_x$ without using a regulator for the Coulomb interaction and their result agrees with that given by GT. Svendsen and von-Barth\cite{Svendsen1995}   claimed to provide an explanation for the discrepancy between the value calculated by Sham\cite{Sham1971} and that obtained by GT, KA, KL and EV; they found that their integrals, which contribute to the calculation of $b_x$, fully converge to the value obtained by the latter when the bare-Coulomb potential is used in the integrands. They also obtained Sham's value when the Yukawa-like potential is used and taking the limit of the regulator $\lambda$ to zero at the end of the calculation. They concluded (which is incorrect as we will argue below) that the correct value of $b_x$ is the one obtained by GT, AK, KL, and EV. This conclusion has influenced the authors who derived the PBE\cite{PhysRevLett.77.3865} and the PBEsol\cite{PhysRevLett.100.136406-PBEsol} functional.

The main objective of the present work is to resolve this controversy. We used
various ways to apply the necessary regularization of the Coulomb interaction potential.
We found that when the regulator is set to zero at the end of the calculation, only the combination of the exchange plus correlation contribution is regularization-scheme independent and thus has a unique value. On the contrary, the exchange or the correlation contributions separately do not have regularization-scheme independent values even though the regulator is set to zero at the end of the calculation, and therefore, it is impossible to define such a separation meaningfully. In what follows, we clarify what this means and, in particular, why a regulator is necessary.

First, it is well-known that the Fourier transform ${\tilde V}_0(q)$ of the Coulomb interaction $V_0(r)=e^2/r$ does not exist, i.e.,
\begin{eqnarray}
  {\tilde V}_0(q) \equiv \int d^3r{{e^2} \over r} e^{i \vec q \cdot \vec r}  = {{4 \pi e^2} \over {q^2}} \int_0^{\infty}dx \sin(x)
\end{eqnarray}
and the last integral does not have a definite value.

What is usually done is to redefine the problem, i.e., the Coulomb interaction as:
\begin{eqnarray}
  V_{\lambda}(r) = e^2 {{e^{-\lambda r}} \over {r}},
  \label{yukawa1}
\end{eqnarray}
where we take the
limit $\lambda \to 0$ at the end of the calculation. Namely, the two operations, i.e., the limit $\lambda \to 0$ and the integration, do not commute.
If $\lambda$ is kept finite, we find:
\begin{eqnarray}
  {\tilde V}_{\lambda}(q)  = {{4 \pi e^2} \over {q^2+\lambda^2}}. \label{yukawa2}
\end{eqnarray}
In many textbooks, after this result is reached, the value of $\lambda$ is set to zero and it is claimed that the Fourier transform of the Coulomb interaction is $4\pi e^2/q^2$.
This is fine if our calculation stops at this point. However, one has to keep in mind two facts: a) To obtain this result, a regulator was necessary at the beginning of the
calculation,
which is set to zero at end of the calculation. b) When we set $\lambda = 0 $ at the end of the calculation of the Fourier transform, the original pathology of the bare-Coulomb interaction,
i.e., that it is of infinite range, is still present in the form of the Fourier transform; namely, in the long-wavelength limit ($q \to 0$), ${\tilde V}_0(q)$ diverges.

Therefore, if we need to use the above result to continue the evaluation of other integrations which require ${\tilde V}_{\lambda}(q)$, we need to continue the calculation keeping $\lambda$ finite, and set $\lambda=0$ after the integrations are done. Otherwise
we might have a problem evaluating the integrals because of the $q \to 0$ singularity.

The conclusion of the previous very elementary but crucial discussion is that we need to use a regulator when we use the Coulomb interaction to calculate various diagrammatic
contributions to any quantity. The regulator should be set to zero at the very end of the calculation and the result is meaningful if and only if it does not depend on
the form of the regulator that we used.

In the present work, we evaluate the exchange ($b_x$) and correlation ($b_c$) coefficients by using the general form of the regularized-Coulomb interaction given by Eq.~\ref{yukawa1} with $\lambda = \lambda_c \beta(k_F)$ and taking $\lambda_c \to 0$ limit at the end of the calculation. This 
function $\beta(k_F)$ is a general function of $k_F$ which includes the case where $\beta(k_F)$ is a constant and the case where $\beta(k_F) = 0$. 

First, we verify that, in our case, not using a regulator, i.e., setting $\lambda=0$ before
the integrations are completed, leads to divergent or ill-defined integrals. Thus, a regulator must be used inside the integrands and should be set to zero after the integrations are performed. We also show that the results for the coefficients $b_x$ and $b_c$ depend on the choice of the function $\beta(k_F)$.
This is done by using different choices of $\beta(k_F)$ in the calculation of $b_x$ and $b_c$, and then taking the $\lambda_c \to 0$ limit.

We found that we can reproduce the value of $b_x$ reported by Sham\cite{Sham1971} and the value of $b_c$ reported by MB, when $\beta(k_F)$ is set to a constant.
When choosing $\beta(k_F)= k_F$, however, this reproduces KT's reported value for $b_c$, which agrees with their regularized potential by rescaling the regulator of the Yukawa-like potential by a $k_F$ factor. When such $\beta(k_F)$ function is used, this yields a value of $b_x$ which is three times the value reported by Sham.
When we set $\beta(k_F)=q_{TF}$, where $q_{TF}$ is the Thomas-Fermi wavevector $q_{TF}$, it leads to $b_c=0$ 
and to a value for $b_x$ twice the value reported by Sham. Lastly, when setting $\beta(k_F)=k_F^{-3/10}$, we reproduce the value of $b_x$ obtained by GT, KA, KL and EV,
and a value for $b_c$ that is $8/5$ times the value reported by MB. 

Therefore, we demonstrate that the coefficients $b_x$ and $b_c$ are regulator-dependent.
As a result, the exchange $B_x s^2$ and correlation $B_c s^2$ terms of the GEA functional do not have a unique value.  We find,  however,
that when we add $b_x$ and $b_c$ to obtain $b_{xc}$, it always yields the same regulator-independent value. This implies that it is incorrect to separate the exchange and correlation terms when constructing a GGA contribution to the density functional by applying various constraints on each part independently.

The paper is organized as follows. In Sec.~\ref{Gradient_approx}, we discuss the GEA for the inhomogeneous electron gas in the limit of smooth-density variation, i.e., when the parameter $s$ introduced above is small, where the coefficients 
$b_{xc}$ of the $q^2$ term and the coefficient $B_{xc}$ of the $s^2$ term appear. In Sec.~\ref{correct_calculation}, we carry out the main steps of the calculation of the proper-polarization function $\Pi^{xc}(q,0)$ that yields the leading contributions (in an $r_s$ expansion) to the $b_{xc}q^2$ term by separating it into two parts, the $b_xq^2$ term due to the exchange and the $b_c q^2$ due to correlations. In this approach, the high-dimensional integrals that contribute to these coefficients are calculated using the regularized-Coulomb potential $\tilde{V}_{\lambda}$ and the regulator is removed
at the end of the calculation. In Sec.~\ref{summary} we summarize the main results of this lengthy calculation
and discuss how it resolves the historical controversy about the correct value of the coefficients. In Sec.~\ref{conclusions} we discuss the implications of the calculation to the constraints that need to
be imposed on any GGA functional.

\section{The Gradient-Expansion Approximation in the inhomogeneous Electron gas}
\label{Gradient_approx}
A well-known extension of the LDA exchange-correlation energy functional is obtained by means of the gradient-expansion approximation (GEA). GEA introduces gradient terms of the electron density $n(\vec{r})$ in the slowly varying density limit and the exchange-correlation energy functional is approximated by:
\begin{equation}
E_{xc}[n(\vec r)] =\int d^3r' \Bigl [ A_{xc}[n({\vec r}\,')] + B_{xc}[n({\vec r}\,')]|{\bf s}(\vec r\,')|^2 \Bigr ],
\label{xc_GE_functional}    
\end{equation}
where ${\bf s}(\vec r)=\nabla n(\vec{r})/(2k_Fn(\vec{r}))$ and $A_{xc}[n({\vec r})]$ is the exchange-correlation energy-density functional found in the LDA.
In the present paper, the coefficient $B_{xc}[n]$ is the quantity of our interest that needs to be calculated within the linear response of the many-electron system when subjected to some weak external field.

This approach can be pursued using the interacting electron gas
because of the universal character of the exchange-correlation functional. However, only in the slowly varying density limit, a linear response treatment
of the electron gas is valid. Within a time-independent linear response, one adds a static weak external potential $\phi^{ext}(\vec{r})$ acting on the electron gas. We begin by adding to the Hamiltonian the following perturbation:
\begin{equation}
\hat{H}^{ext} = \sum_{\vec{q}} \Bigl [\phi^{ext}_{\vec{q}} \hat{n}_{-\vec{q}} \Bigr ].
\label{H_external}    
\end{equation}
where $\phi^{ext}_{\vec{q}}$ is the Fourier transform of the weak external potential and $\hat{n}_{\vec{q}}$ is the Fourier transform of the density operator. The energy functional of the density, i.e.,  $E[n(\vec{r})]$, is written as:
\begin{equation}
E[n(\vec{r})] = F[n(\vec{r})] + \int d^3r n(\vec{r})\phi^{ext}(\vec{r}),
\label{E_functional}    
\end{equation}
where $F[n(\vec{r})]$ is defined as:
\begin{equation}
F[n(\vec{r})] = T_s[n(\vec{r})] + \frac{1}{2}\int d^3r \int d^3r' \frac{n(\vec{r}) n(\vec{r}\,')}{|\vec{r}-\vec{r}\,'|}+E_{xc}[n(\vec{r})],
\label{F}    
\end{equation}
and $T_s[n(\vec{r})]$ is the kinetic energy functional that corresponds to an equivalent non-interacting system, which is given by the Kohn-Sham states. Due to this non-interacting system, the second-order functional derivative of the kinetic energy functional $T_s[n(\vec{r})]$, is given by:
\begin{equation}
\frac{\delta^2T_{s}[n]}{\delta n(\vec{r}) \delta n(\vec{r}\,')} = -\Pi^{-1}_0(\vec{r},\vec{r}\,',0).
\label{density_response_Ts}    
\end{equation}
A functional Taylor expansion of the universal functional up to second order in the density, yields:
\begin{eqnarray}
  F[n] =  F[n^0]+{{1}\over {2}} \int d{\vec r}_1 d{\vec r}_2 {{\delta^2F[n]}\over {\delta n(\vec{r}_1)\delta n(\vec{r}_2)}} \delta n(\vec{r}_1) \delta n(\vec{r}_2),
\label{F_taylor}    
\end{eqnarray}
where $n^0$ is the homogeneous part of the electron density and the first functional derivative term at $n=n^0$ must yield zero, since the universal functional must be minimum at $n=n_0$. By substituting this functional form in the expression of the energy functional given by Eq.~\ref{E_functional}, and by minimizing the energy functional, we obtain:
\begin{equation}
\frac{\delta E[n]}{\delta n(\vec{r})} = \left. \int d^3r' \frac{\delta^2 F[n]}{\delta n(\vec{r}) \delta n(\vec{r}\,')} \right|_{n=n^0} \delta n(\vec{r}\,') +\phi^{ext}(\vec{r}) = 0,    
\label{minimizing_E}
\end{equation}
where we conclude that the external potential can be written in terms of the second-order functional derivative of the universal functional $F[n]$ as follows:
\begin{equation}
\phi^{ext}(\vec{r}) = \left. -\int d^3r'\frac{\delta^2 F[n]}{\delta n(\vec{r}) \delta n(\vec{r}\,')}\right|_{n=n^0} \delta n(\vec{r}\,').
\label{ext_potential}    
\end{equation}
Within the linear response, the variation of the density $\delta n(\vec{r})$ is written in terms of the density-response function $\chi(\vec{r},\vec{r}\,',0)$ as follows:
\begin{equation}
\delta n(\vec{r}) = \int d^3r' \chi(\vec{r},\vec{r}\,',0) \phi^{ext}(r'),
\label{linear_response}    
\end{equation}
which, by comparing the expression of the second-order functional derivative of the universal functional $F[n(\vec{r})]$ with the condition on the external potential that  the ground-state energy is extremized, yields:
\begin{equation}
\chi^{-1}(\vec{r},\vec{r}\,',0) = \left. -\frac{\delta^2 F[n]}{\delta n(\vec{r}) \delta n(\vec{r}\,')} \right|_{n=n^0}.
\label{density_response_inverse}    
\end{equation}
We can express the ground-state energy in terms of the density-response function in wave-vector space as follows:
\begin{equation}
E[n] = F[n^0] + \frac{1}{2}\sum_{\vec{q} \neq 0}\chi^{-1}(q,0) \delta n_{\vec{q}}\, \delta n_{-\vec{q}},
\label{energy_functional_extremized}
\end{equation}
where we have also used the conservation of the total number of particles, which leads to also having $\phi^{ext}(q=0)=0$, since $\delta n_{\vec{q}=0}=0$. 

\begin{figure*}[htp]
   \begin{center}
   \includegraphics[scale=0.37]{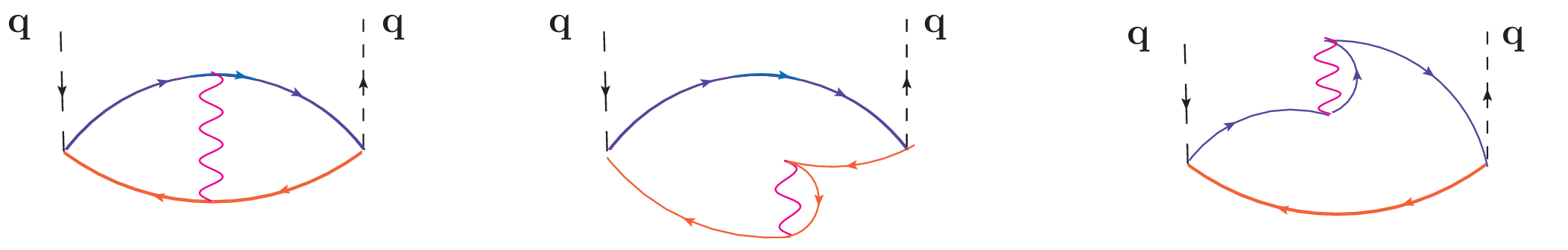} 
   \end{center}
   \caption{Diagrammatic contributions to the irreducible polarization function up to first order of an expansion in $\tilde{V}_{\lambda}(p)$ given by Eq.~\ref{yukawa2}.}
   \label{GA_Fock}
\end{figure*}

\begin{figure*}[htp]
   \begin{center}
   \includegraphics[scale=0.26]{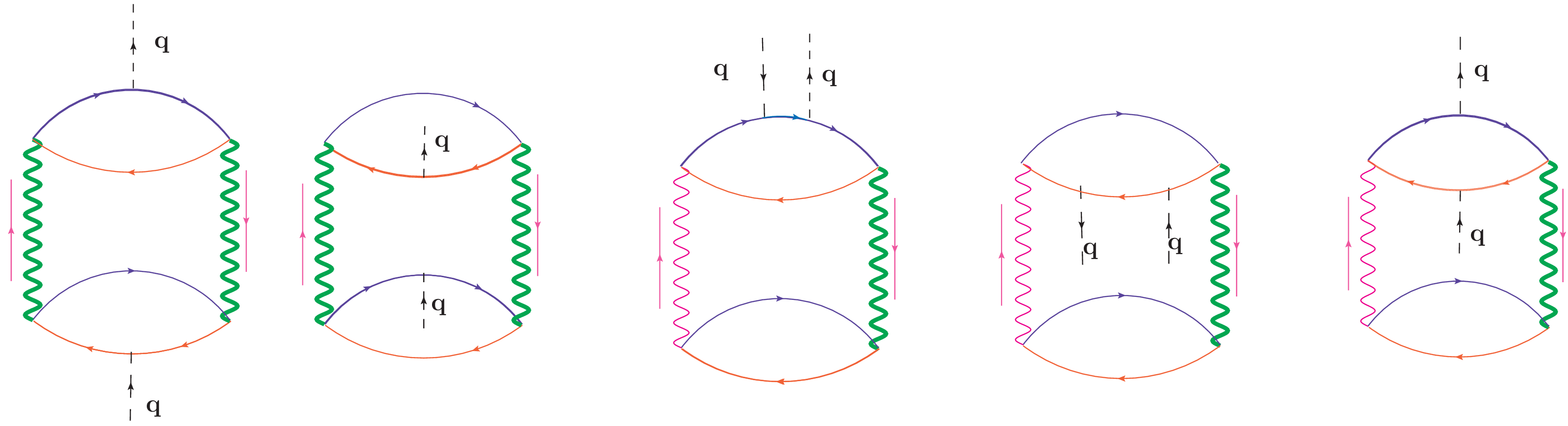} 
   \end{center}
   \caption{The rest of the diagrams contributing to $\Pi^{xc}(q,0)$ that need to be combined with the diagrams from Fig.~\ref{GA_Fock}. These diagrams contribute to the same order in $r_s$ in the long-wavelength limit. The fuchsia color represents the Regularized-Coulomb interaction line, while the green color represents the RPA renormalized interaction line. The solid blue (red) lines represent the fermionic non-interacting electron (hole) propagator, while the dashed line represents the insertion due to the weak external potential.}
   \label{Pi_c}
\end{figure*}

By taking a double functional derivative of the $F[n]$ functional using Eq.~\ref{F}, we obtain an expression that relates the density-response function, the lowest order proper-polarization function $\Pi_0(\vec{r},\vec{r}\,')$, the Coulomb potential $V_0(\vec{r}-\vec{r}\,')$ and the second-order functional derivative of the exchange-correlation functional $E_{xc}[n]$ as follows:
\begin{equation}
K_{xc}(\vec{r}-\vec{r}\,')= -\chi^{-1}(\vec{r},\vec{r}\,',0) + \Pi^{-1}_0(\vec{r},\vec{r}\,',0)-V_0(\vec{r}-\vec{r}\,'),
\label{2nd_deriv_Exc}    
\end{equation}
where $K_{xc}(\vec{r}-\vec{r}\,')$ is the kernel that includes the correction to the exchange-correlation energy functional from the inhomogeneous electron gas, which is given by:
\begin{equation}
K_{xc}(\vec{r}-\vec{r}\,') = \frac{\delta^2 E_{xc}[n]}{\delta n(\vec{r}) \delta n(\vec{r}\,')}.
\label{Kxc_kernel}    
\end{equation}
The density-response function in wavevector space is found as a summation of a geometric series with the proper-polarization function $\Pi^*(q,0)$ as its ratio:
\begin{equation}
\chi(q,0) = \frac{\Pi^*(q,0)}{1-\tilde{V}_0(q)\Pi^*(q,0)}.
\label{chi_geometric}
\end{equation}
By means of a Fourier transform in Eq.~\ref{2nd_deriv_Exc}, and using the expression of the density response from Eq.~\ref{chi_geometric}, we can express the exchange-correlation kernel of the functional as follows:
\begin{equation}
K_{xc}(q,0) = -\left(\frac{1}{\Pi^*(q,0)}-\frac{1}{\Pi_0(q,0)}\right). 
\label{exc_kernel_fourier}    
\end{equation}
Lastly, the correction to the exchange-correlation energy due to the inhomogeneity effects of the electron gas is given by the following expression:
\begin{equation}
E_{xc}[n] = E_{xc}[n^0]-\frac{1}{2}\sum_{q}\left(\frac{1}{\Pi^*(q,0)}-\frac{1}{\Pi_0(q,0)} \right)\delta n_{\vec{q}}\, \delta n_{-\vec{q}},
\label{Exc_inhomoegeneous_final}    
\end{equation}
where the main focus of this work is to calculate the proper-polarization function $\Pi^*(q,0)$ up to a certain order of expansion in the long-wavelength ($q \to 0$) limit and to calculate
the difference of the reciprocal terms given in the parentheses in Eq.~\ref{Exc_inhomoegeneous_final}. By expanding the proper-polarization function in small $q$, the term in the sum from Eq.~\ref{Exc_inhomoegeneous_final} can be written as:
\begin{equation}
\frac{1}{\Pi^{*}(q,0)} -\frac{1}{\Pi_{0}(q,0)} \approx c_0+c_2 q^2, 
\label{polarization_q_expansion}    
\end{equation}
where $c_0$ is an extra contribution to the LDA correlation energy
while $c_2$ corresponds to the coefficient of the gradient-squared term of Eq.~\ref{xc_GE_functional}.
In order to obtain the coefficients $c_i$, we have to calculate the proper-polarization functions $\Pi^{*}(q,0)$ and $\Pi_0(q,0)$, where the latter term is found from the Lindhard function. At any order of the perturbative expansion of $\Pi^*(q,0)$, we can always factorize a factor of $m^2 e^2/\pi^3$, however, the higher-order terms of $\Pi^*(q,0)$ are given by higher-order powers of the Wigner-Seitz radius $r_s$. In our notation, the proper-polarization function $\Pi_0(q,0)$  in the $q \to 0$ limit, is expressed as follows:
\begin{equation}
\Pi_0(q,0) = \frac{e^2 m^2}{\pi^3}(\tilde{a}_0 + \tilde{b}_0 q^2), 
\label{polarization_proper_twiddles}
\end{equation}
where the coefficients $\tilde{a}_0$ and $\tilde{b}_0$ are given by:
\begin{eqnarray}
\tilde{a}_0 &=& -\frac{\pi}{\alpha r_s},\label{a0_twiddle}\\
\tilde{b}_0 &=& \frac{\pi \alpha r_s}{12}.\label{b0_twiddle}
\end{eqnarray}
In this work, we calculated the sum of the proper-polarization functions, which are listed in Fig.~\ref{GA_Fock} and Fig.~\ref{Pi_c}, and call it $\Pi^{xc}(q,0)$. These diagrams contribute as $r_s^2$ for the $c_2$ coefficient. Fig.~\ref{GA_Fock} illustrated the diagrams that contribute to $b_x$, while Fig.~\ref{Pi_c}
gives the diagrams that contribute to $b_c$ and have been worked previously by Ma-Brueckner. After performing the sum of these proper-polarization functions, we can express the coefficient $c_2$ as:
 \begin{equation}
 c_2 = \frac{\pi^3}{e^2m^2}\frac{(2 \tilde{a}_x \tilde{b}_0-\tilde{b}_{xc} \tilde{a}_0)}{\tilde{a}^3_0},
 \label{c2}
 \end{equation}
where $\tilde{a}_x$ is the zeroth order of the expansion in the $q \to 0$ limit when a $e^2m^2/\pi^3$ has been factored out, and is a unique value. This coefficient has been used by Sham, AK, KL and EV to obtain the exchange contribution to the $B_{xc}$. The value of $\tilde{a}_x$ is given by:
 \begin{equation}
 \tilde{a}_x = -1.
 \label{ax_twiddle}
 \end{equation}  
The coefficient $\tilde{b}_{xc}$ is given by $\tilde{b}_{xc}=\pi^3 b_{xc}/e^2 m^2$, where $b_{xc}$ is related to the coefficient of the $q^2$ order of $\Pi^{xc}(q,0)$. The $s^2$ coefficient $B_{xc}[n]$ is given in terms of the $c_2$ coefficient by the expression below:
\begin{equation}
B_{xc} = -2(3\pi^2)^{\frac{2}{3}}n^{\frac{8}{3}}c_2.
\label{Bxc}
\end{equation}

In Refs.~\cite{Geldart-Taylor1970,Sham1971,Kleinman-Antoniewicz,Kleinman-Lee1988,Engel-Vosko1990,Svendsen1995} only the exchange part $B_x$ to the $B_{xc}$ coefficient was calculated, while in Refs.~\cite{Ma-Brueckner,Geldart-Rasolt,Kleinman-Tamura}  only the correlation contribution $B_c$ to the density functional was calculated. In our notation, the expressions of these two coefficients are expressed as follows:
\begin{eqnarray}
B_x &=& -2 (3\pi^2)^{\frac{2}{3}}n^{\frac{8}{3}}c_{21},\label{Bx}\\
B_c &=& -2 (3\pi^2)^{\frac{2}{3}}n^{\frac{8}{3}}c_{22},\label{Bx}
\end{eqnarray}
where the coefficients $c_{21}$ and $c_{22}$ are given by the following expressions:
\begin{eqnarray}
c_{21} &=& \frac{\pi^3}{e^2m^2}\frac{(2 \tilde{a}_x \tilde{b}_0-\tilde{b}_{x} \tilde{a}_0)}{\tilde{a}^3_0},\label{c21}\\
c_{22} &=& -\frac{\pi^3}{e^2m^2}\frac{\tilde{b}_{c} \tilde{a}_0}{\tilde{a}^3_0},\label{c22}
\end{eqnarray}
where $\tilde{b}_x$ ($\tilde{b}_c$) is found from the coefficient $b_x$ ($b_c$) where $e^2m^2/\pi^3$ has been factored out. The term $b_x$ ($b_c$) is found from the $q^2$ order of the Taylor expansion of the sum of all the diagrams illustrated in Fig.~\ref{GA_Fock} (Fig.~\ref{Pi_c}) in the $q \to 0$ limit. In the next section, however, we explain how we calculate the $\Pi^{xc}(q,0)$.

\section{Calculation of the exchange-correlation coefficient}
\label{correct_calculation}
The diagrams illustrated in Fig.~\ref{GA_Fock} have been the main focus of calculation for decades. Two different approaches have been taken when calculating the $q^2$ term of the sum of these proper-polarization functions; they differ in the usage of a regulator for the Coulomb interaction. A regulator was first used by Sham by using a Yukawa-like potential ${\tilde V}_{\lambda}(q)$ given by Eq.~\ref{yukawa2}, where $\lambda$  is taken to zero at the end of the calculation.
Kleinman and Tamura (KT)\cite{Kleinman-Tamura} use an approach which effectively corresponds to using $\lambda =\lambda_c k_F$ in Eq.~\ref{yukawa2} and at the end of the calculations the $\lambda_c \rightarrow 0$ limit is taken.  KT's treatment of the regulator was used only for the calculations of two specific coefficients that contribute to  $b_c$,
labeled by MB\cite{Ma-Brueckner} as $b'$ and $b''$. Their result disagrees with the value of $b'$ reported by MB\cite{Ma-Brueckner}.
The exchange part of the coefficient of the functional in GEA, i.e., $b_x$ (which corresponds to $B_x$ and the diagrams from Fig~\ref{GA_Fock}) was calculated by other authors using  $\tilde{V}_0(q)$ from the start of the calculation\cite{Geldart-Taylor1970,Kleinman-Antoniewicz,Kleinman-Lee1988,Engel-Vosko1990}. 

The results vary depending on which approach is used, which is a demonstration that the integral expressions associated with the diagrams in Fig.\ref{GA_Fock} are sensitive to the form of the regulator. As we will see in this Section, this is the origin of the controversy. Interestingly,  Kleinman's value for $b_x$, which is $10/7$ times larger than the coefficient obtained by Sham, seems to have influenced the derivation of the PBE, the PBEsol\cite{PhysRevLett.100.136406-PBEsol} and other functionals.

In the following, we present the calculation of the coefficients of $q^2$
for the general case. We use the following notation
\begin{equation}
\tilde{V}(k) = \frac{4 \pi e^2}{k^2+(\lambda_c \beta(k_F))^2}.
\label{V_generic}    
\end{equation}
for the regularized-Coulomb potential when it appears
inside the necessary integrals, where $k$ is a dummy integration variable.
Notice, that we have used the notation: ${\tilde V}(k)$, i.e.,
we have dropped the explicit dependence on the
regulator $\lambda_c \beta(k_F)$ for simplicity. 
After concluding the calculations which can be done without specifying the form of $\title V(k)$,  specific expressions for $\beta(k_F)$ are considered;
 we show that particular forms of $\beta(k_F)$ reproduce the values that has been obtained in previous works for $b_x$\cite{Geldart-Taylor1970,Sham1971,Kleinman-Antoniewicz,Kleinman-Lee1988,Engel-Vosko1990}, and for $b_c$\cite{Ma-Brueckner,Geldart-Rasolt,Kleinman-Tamura}.

\subsection{The $q^2$ coefficients of the proper-polarization function $\Pi^{xc}(q,0)$}
We organize the sum of all diagrams from Fig.~\ref{GA_Fock} and Fig.~\ref{Pi_c} as:
\begin{equation}
\Pi^{xc}(q,0) = \sum_{i=1}^{3} \Pi^{xc}_i(q),
\label{Pi_xc_reorganized}    
\end{equation}
where $\Pi^{xc}_1(q,0)$ is the sum of the vertex bubble on the left side of Fig.~\ref{GA_Fock} and the last term on the right of Fig.~\ref{Pi_c}. $\Pi^{xc}_2(q,0)$ is the sum of the rest of the diagrams from Fig.~\ref{GA_Fock} with the third and fourth diagram in Fig.~\ref{Pi_c} from left to right. We denote the sum of the remaining two diagrams in Fig.~\ref{Pi_c} as $\Pi^{xc}_3(q,0)$. For the latter term, we only calculate one of the two diagrams, and then multiply by two since the proper-polarization function can be written in even powers of $q$ and the other contribution to $\Pi^{xc}_3(q,0)$ is obtained by the other term by flipping the direction of vector $\vec{q}$. The term labeled as $\Pi^{xc}_{31}(q,0)$, which is
calculated next, corresponds to the first diagram on the left in Fig.~\ref{Pi_c}. The integral expressions of these terms are given by:
\begin{eqnarray}
\Pi^{xc}_1(q) &=& 2\int d[{\bf p}] d[{\bf p}'] M(p^{\mu},p'^{\mu},q^{\mu})\frac{\tilde{V}(\vec{p}\,'-\vec{p})}{\epsilon(p'^{\mu}-p^{\mu})},
\label{pixc_1}    \\
  \Pi^{xc}_2(q) &=& -2i\int d[{\bf p}] (G^0(p^{\mu}))^2\Sigma_{GW}(p^{\mu})\nonumber \\
  &\times&\left(G^0(p^{\mu}+q^{\mu})+G^0(p^{\mu}-q^{\mu})\right),
\label{pixc_2}    \\
  \Pi^{xc}_{31}(q) &=& -4i \int d[{\bf p}]  d[{\bf  p}'] d[{\bf  k}] M(p^{\mu},p'^{\mu},q^{\mu})\nonumber \\
  &\times&A(p^{\mu},p'^{\mu},k^{\mu})\frac{\tilde{V}(\vec{k}-\frac{\vec{q}}{2})\tilde{V}(\vec{k}+\frac{\vec{q}}{2}) }{\epsilon(k^{\mu}-\frac{q^{\mu}}{2}) \epsilon(k^{\mu}+\frac{q^{\mu}}{2})}, 
\label{pixc_31}    
\end{eqnarray}
where $d[{\bf p}] \equiv d^4 p/(2 \pi)^4$, and the four-momentum convention is used as a shorthand notation $p^{\mu}=(p^0,\vec{p})$, where in all of these expressions we have $q^{\mu}=(0,\vec{q})$. The function $M(p^{\mu},p'^{\mu},q^{\mu})$
is the product
\begin{equation}
G^0(p^{\mu}-\frac{q^{\mu}}{2})G^0(p^{\mu}+\frac{q^{\mu}}{2}) G^0(p'^{\mu}+\frac{q^{\mu}}{2})G^0(p'^{\mu}-\frac{q^{\mu}}{2}),
\label{M}    
\end{equation}
and 
\begin{equation}
A(p^{\mu},p'^{\mu},k^{\mu}) = G^0(p^{\mu}-k^{\mu}) G^0(p'^{\mu}+k^{\mu}),
\label{B}    
\end{equation}
where $G^0(p^{\mu})$ is the non-interacting Green's function. And the term $\Sigma_{GW}(p^{\mu})$ corresponds to the well-known GW self-energy. It is usually convenient to break into two parts the expression of $\Sigma_{GW}(p^{\mu})$ to be consistent with the time-ordering of the fermionic propagator. The two terms that arise is the Fock self-energy $\Sigma_F(p^{\mu})$ and the leftover term is the self-energy $\Sigma_r(p^{\mu})$ which contains the contribution from the ring-like diagrams\cite{benites2024}.
The expressions of these self-energy terms are very well-known and are defined by the integral expressions given in Appendix~\ref{Pi_xc_calculations}, by Eqs.~\ref{self-energy_Fock}--\ref{self-energy_ring}. The ring-series self-energy $\Sigma_r(p^{\mu})$, is the only term in $\Sigma_{GW}(p^{\mu})$ that contains the renormalized-interaction potential within RPA $V_{e}(k^{\mu})$, which is given by:   
\begin{equation}
\tilde{V}_{e}(k^{\mu}) = \frac{\tilde{V}(k)}{\epsilon(k^{\mu})},
\label{V_effective}    
\end{equation}
where the denominator is the dielectric function within the RPA, and depends on the lowest-order polarization function $\Pi_0(k^{\mu})$ given by:
\begin{equation}
\epsilon(k^{\mu}) = 1-\tilde{V}(k)\Pi_0(k^{\mu}).
\label{dielectric}    
\end{equation}
A separation of terms, similar to what was done for the GW self-energy, is done for the specific vertex function $\Lambda^{GW}_2(p^{\mu})$ given by Eq.~\ref{lambda2_GW_definition};
it can be separated into two terms $\Lambda_2(p^{\mu})$ and $\Lambda^r_2(p^{\mu})$ as explained in Appendix~\ref{Pi_xc_calculations}. The calculations of these proper-polarization terms from Eq.~\ref{pixc_1}, Eq.~\ref{pixc_2}, and Eq.~\ref{pixc_31} rely on a Taylor expansion in powers of the small-wavevector $q$. In that expansion, some of the terms that emerge can be expressed in terms of these two vertex functions, where the integral expression of $\Lambda_2(p^{\mu})$ involves the potential $\tilde{V}(k)$, while $\Lambda^{r}_2(p^{\mu})$ involves the difference of $\tilde{V}(k)$ from $V_{e}(k^{\mu})$ inside the integrand. Within this expansion, it also emerges a specific term that involves two RPA-renormalized interaction $V_{e}(k^{\mu})$ terms in the integrand, and it is not convenient to separate it into several terms. 

After doing the Taylor expansion inside these integral expressions, as explained in the Appendix~\ref{Pi_xc_calculations}, by separating the terms of $\Sigma_{GW}(p^{\mu})$ and $\Lambda^{GW}_2(p^{\mu})$, and by using a set of identities of high-order powers of partial derivatives of the non-interacting Green's function $G^0(p^{\mu})$ given in Appendix~\ref{Identities Green's function}, we extract the coefficients of the $q^2$ $b^i_{xc}$ from the parts of the proper-polarization functions $\Pi^{xc}_i(q,0)$ (for $i=1,2,3$).
The corresponding expressions from these coefficients are given by Eqs.~\ref{b1_xc}, \ref{b2_xc} and \ref{bxc_3_combination}. At last, by reorganizing the sum of the coefficients $b^i_{xc}$ (for $i=1,2,3$) into four main parts, $b'_{xc}$, $b^{''}_{xc}$, $b^{'''}_{xc}$, and a term that depends on the value of $\Sigma_{GW}(k_F,0)$, we have:
\begin{equation}
\sum_{i=1}^3 b^i_{xc} = b^{'}_{xc}+b^{''}_{xc}+b^{'''}_{xc}-\frac{m^2 \Sigma_{GW}(k_F,0)}{12 \pi^3 k_F^3},
\label{b_xc_sum}    
\end{equation}
where $b'_{xc}$ and $b^{''}_{xc}$ have to be separated into two terms, given that we are separating the terms of the vertex function $\Lambda^{GW}(p^{\mu})$ and the self-energy $\Sigma_{GW}(p^{\mu})$. After some work which is summarized in Appendix~\ref{derivation_general_expressions_b_xc_primed},  the final expressions of the $b_{xc}$-primed coefficients are given by:
\begin{equation}
b'_{xc} = b^{'}_x+b^{'}_c, \quad b^{''}_{xc} = b^{''}_x + b^{''}_c,
\label{bxc}   
\end{equation}
where the $b$-primed coefficients are given by:
\begin{equation}
b^{'}_x = b^{'}_{x,1}+b^{'}_{x,2},
\label{b'_x}
\end{equation}
\begin{equation}
b^{'}_{x,1} = \frac{i}{m} \frac{\partial}{\partial \mu_0} \int d[{\bf p}] \Sigma_F(p^{\mu}) \frac{1}{2}\frac{\partial^2 G^0(p^{\mu})}{\partial \mu^2_0},
\label{b'_x1}    
\end{equation}
\begin{equation}
b^{'}_{x,2} = -\frac{i}{m}\frac{\partial}{\partial \mu_0} \int d[{\bf p}] \Sigma_F(p^{\mu}) \frac{\epsilon^0_p}{9}\frac{\partial^3 G^0(p^{p^{\mu}})}{\partial \mu^3_0},
\label{b'_x2}    
\end{equation}
\begin{equation}
b^{''}_x = -\frac{i}{6m} \int d[{\bf p}] \Sigma_F(p^{\mu}) \frac{\partial^3 G^0(p^{\mu})}{\partial \mu^3_0},
\label{b''_x}    
\end{equation}
\begin{eqnarray}
  b^{'}_c &=& -\frac{i}{m} \frac{\partial}{\partial \mu_0} \int d[{\bf k}] \left[V_{e}(k^{\mu})-\tilde{V}(k)\right] \nonumber \\
  &\times&\left(\frac{1}{2}I_1(k^{\mu}) -\frac{1}{9}I_3(k^{\mu}) \right), 
\label{b'_c}    
\end{eqnarray}
\begin{equation}
b^{''}_c = \frac{i}{6m} \int d[{\bf k}] \left[V_{e}(k^{\mu})-\tilde{V}(k) \right]I_2(k^{\mu}), 
\label{b''_c}    
\end{equation}
 where $\mu_0$ is the non-interacting chemical potential $\mu_0$ and finally, the last expression that is not separated into two terms is given by:
\begin{eqnarray}
  b^{'''}_{xc} &=& \frac{i}{24}\int d[{\bf k}] \left(\frac{\partial \Pi_0(k^{\mu})}{\partial \mu_0} \right)^2 \nonumber \\
  &\times&\left[V_{e}(k^{\mu})\nabla^2_k V_{e}(k^{\mu})-\left(\frac{d}{dk}V_{e}(k^{\mu}) \right)^2 \right],
\label{b'''_xc}    
\end{eqnarray}
where this last expression of $b^{'''}_{xc}$ agrees with the coefficient $b^{'''}$ that MB obtained in their work\cite{Ma-Brueckner}. The functions $I_i(k^{\mu})$ (for $i=1,2,3$) are given by the following expressions:
\begin{equation}
I_1(k^{\mu}) = -i\int d[{\bf p}]G^0(p^{\mu}+k^{\mu}) \frac{\partial^2}{\partial \mu_0^2} G^0(p^{\mu}),   
\label{I1}    
\end{equation}
\begin{equation}
I_2(k^{\mu}) = -i\int d[{\bf p}]G^0(p^{\mu}+k^{\mu}) \frac{\partial^3}{\partial \mu_0^3} G^0(p^{\mu}),   
\label{I2}    
\end{equation}
\begin{equation}
I_3(k^{\mu}) = -i\int d[{\bf p}]G^0(p^{\mu}+k^{\mu}) \frac{\partial^3}{\partial \mu_0^3} G^0(p^{\mu}) \epsilon^0_{\vec{p}}.   
\label{I3}    
\end{equation}
The expressions found so far are valid for any general function $\tilde{V}(k)$. From this point, we start discussing the calculation of the integrals of all the $b_{xc}$-primed coefficients by using a specific expression for the potential $\tilde{V}(k)$.
This is as far as we can reach regarding the calculation of the
integral expressions of the $q^2$ coefficients of $\Pi^{xc}(q,0)$ by not
specifying the form of the potential $\tilde{V}(k)$, which is
explicitly given by Eq.~\ref{V_generic}. Once the calculations of these $q^2$ coefficients are performed, we calculate their values by using different expressions of the function $\beta(k_F)$; this is illustrated, next, in Sec.~\ref{using-regulator} where we focus on calculating the total contribution to $b_{xc}$ given by  Eq.~\ref{b_xc_sum}.

\subsection{Calculation of the $b_{xc}$ coefficient by using a regulator}
\label{using-regulator}
We proceed to calculate all the $b_{xc}$-primed coefficients by using the generic potential $\tilde{V}(k)$ given by Eq.~\ref{V_generic}. This means that we have
used a $k_F$-dependent regulator by using a Yukawa-like potential from the beginning of the calculation of the proper-polarization function $\Pi^{xc}(q,0)$ in the static limit, as it should be done. 

By using the generic expression of the potential $\tilde{V}(k)$, one calculates the integrals that yield both self-energy terms: $\Sigma_{F}(p)$ and $\Sigma_{r}(p^{\mu})$, given by Eqs.~\ref{self-energy_Fock} and \ref{self-energy_ring} respectively. These expressions are necessary for the calculation of the $b_{xc}$-primed coefficients and, as we will show, are regulator dependent. Superscripts were added in all expressions of the self-energy indicating that we will be using the constant part of the regulator $\lambda_c$ at the calculation level of the $b_{xc}$-primed coefficients.  

The frequency integrals in the expressions of $b^{'}_{x}$ and $b^{''}_x$ are straightforward to calculate, where the non-zero terms arise from the Dirac delta term $\delta(p^0)$; more details are given in Appendix~\ref{explicit_calculation_bx}. These steps yield three-dimensional expressions for both $b_x$-primed coefficients given by Eq.~\ref{b'_x_3d} and Eq.~\ref{b''_x_3d}. The three-dimensional integral expressions are given by:
\begin{eqnarray}
  b^{'}_x &=& -\frac{1}{m}\frac{\partial}{\partial \mu_0}\int d[{\bf p}]\Sigma^{\lambda}_F(p)\nonumber \\
 &\times& \left[\frac{1}{2} \frac{\partial^2\Theta(\mu_0 - \epsilon^0_p)}{\partial \mu^2_0}-\frac{\epsilon^0_p}{9}\frac{\partial^3}{\partial \mu^3_0}\Theta(\mu_0-\epsilon^0_p)\right],
\label{b'_x_3d}\\    
b^{''}_x &=& \frac{1}{6m}\int d[{\bf p}]\Sigma^{\lambda}_F(p)\frac{\partial^3}{\partial \mu^3_0}\Theta(\mu_0-\epsilon^0_p),
\label{b''_x_3d}    
\end{eqnarray}
where in this particular case we have used the following notation
$d[{\bf p}]=d^3p/(2\pi)^3$.
These integrals agree with Sham's expressions labeled as $I_1$,$I_2$, and $I_3$ obtained in \cite{Sham1971}. It is straightforward to express these integrals in a more convenient form by using the identity $\partial_x \delta(x-y) = -\partial_y \delta(y-x)$, as explained in Appendix~\ref{explicit_calculation_bx}. 

At this point of the calculation, $b_x$ can be written as the sum of $b^{'}_x$ with $b^{''}_x$ and a term proportional to $\Sigma_{F}(k_F,0)$ that comes from the
separation of terms in the sum given in Eq.~\ref{b_xc_sum}:
\begin{equation}
b_x = b^{'}_x+b^{''}_x-\frac{m^2\Sigma^{\lambda}_F(k_F,0)}{12 \pi^3 k_F^3},
\label{b_x_sum}    
\end{equation}
where the values that we obtain for each coefficient and the self-energy term in the $\lambda_c \rightarrow 0$ limit are dependent on the $k_F$-dependent function $\beta(k_F)$:
\begin{eqnarray}
b^{'(\lambda_c \rightarrow 0)}_x = -\frac{5m^2e^2}{36 \pi^3 k_F^2}\left[1-k_F\frac{\partial}{\partial k_F}\ln(\beta(k_F))\right],
\label{b'_x_final} 
\end{eqnarray}
\begin{eqnarray}
b^{''(\lambda_c \rightarrow 0)}_x &=& \frac{m^2 e^2}{8 \pi^3 k_F^2}
\label{b''_x_final}  \\  
\Sigma^{\lambda_c \rightarrow 0}_F(k_F,0) &=& -\frac{e^2k_F}{\pi}.
\label{fock_kf_value_lambda}    
\end{eqnarray}
Therefore, the net result for the $q^2$ coefficient $b_x$ given by Eq.~\ref{b_x_sum} when we take the limit $\lambda_c\rightarrow 0$  limit at the end of the calculation, is given as:
\begin{equation}
b^{(\lambda_c \rightarrow 0)}_x = \frac{5 m^2 e^2}{72 \pi^3 k_F^2}\left[1+2k_F\frac{\partial}{\partial k_F}\ln(\beta(k_F))\right],
\label{b_x_final}    
\end{equation}
where the superscript indicates that the constant term of the regulator $\lambda_c$ was kept at every calculation step and it is set to zero only at the end of the
integration.

If we consider the regulator to be a constant independent of $k_F$, the above value of the coefficient $b^{(\lambda_c \rightarrow 0)}_x$ agrees with that obtained by Sham:
\begin{equation}
b^{S}_x = \frac{5 e^2 m^2}{72\pi^3k_F^2},
\label{b_x_Sham}    
\end{equation}
where the superscript stands for Sham.

We also explored what happens when we use the same choice of regulator that Kleinman-Tamura (KT) used in Ref.~\cite{Kleinman-Tamura} for the calculation of the $q^2$ coefficients $b'_c$ and $b^{''}_c$.  KT keep the ratio $\lambda'=\lambda/k_F$  fixed, which implies that the $k_F$-dependent part of the regulator $\beta(k_F)$ is proportional to $k_F$.
Using such expression of $\beta(k_F)$ in Eq.~\ref{b_x_final}, we obtain that $b_x$ becomes three times Sham's value, i.e., 
\begin{equation}
b^{KT}_{x} = \frac{5 e^2m^2}{24 \pi^3 k_F^2} = 3 b^S_x,
\label{b_x_KT}    
\end{equation}
where the superscript means that the KT's choice of the regulator was used in the calculation of $b_x$. Note that this is not a value that was worked out by KT since their choice of regulator for the potential was only applied to the contributions of the $b_c$ coefficient.

Using a regulator of the form $\beta(k_F)=k_F^{-3/10}$, we obtain
\begin{equation}
b^{*}_{x} = \frac{e^2m^2}{36\pi^3k_F^2},
\label{b_x_GT_KA}    
\end{equation}
where the * is used to note that  
this value has been reported by Geldart-Taylor (GT), Antoniewicz-Kleinman (AK), Kleinman-Lee, and Engel-Vosko (EV)\cite{Geldart-Taylor1970,Kleinman-Antoniewicz,Kleinman-Lee1988,Engel-Vosko1990}.
These authors claim that their results fully converge by using
${\tilde V}_0(q)$ for the Coulomb interaction, i.e.,  performing the necessary integrations with $\lambda=0$. However, what they reported was the Cauchy principal value of their nested integrals; some of their calculations were numerical, and a proof of convergence of these integrals was not given in previous works. This implies that the work reported in Refs.~\cite{Geldart-Taylor1970,Kleinman-Antoniewicz,Kleinman-Lee1988,Engel-Vosko1990} is equivalent to using an effective regulator at the integral level over the momentum variables.

The reason for this contradiction is that the necessary integrals should be redefined by using a regulator; otherwise, they do not have a definite value.
If we were to substitute $\beta(k_F)=0$ in Eqs.~\ref{b'_x_final}--\ref{b''_x_final} it would lead to a divergent $b_x$. 
The expressions indeed diverge, but this conclusion can not be rigorously reached by such a simple substitution.
The calculation for this case, $\lambda=0$, is explained in detail in Subsection~\ref{setting_lambdac_0_explanation}.

Next, we focus our discussion on the results that we have obtained for the contribution to $b_c$ by using our choice of the regulator used in Eq.~\ref{V_generic}. Where $b_c$, in our notation, is given by the sum of the three coefficients:
\begin{equation}
b_c = b^{'}_c+b^{''}_c+b^{'''}_{xc}.
\label{b_c}
\end{equation}

First, the calculations of the $b_c$-primed coefficients require the calculation of the functions $I_i(k^{\mu})$ (for $i=1,2,3$), which are presented in Appendix~\ref{explicit_calculation_bc}. One way we can keep track of these integrals is to first write the higher-order partial derivatives of the non-interacting Green's function with respect to $\mu_0$ in terms of partial derivatives of the Green's functions with respect to the frequency variable.
When this is done, there are Dirac delta functions that must be tracked, as explained in Appendix~\ref{explicit_calculation_bc}. Also, the integrals from Eqs.~\ref{b'_c} and \ref{b''_c} are over these functions $I_i(k^{\mu})$, which are multiplied by an even function of the frequency variable such as the renormalized potential
within RPA, $\tilde{V}_e(k^{\mu})$. Because of this, it is convenient to rewrite the functions $I_{i}(k^{\mu})$ in terms of an even and odd parts of these integrals with respect to the frequency variable, where only the even part of the integral yields a non-zero value when carrying out the integrals in Eq.~\ref{b'_c} and Eq.~\ref{b''_c}. These even integrals
are labeled as $I^{S}_i(k^{\mu})$ (for $i=1,2,3$) and are given by Eqs.~\ref{Is_1}, \ref{Is_2} and \ref{Is_3}.

The next step involves the separation of the parts of the partial derivative terms of the functions $\Pi_0(k^{\mu})$ and $J(k^{\mu})$ that yield terms that contain $\delta(k^0)$. By doing this, one can prove that there is a cancellation between all the terms which contain Dirac deltas in Eqs.~\ref{Is_1},~\ref{Is_2},~\ref{Is_31} and~\ref{Is_32}. The surviving terms come from the real part of the derivatives of these functions and the frequency integral can be handled now (due to the cancellation of $\delta(k^0)$ terms) by mapping it into an integral along the imaginary line; we have set $k^0=i\nu$.

After this step, we rescale the momentum and frequency variables into dimensionless variables $\vec{q}=k_F \vec{q}\,'$ and $\nu = k_F^2 \nu'/m$. Next, we change  variables
as $\nu'=q'y$ and  $q'=2x$. A similar treatment is also done in the integral expression of $b^{'''}_{xc}$, as explained in Appendix~\ref{b'''_xc_calculations}. After such steps, we obtained that the two-dimensional integral expressions of the primed terms contributing to the $b_c$ coefficient are given by the following terms:
\begin{eqnarray}
  &b^{'}_c&= \frac{\partial}{\partial \mu_0} \int^{\infty}_0dx \int^{\infty}_0 dy Z_{\lambda'}(x,y) C(x,y),\label{b'_c_2d}\\
&Z_{\lambda'}(x,y)&\equiv -{{2e^2m}\over {\pi^4}} \frac{x}{x^2+{\frac{\lambda'^2}{4}}} \left(\frac{x^2+\frac{\lambda'^2}{4}}{\overline{\epsilon_{\lambda}}(x,y)}-1 \right),\label{Z_function}\\
&b^{''}_c& = \frac{m }{3 k_F^2}\int_0^{\infty} dx \int_{0}^{\infty}dy Z_{\lambda'}(x,y)I'_2(x,y),\label{b''_c_2d}\\
  &b^{'''}_{xc}& = \frac{m^3 e^4}{96 \pi^5 k_F^3}\int^{\infty}_0 \frac{dx}{x}\int_0^{\infty} dy \left(\frac{g(x,y)}{\overline{\epsilon_{\lambda}}(x,y)}\right)^2\nonumber \\
&\times&  \left[1-\frac{\alpha r_s M_9(x,y)}{2 \pi \overline{\epsilon_{\lambda}}(x,y)}-\frac{(\alpha r_s M_{10}(x,y))^2}{8 \pi^2 (\overline{\epsilon_{\lambda}}(x,y))^2} \right],\label{b'''_c_2d}        
\end{eqnarray}
where the symbol $\lambda'=\lambda_c\beta(k_F)/k_F$ is used as a reminder that the function is dependent on $\lambda$, the functions $C(x,y)$, $I'_2(x,y)$ are given in Appendix~\ref{explicit_calculation_bc}, while $M_9(x,y)$ and $M_{10}(x,y)$ are given in Appendix~\ref{b'''_xc_calculations}. The function $\overline{\epsilon_{\lambda}}(x,y)$ comes from the product of the denominator term of ${\tilde V}(k)$ used, times the dielectric function by factorizing a factor of $4k_F^2$.   We obtained:
\begin{equation}
\overline{\epsilon_{\lambda}}(x,y) = x^2+\frac{\lambda'^2}{4}+\frac{\alpha r_s Q(x,y)}{4 \pi},
\label{epsilon_bar}    
\end{equation}
where $Q(x,y)$ is the Lindhard function for imaginary frequency, given by the following expression:
\begin{eqnarray}
  Q(x,y) &= &2+\left( \frac{1+y^2-x^2}{2x} \right)g(x,y)-g_2(x,y),
\label{Q function}    
\end{eqnarray}
where $g(x,y)$ and $g_2(x,y)$ are given by the expressions below:
\begin{eqnarray}
g(x,y) &=& \ln \left|\frac{(1+x)^2+y^2}{(1-x)^2+y^2} \right|,\label{g(x,y)}\\[6pt]
g_2(x,y) &=& 2y\left[\tan^{\!-1}\!\left(\frac{1+x}{y}\right)+\tan^{\!-1}\!\left(\frac{1-x}{y}\right)\right].\label{g2}
\end{eqnarray}

We performed numerical integrations to calculate all of the $b_c$-primed coefficients by using Eqs.~\ref{b'_c_2d}, \ref{b''_c_2d}, and \ref{b'''_c_2d}, by approaching $\lambda_c \rightarrow 0$. The numerical results yield the same values that Ma-Brueckner obtained in Ref.~\cite{Ma-Brueckner}. The integral expression of $b^{'}_c$ is the only term that is sensitive to the regulator because of the partial derivative with respect to $\mu_0$ acting over the two-dimensional integral. Also, the usage of the regulator guarantees the existence of the integral in the expressions for $b^{'}_c$ and $b^{''}_c$, as explained below in this ection. The expressions of $b^{'''}_{xc}$ give the same value as that reported by MB for $b^{'''}$ in Ref.~\cite{Ma-Brueckner}, which indicates that the regulator does not play a significant role in the integral given in Eq.~\ref{b'''_c_2d}. The  numerical values of the coefficients are given by:
\begin{eqnarray}
b^{'(\lambda_c \rightarrow 0)}_c = &\frac{5 e^2 m^2}{72\pi^3 k_F^2}&\left[1-2k_F\frac{\partial}{\partial k_F}\ln(\beta(k_F)) \right],\label{b'_c_final_result}\\
&b^{''(\lambda_c \rightarrow 0)}_c& = 0.82872\frac{e^2 m^2 }{(2 \pi)^3 k_F^2},\label{b''_c_final_result}\\    
&b^{'''(\lambda_c \rightarrow 0)}_{xc}& =  0.59136 \frac{e^2 m^2}{(2 \pi)^3 k_F^2},\label{b'''_c_final_result}    
\end{eqnarray}
where the superscript indicates that the regulator $\lambda_c$ is treated as a constant and then is taken in the $\lambda_c \rightarrow 0$ limit at the very end of the calculation. These final expressions, obtained in this work for the leading terms in $r_s$ of both, $b_c$-primed and $b^{'''}_{xc}$ coefficients, were found by focusing on the region of integration $x \in (0,1/2)$. Such region of integration corresponds to the wavevector integration done within $k \in (0,k_F)$.

Sham's regulator, i.e., taking $\beta(k_F)$ as a constant, is consistent with GR's\cite{Geldart-Rasolt} insertion of an infrared (IR) cutoff in the integral expression of $b^{'}_c$,
which in their work is labeled as $b^{'}$. GR's value of the leading term of $b^{'}$ agrees with MB's value reported in Ref.~\cite{Ma-Brueckner}. They claimed in their work that the IR cutoff is necessary for the integral to exist. However, in MB's calculation of $b'$ there is no explicit use of an IR regulator. The fact that they found a finite value for this coefficient suggests that a regulator
was implicitly used, which is consistent with what GR used; this is further discussed in the next subsection.
Our calculation of $b^{'}_c$ agrees with GR's value, when we set $\beta(k_F)$ to a constant and it is given by:
\begin{equation}
b^{'(MB/GR)}_c =  \frac{5e^2m^2}{72\pi^3k_F^2},
\label{b'_c_MB_GR}    
\end{equation}
where the superscript of this coefficient stands for Ma-Brueckner and Geldart-Rasolt.

Our expression of $b^{'}_c$, when setting $\beta(k_F)=k_F$, yields the value that is consistent with the calculation done by KT for the sum of all $b$-primed coefficients found in Ref.~\cite{Kleinman-Tamura}, where the ratio $\lambda'=\lambda/k_F$ is kept fixed. Such term, despite the fact that the expression of the $b'$ coefficient was not explicitly
reported by KT, has the following expression:
\begin{equation}
b^{'(KT)}_c =  -\frac{5e^2m^2}{72\pi^3k_F^2}.
\label{b'_c_MB_GR}    
\end{equation}

By using the regularized-Coulomb potential $\tilde{V}(k)$, given by Eq.~\ref{V_generic}, the leading term of the $b^{''}_c$ coefficient that we obtain agrees with MB's reported value~\cite{Ma-Brueckner}, where in their work is labeled as $b^{''}$. Even if GR were aware that the convergence of the integrals associated with $b^{'}$ relied on using an IR regulator, they did not applied this to their two-dimensional expression of the $b^{''}$ coefficient and claimed that they agreed with MB's value for such term. Such integral expression is not convergent, as it is explained later by Kleinman-Tamura (KT) in Ref.~\cite{Kleinman-Tamura}, where a regulator is required for the integral to converge. We explain more details about this later in this work.

The reason why our calculated value for $b^{''}_c$ is independent of $\beta(k_F)$, and not for $b^{'}_c$ is because of a partial derivative with respect to $\mu_0$ that acts on the two-dimensional integral expression of $b'_c$, given by Eq.~\ref{b'_c_2d}. This also explains why the calculation of $b^{'''}_c$ yields a value that is independent of $\beta(k_F)$, which in fact agrees with MB's reported value of $b^{'''}$~\cite{Ma-Brueckner}. We point out that GR claims to agree with MB's value of $b^{'''}$, but in their expression their is a typographical error where they are missing a global minus sign in their expression given in Ref.~\cite{Geldart-Rasolt}.  

The coefficient $b_x$ that we obtained in Eq.~\ref{b_x_final} is dependent on the choice of the expression used for $\beta(k_F)$. Such coefficient is mainly attributed as the main contribution to the exchange part of the gradient of the density squared coefficient of the functional. The fact that this coefficient is regulator dependent, it implies
that $b_x$ has no well-defined value. Similarly, the sum of both $b_c$ primed and $b^{'''}_{xc}$ is dependent on $\beta(k_F)$, as seen in in Eq.~\ref{b'_c_final_result}.
This also implies that $b_c$ has a well-defined value. 

In contrast, the total coefficient of the $q^2$ term from the proper-polarization function $\Pi^{xc}(q,0)$, which is necessary to extract the leading contribution in $r_s$ of the total coefficient of the gradient of the density square, is given by:
\begin{equation}
b^{\lambda_c \to 0}_{xc} = 0.316399 \frac{e^2 m^2 }{\pi^3 k_F^2},
\label{b_xc_lambda_final}    
\end{equation}
and it is independent of the choice of the regulator. By using this result in Eq.~\ref{Bxc}, we find that the correct constraint to be used in a density functional within GGA that captures the correct $q \to 0$ and $s \to 0$ limits, is given by the following term:
\begin{equation}
B_{xc} = \frac{0.029116}{r_s^4}.
\label{Bxc_final}
\end{equation} 

\subsection{Setting $\lambda=0$ before calculating the integrals}
\label{setting_lambdac_0_explanation}
In this subsection, we set $\tilde{V}(k)$ as the commonly-known ``bare-Coulomb interaction'' $\tilde{V}_0(k)$ and discuss the problems that arise due to its usage in the integral expressions for the $b_x$ and $b_c$-primed coefficients.

The expressions associated with the $b$ ``primed'' contributions to the coefficients $b_x$ and $b_c$ are obtained by using the integral expressions given in Eq.~\ref{b'_c},\ref{b''_c}, \ref{b'_x_3d}, and \ref{b''_x_3d} and setting $\lambda=0$ before applying any operation or integrals. This means that any term we previously had for the expressions of these contributions to the coefficients can drop their subindex $\lambda$. The Fock self-energy is now obtained from the following expression:
\begin{equation}
\Sigma_F(p) = -\frac{e^2 k_F}{\pi} \left[1+\left(\frac{k_F^2-p^2}{2pk_F} \right) \ln{\left|\frac{k_F+p}{k_F-p} \right|} \right].
\label{self_energy_fock_lambdazero}    
\end{equation}

This term can be used in the expressions we had for the contribution to $b_x$ 
deriving from the diagrams illustrated in Fig.\ref{GA_Fock} given by Eq.~\ref{b'_x_3d_2} and Eq.~\ref{b''_x_3d_2}.
However, the integral of the term involving the second partial derivative  with respect to the energy dispersion and momentum $p$ of the function $p \Sigma_F(p)$ is problematic, since the integrated expression as a factor a Dirac
delta-function $\delta(\mu_0-\epsilon^0_p)$, making the integral diverge. In the previous works done in Refs.~\cite{Geldart-Taylor1970,Kleinman-Antoniewicz,Kleinman-Lee1988,Engel-Vosko1990} a proof that the integral expressions used for the calculation of the exchange coefficient $B_x$ converge, was not given.
One can obtain a finite value by taking the Cauchy principal part of
an otherwise ill-defined integral; this, however,  is equivalent to also
using a regulator.
Later, Svendsen and von-Barth\cite{Svendsen1995} (SvB) revisited the calculation of $b_x$ and concluded that the result reported in Refs.~\cite{Geldart-Taylor1970,Kleinman-Antoniewicz,Kleinman-Lee1988,Engel-Vosko1990} is 
correct. They concluded this because they obtained a finite value of a more compact two-dimensional nested integrals that contributes to the $b_x$ coefficient by setting the regulator to zero before integrating. The expressions for the
two-dimensional integrals, given by SvB in the case where the regulator is set to zero before doing the calculation, are given below:
\begin{equation}
L_1 = \int^1_{-1}\!\!dx \!\int^1_{-1}\!\!dy \frac{(x-y)^2(3x^2+3y^2+7xy-1)}{xy|x-y|},
\label{L1}    
\end{equation}
\begin{equation}
L_2 = \int^1_{-1}\!\!dx \!\int^1_{-1}\!\!dy \frac{2(3xy-1)}{|x-y|}.
\label{L2}    
\end{equation}
Next, we carefully focus on the regions of integration over
$x$ and $y$ near the values of $x$ and $y$ where
the denominator vanishes.  After a change of variables, $y=u+x$,
and limiting the integration over $u$ from $-\eta$ to $\eta$ we obtain
the following contribution to $L_1$:
\begin{equation}
L_1 = \frac{48}{3}+2\int^{\eta}_{-\eta} \frac{du}{u},
\label{L1_problematic}    
\end{equation}
where $\eta \to 0^+$; the first term is the Cauchy principal value of $L_1$ and the second term of the integral is well-known for not converging.
For $L_2$ we obtain:
\begin{eqnarray}
L_2 =-\frac{32}{3}&-&\int^1_{-1} dx 2(3x^2-1)\int^{\eta}_{-\eta} \frac{du}{|u|}\nonumber\\
      &-4&\ln|\eta|\int^1_{-1} dx (3x^2-1),
\label{L2_problematic}    
\end{eqnarray}
where the first term is the principal value of $L_2$, while the second and the third terms do not exist for the following reasons.
The second term yields zero when performing the integration on the $x$ variable first, followed by the integration on the variable $u$ afterwards.
Instead, if the integration on $u$ is performed first, we are faced
with a non-existing integral. The third term has a factor of $\ln(\eta)$
multiplying by an integral that yields zero. This product has no
well-defined value in the limit of
$\eta \to 0$.
This shows that these two-dimensional nested integrals do not exist. Bypassing this issue is equivalent to choosing an effective regulator.

Now, we discuss what happens with the integral expressions of the
$b$``primed'' contributions to $b_c$  and $b^{'''}_{xc}$, which are what Ma-Brueckner originally obtained in Ref.~\cite{Ma-Brueckner}. We can obtain the expressions given by Eqs.~\ref{b'_c_2d}, \ref{b''_c_2d}, and \ref{b'''_c_2d} by setting $\lambda=0$ before calculating the integrals. The two expressions that we would obtain, in the $\lambda=0$ case, for the coefficients $b^{'}_c$ and $b^{''}_c$ are integrals that are ill-defined.

In the case of $b^{'}_c$, the partial derivative with respect to $\mu_0$, is acting on an integral that diverges, as we explain in more detail in appendix \ref{explicit_calculation_bc}. The problematic part of the integral is in the region of integration in the proximity of $x \rightarrow 0$, since it diverges logarithmically. One could assume that, given that this integral is independent of $k_F$, the partial derivative with respect to $\mu_0$ acting on this integral yields zero. If that is the case, the partial derivative only acts on the function $\overline{\epsilon}(x,y)$ in Eq.~\ref{b'_c_2d} (with $\lambda=0$). This yields the following integral expression:
\begin{equation}
b^{'(\lambda=0)}_c = -\frac{e^4 m^3}{2 \pi^5 k_F^3} \int^{\infty}_0 dx x\int^{\infty}_0 dy\frac{Q(x,y)C(x,y)}{\left(\overline{\epsilon}(x,y) \right)^2},
\label{b'_c_2d_lambda=0}    
\end{equation}
where $\overline{\epsilon}(x,y)=x^2+\alpha r_s Q(x,y)/4\pi$. Even though the expression given by Eq.~\ref{b'_c_2d_lambda=0} is an integral that converges, this approach is incorrect, since the partial derivative acts on an integral that
does not converge in the first place; therefore, the value of $b^{'(\lambda=0)}$ does not exist if we set $\lambda=0$ at the integrand level.

We recall that the $q^2$ coefficients $b'_x$, $b^{''}_x$, $b'_c$ and $b^{''}_c$ come from performing a Taylor expansion in small-wavevector $q$ of every contribution to  $\Pi^{xc}(q,0)$, illustrated in Fig.\ref{GA_Fock} and Fig.\ref{Pi_c}. We point out that the problematic coefficients $b^{'}_x$ and $b^{''}_x$ at $\lambda=0$ shares common terms that are third-order partial derivatives with respect to $\mu_0$ acting on the non-interacting Green's function $G^0(p^{\mu})$  in expressions Eq.~\ref{b'_x} and Eq.~\ref{b''_x}. The coefficients $b^{'}_c$ and $b^{''}_c$ also share a common term at their respective integrand level, which depends also on a third-order partial derivative of $G^0(p^{\mu})$ with respect to $\mu_0$, which are contained in the expressions of $I_2(k^{\mu})$ and $I_3(k^{\mu})$, given by Eq.~\ref{I2} and Eq.~\ref{I3}. These third-order derivative terms are the sources of the convergence problem of these coefficients when their respective integrals are calculated separately. This means that if $b'_x$($b^{'}_c$) do not converge, then the other coefficients $b^{''}_x$($b^{''}_c$) cannot exist either when no regulator is used. Now, we proceed to point out the problematic term in the integral expression of $b^{''}_c$ in the $\lambda=0$ case.   

The integral associated with $b^{''}_c$ is obtained from Eq.~\ref{b''_c_2d} by setting $\lambda=0$ at the integrand level. The issue here is that the two-dimensional integral does not exist, as it was previously pointed out by Kleinman\cite{Kleinman-Tamura}. The dominant term of the integral is given within the region of integration $x\in(0,1/2)$, which is equivalent to  the region given by: $k\in(0,k_F)$, where $k$ is the magnitude of the momentum variable. This is the same region of integration that Ma-Brueckner used in Ref.~\cite{Ma-Brueckner} to obtain an analytical calculation for $b^{''}_c$. Within this region of integration, the problematic part of the integral from Eq.~\ref{b''_c_2d} that is given in Appendix~\ref{explicit_calculation_bc} by Eq.~\ref{b''_lambda_0_2}. In this expression, since the region of integration is in $x \in (0,\frac{1}{2})$, one keeps the lowest order of a Taylor expansion in variable $x$ for the function $I'_2(x,y)$ given by Eq.~\ref{I'_2}. The problem with Eq.~\ref{b''_lambda_0_2} is that if we first calculate the integral over the $y$ variable, it yields zero, however, if we first integrate over the $x$ variable, then it diverges logarithmically. Therefore, this two-dimensional integral does not exist due to this ambiguity. If we had taken $\lambda$ to be a finite non-zero value, regardless of the choice of the $\beta(k_F)$ function, then this indeterminate case is resolved; this makes the integral over the $y$ variable yield zero and it becomes finite when we remove the regulator after the integration.

Lastly, the two-dimensional integral given in Eq.~\ref{b'''_c_2d} has no convergence issue by removing the $\lambda$ regulator and yields numerically the same value that Ma-Brueckner reported for the $b^{'''}$ coefficient in \cite{Ma-Brueckner}. This is the only coefficient of $q^2$ that has no convergence issue and yields the same value regardless of whether a regulator is used in the potential expression $\tilde{V}(k)$.

We found previously, that the leading terms of the $q^2$ coefficients of $\Pi^{xc}(q,0)$, $b^{(\lambda_c \rightarrow 0)}_x$ and $b^{'(\lambda_c\rightarrow 0)}_{c}$, have a dependence on the $\beta(k_F)$ function, which exposes the regulator dependence of the exchange $B_x$ and correlation $B_c$ coefficients of the density functional. However, when we take the sum of the two terms, we found a cancellation among the separate terms converging to a unique value for $b^{\lambda_c\rightarrow 0}_{xc}$, which means that the combined coefficient is regulator-independent. However, it is not possible to determine a value when we set $\lambda=0$
before the integrations, because each of the contributions to the $b_{xc}$ coefficient does not exist; this is the reason for introducing the regulator.
This divergence issue exposes the illness of the integrals associated to the $q^2$ coefficients $b_x$ and $b_c$ when using no regulator in the Coulomb potential, and that a regularization procedure must be used to calculate them separately.
However, it is not surprising the fact that $b_{xc}$ is regulator-independent. This is because the high-dimensional integral expressions for $\Pi_{xc}(q,0)$ involve
not the bare but the RPA-renormalized potential $\tilde{V}_e(k^{\mu})$; the Lindhard function acts as a regulator making the integrals in Eqs.~\ref{pixc_1},~\ref{pixc_2} and~\ref{pixc_31} convergent.

\section{Summary of our results}
\label{summary}

We have found that by carrying out a Taylor expansion in the $q \to 0$ limit of the proper-polarization function $\Pi^{xc}(q,0)$, we can keep track of the high-dimensional integrals that contribute to the net coefficient $b_{xc}$ of all the terms of order $q^2$. This was done by using the regularized potential at the integrand level, where we have used a general $k_F$-dependent regulator $\lambda = \lambda_c \beta(k_F)$. We found that the calculation of the contribution
$b_x$ to this coefficient from the exchange diagrams is written as a sum of three terms, which correspond to three different contributions of the proper-polarization function illustrated in Fig.~\ref{GA_Fock}. One diagram corresponds to the vertex bubble, which includes the electron-hole interaction, and in one of the other two diagrams the Fock self-energy is inserted in one of the two fermionic propagator lines of the bubble. The contribution $b_c$ to the coefficient $b_{xc}$ from correlations is obtained from the higher-order  proper-polarization diagrams illustrated in Fig.~\ref{Pi_c} by collecting the respective same order contributions, in powers of the Wigner-Seitz radius $r_s$, to those contributing to $b_x$. In this work we focused only up to order $r^2_s$ for such coefficient. Our calculations reproduce the expressions for both $b_{x}$ and $b_c$ that have been obtained previously by several authors.

Svendsen and von-Barth\cite{Svendsen1995} have also calculated analytically the contribution of the exchange in the gradient expansion, i.e., the $b_x q^2$, as $B_x s^2$. They claim that the two nested integrals that contribute to $B_x$ do not commute and that by performing the $x$ and $y$ variable integrals given by Eq.~\ref{L1_problematic} and Eq.~\ref{L2_problematic}, and then taking the $\lambda \to 0$ limit yields the value reported by Sham\cite{Sham1971}. However, they claimed that the correct result of $B_x$ is when $\lambda$ is set to zero before the integration is carried out and this corresponds to the value reported in Refs.~\cite{Geldart-Taylor1970,Kleinman-Antoniewicz,Kleinman-Lee1988,Engel-Vosko1990}. This value is $10/7$ times the size of the value obtained by Sham\cite{Sham1971}. This result seems to have been accepted by many authors. However, setting the regulator to zero at the integrand level is incorrect, since we obtained that the integrals that contribute to $b_x$ do not have a well-defined value.
In this work,  we demonstrated that
the results reported in Ref.~\cite{Svendsen1995} correspond to the Cauchy principal value of their integrals ignoring the fact that
some integrals are ill-defined in the region of integration where the denominator of the integrand vanishes; as we show this leads to ill-defined
values for these integrals.

The coefficient $B_c$  of the $B_c s^2$ term which  corresponds to the
$b_c q^2$ term  and which includes the contributions  of the so-called
correlation diagrams  illustrated in Fig.~\ref{Pi_c}, was  defined and
calculated by Ma and Brueckner\cite{Ma-Brueckner} (MB) in the $r_s \to
0$,  $s \to  0$ and  $q \to  0$  limits. In  order to  find $B_c$,  MB
separated the  exchange $B_x$ and the  correlation $B_c$ contributions
to the coefficient $B_{xc}$. Their work focused on finding the leading
part of $B_{c}$ by calculating  the terms of order $q^2$ corresponding
to the above-mentioned five proper-polarization  functions within the
large-density and long-wavelength limit. Those five terms (illustrated
in  Fig.~\ref{Pi_c}) arise  within a  resummation of  the polarization
bubbles, which corresponds to renormalizing  the Coulomb  interaction
within the random phase approximation (RPA).  Their  inclusion  is
consistent    with     our    previously    reported\cite{benites2024}
reorganized-sum of Goldstone diagrams in  terms of  the renormalized
interaction line within the  RPA.  MB  expressed  the  sum  of  the
contributions of  these five diagrams as  a sum of three  terms, i.e.,
$b_c =  b^{'}+b^{''}+b^{'''}$, where  each coefficient  is given  by a
four-dimensional integral\cite{Ma-Brueckner}. Here,  we point out that
the analytical expressions of these three contributions require a
regulator at the Coulomb potential expression for the integrals to exist, 
which should be set to zero after the integration is done. 
We have shown that the integral expression derived by MB for $b^{'}$
does not lead to the results they obtained. This is because
the  MB  expression has a partial derivative with respect to $\mu_0$  acting on a non-converging integral; this was
previously pointed out by Geldart-Rasolt (GR)\cite{Geldart-Rasolt}. 
The fact that MB found a finite value for this coefficient implies
that a regulator was implicitly used in the integrand. Nevertheless, 
if we use the regulator as Sham did in Ref.~\cite{Sham1971}, carry 
out the integrations, and then set the regulator to zero, we obtain MB's
 reported value for $b'$. However, if we use a different regulator, namely, 
 $\beta(k_F)=k_F$ and   then  set   $\lambda_c$  to  zero   after  the integration, 
 we obtain  the same value but with opposite  sign to that obtained  by 
 Ma-Brueckner\cite{Ma-Brueckner}. This means that $b_c$ is
regulator dependent,  which explains the different results obtained
by different authors. In addition, this regulator dependence of these
separate contributions implies that they do not have a  well-defined
value. On the contrary, we find that the physically well-defined contribution,
which consists of the sum of the exchange and correlation contributions,
is regulator independent.

\setlength{\tabcolsep}{4pt}
\def\arraystretch{1.5}
\begin{table}[ht]
    \centering
    \begin{tabular}{|c|c|c|c|c|c|c}
        \hline
         & Case 1 & Case 2  & Case 3 & Case 4 & Case 5 \\
        \hline
        $b_x$ & $5/72$ & $5/24$ & $5/36$ & $1/36$ &  Diverges \\
        \hline
        $b^{'}_c$ & $5/72$ & $-5/72$ & $0$ & $1/9$ & Do~not~exist\\
        \hline
        $b^{''}_c$ & $0.10359$ & $0.10359$ & $0.10359$ & $0.10359$ & Do~not~exist\\
        \hline
        $b^{'''}_{xc}$ & $0.07392$ & $0.07392$ & $0.07392$ & $0.07392$ & $0.07392$ \\
        \hline
        $b_{xc}$ & $0.3164$ & $0.3164$ & $0.3164$ & $0.3164$ &  undetermined\\
        \hline
    \end{tabular}
    \caption{Leading coefficients of the $q^2$ terms corresponding to $\Pi^{xc}(q,0)$ found from the sum of the diagrams illustrated in Fig.\ref{GA_Fock} and Fig.\ref{Pi_c}. The values in the table are given in units of a common factor of
      $\tau = e^2m^2/\pi^3k_F^2.$}
    \label{table_exchange_correlation}
\end{table} 

Here, we also find that the integral expression for $b^{''}$ does not exist, which has already been previously explained by Kleinman and Tamura\cite{Kleinman-Tamura} (KT). One of the terms in the integral expression of $b^{''}$ over the momentum tends to infinity, while the integral over the frequency yields zero. This indefinite case for $b^{''}$ is resolved by using a regulator in the interaction potential and the result agrees with the value that Ma-Brueckner claims to obtain; this occurs despite the fact that the value of the integral of their original expression is not well-defined. This indicates that their analytical work for finding the dominant term for $b^{''}$ in the large density limit is consistent with implicitly using a regulator.

The coefficient $b^{'''}$, however, is the only term that is not regulator dependent, and it yields the same value as reported by MB, even if we set the regulator to zero before calculating the integrals. All of these coefficients were obtained by keeping only their dominant contribution of the expansion in powers of $r_s$, which is of the $r_s^2$ order. We point out that the values of $b^{''}$ and $b^{'''}$ are regulator independent, and we obtain the same results to those reported by MB in Ref.~\cite{Ma-Brueckner}. It is only the $b^{'}$ coefficient that is
regulator dependent. This also implies that $b_c$ does not have a unique value.

GR found a general two-dimensional integral expression (for any order of $r_s$)
for every contribution to $b_c$ in the long-wavelength limit, and they claim that their expressions reproduce the MB results in the large-density limit. They
introduced an infrared (IR) cutoff as the lower bound of the momentum integral in the expression of $b'$ in order for the integral to converge.
The inclusion of the cutoff gives rise to an extra contribution to the value of $b^{'}$ when taking the partial derivative with respect to $\mu_0$, agreeing with MB's value of the $b^{'}$ coefficient. The inclusion of such infrared regulator makes their calculation equivalent to our treatment of the regulator $\lambda$ when it is taken to be a constant.
Even though GR seem to have appreciated the importance of including an IR cutoff, which is  equivalent to our usage of a regulator, they did not include a
cutoff or regulator for the integral expression of $b^{''}$, which is needed
to take care of the singularity in the $q \to 0$ limit.

In addition, we notice a mistake in the simplified expressions of GR's four-dimensional integrals
of all of the $b$-primed coefficients. Namely, we have found additional higher-order terms in the expansion in powers of $r_s$; these arise from terms that
contain the Dirac delta functions in the integrand.
The combination of such terms cancels each other out, implying that the other higher-order $r_s$ terms reported by GR are
incorrect. We calculated all of these $b$-primed coefficients mostly
analytically using our regularization of the Coulomb interaction, and verified that the analytical results agree with the numerical results
of these four-dimensional integrals in the $r_s \to 0$ limit
and the ($\lambda_c \to 0$) limit.
These numerical and analytical results illustrate that the extrapolated value of
 $b^{'}$  is regulator dependent.
 We also point out a typographical error in GR's two-dimensional integral expression for $b^{'''}$ where they have an extra global minus sign factor.

 Table~\ref{table_exchange_correlation} contains the leading terms of $b_x$ and the $b_c$-primed coefficients of $q^2$ that we have obtained for a few choices of the regulator $\beta(k_F)$ function used in this work, and are consistent with the regulator treatments used in some previous works\cite{Sham1971,Kleinman-Tamura}. A novel scenario is also considered where the regulator is rescaled in terms of the Thomas-Fermi wavevector $q_{TF}=\sqrt{4k_F/\pi a_0}$, where $a_0$ is the Bohr radius.  Below is a discussion of the results of Table~\ref{table_exchange_correlation}. 

Case 1 is obtained by treating $\beta(k_F)$ as a constant independent of $k_F$, which reproduces the choice used by Sham for $b_x$\cite{Sham1971}, and is the ``equivalent'' to the regulator choice that Geldart-Rasolt made in Ref.~\cite {Geldart-Rasolt}. This case reproduces also the value of the leading term in $r_s$ for the $b^{'}_c$ coefficient found by Ma-Brueckner\cite{Ma-Brueckner}.

Case 2 is obtained by setting $\beta(k_F)=\beta_0 k_F$. This corresponds to treating the ratio $\lambda'=\lambda/k_F$ as a constant and is the same choice of the regulator that Kleinman-Tamura used in Ref.~\cite{Kleinman-Tamura}. This case yields a value of $b'_c$ which is opposite in sign to the result
reported by Ma-Brueckner.  To the best of our knowledge, this value of $b_x$,
which is three times the value reported by Sham, has not been reported before.

Case 3 corresponds to a novel case where the regulator is proportional to the Thomas-Fermi wavevector $q_{TF}$, i.e.,  $\beta(k_F)=q_{TF}$.
This case yields a value for $b_x$ twice the value reported by Sham;
in addition,  it yields zero for the $b'_c$ coefficient!
By using Eq.~\ref{b_x_final}, we can find an equivalent scenario which results
in a zero value for the coefficient $b_x$. Such a scenario is realized when $\beta(k_F)=k_F^{-1/2}$ and results in a value of $b'_c$
which is twice the value reported by MB.

Case 4 is found by answering the question: Which form of the regulator
is needed to reproduce the value of $b_x$ that was claimed to be obtained by Geldart-Taylor~\cite{Geldart-Taylor1970}, Antoniewicz-Kleinman~\cite{Kleinman-Antoniewicz}, Kleinman-Lee~\cite{Kleinman-Lee1988} and Engel-Vosko~\cite{Engel-Vosko1990}? It turns out that setting $\beta(k_F)= k_F^{-3/10}$ reproduces it.
In Table~\ref{table_exchange_correlation}, we have also reported the corresponding values of both $b_c$ primed and $b^{'''}_{xc}$ coefficients.

Case 5 (last column in Table~\ref{table_exchange_correlation}) is the nonsensical scenario where we set $\lambda=0$ after calculating the Fourier transform of the regularized-Coulomb interaction.
In Table~\ref{table_exchange_correlation}, we report as well the $q^2$ coefficient terms of $\Pi^{xc}(q,0)$ that has a convergence issue. These results expose
the fact that the calculation of these coefficients is compromised due to the long-wavelength nature of the Coulomb potential. (See our discussion in the Introduction of the paper).  

In all of these cases, we always obtain the same value for $b^{'''}_{xc}$.  For $b^{'}_{c}$ and $b^{''}_c$, however, the $\lambda=0$ case is the only scenario where these two terms do not converge. Table~\ref{table_exchange_correlation} also helps to illustrate of how we obtain a unique value of the total coefficient $b_{xc}$. This uniqueness in value for $b_{xc}$ is what implies the existence of an exchange-correlation coefficient within the gradient expansion; it also proves that the exchange and correlation contributions to the coefficient $b_{xc}$
do not exist separately. This unique value of $b_{xc}$ was used in order to find the $s^2$ coefficient value of $B_{xc}$ given in Eq.~\ref{Bxc_final}.
\section{Conclusions}
\label{conclusions}
We have revisited the calculations of the coefficient $b_{xc} = b_x + b_c$ of the $b_{xc}q^2$ term in the  ${q} \to 0$ limit, deriving from  the diagrammatic contributions to $\Pi^{xc}(q,0)$ illustrated in Fig.~\ref{GA_Fock} and Fig.~\ref{Pi_c}. We found that the coefficient $b_{x}$ and the $b^{'}$ part of $b_c$ depend on the regulator $\beta(k_F)$;
their dependence on $\beta(k_F)$ is  given by Eq.~\ref{b_x_final} and Eq.~\ref{b'_c_final_result} respectively. We showed that when we  add them together 
their $\beta(k_F)$ dependence cancels out, which implies that the leading $r_s$ contribution to $b_{xc}$ has a unique value given by Eq.~\ref{b_xc_lambda_final}. Therefore,
only the coefficient $B_{xc}$ of the $B_{xc} s^2$ term of the GEA functional has a unique value given by Eq.~\ref{Bxc_final} and not the separate contributions $B_x$ and $B_c$.   

In Table~\ref{table_exchange_correlation}, we also gave our results for the $b_x$ and $b_c$ coefficients by using five different forms of $\beta(k_F)$. Notice that the sum $b_{xc} = b_x + b_c$ of the two
contributions  have a unique value when we keep $\lambda_c\neq 0$ during the integration procedure and we let $\lambda_c \to 0$ after the integrations.

The first case (listed in the second column of Table~\ref{table_exchange_correlation}) is obtained by taking the function $\beta(k_F)$ to be a constant and yield the same value for $b_c$ as reported by Ma and Brueckner (MB)\cite{Ma-Brueckner}, while
the value of $b_x$ agrees with the value reported by Sham\cite{Sham1971}.
While this value is the same as that reported by MB, it can only be obtained with this specific choice of
the regulator, which is equivalent to the infrared cutoff applied by Geldart and Rasolt\cite{Geldart-Rasolt} to calculate the $b'_c$ coefficient. In the second case (third column of Table~\ref{table_exchange_correlation}), we obtained the same value of $b'_c$ reported by Kleinman and Tamura\cite{Kleinman-Tamura}. In this case, we found a value for $b_x$ that was not reported previously, and it is three times the value reported by Sham.
The third case corresponds to a regulator proportional to the Thomas-Fermi wavevector, which yields a vanishing $b'_c$ coefficient, while the fourth case recovers the exchange $b_x$ value found in Refs.~\cite{Geldart-Taylor1970,Kleinman-Antoniewicz,Kleinman-Lee1988,Engel-Vosko1990}. Notice that while the values of $b_x$ and $b_c$ are different
for the three different regulator choices, the sum $b_{xc}=b_x+b_c$ is the same
for all three cases. Therefore, our results explain the controversy between the different reported values of the coefficients $b_x$ and $b_c$ and demonstrate that only the combination is the quantity with a well-defined value.
In the RPA-renormalized Coulomb interaction, the Lindhard function acts as a ``natural" regulator  in every integral expression of $\Pi^{xc}(q,0)$. In the case of the last column,
the integrals that define $b_x$ and $b_c$ do not converge because a regulator is needed.
The  exchange and  correlation contributions are obtained by
making the following separation: ${\tilde V}_0(k)/\epsilon(k,k^0) = {\tilde V}_0(k)+{\tilde V}_0(k)[1/\epsilon(k,k^0)-1]$; as a result, it is no surprise that the exchange contribution, i.e.,
the contribution from $\tilde{V}_0(k)$ alone is pathological as discussed
in the Introduction Section.
Since the net contribution is convergent, because the pathologies of the
bare-Coulomb interaction are screened, the other term in this artificial separation should also carry the same pathology.

Most of the GGA part of the density functionals were constrained to agree with the GEA corrections in the limits of $s \to 0$ and $r_s \to 0$.
In addition, these works separate the exchange and correlation
part of the GGA contribution and several constraints were used independently
for each contribution.  However, in the present work, we have shown that
these coefficients in the limits of $s \to 0$ and $r_s \to 0$ are regulator dependent, which implies that such separation of the functionals between exchange and correlation is an invalid assumption.  

One of the well-known functionals was developed by Perdew, Burke, and Erzenhof (PBE)\cite{PhysRevLett.77.3865}, where they add GGA corrections to the LDA part of the functional provided by Perdew-Wang\cite{PhysRevB.45.13244}. The PBE functional separates the coefficients $B_c$ and  $B_x$. They also use the value of $B_c$ obtained by MB and incorrectly assume a value for $B_x$ so that to cancel the contribution of $B_c$ in the $r_s \to 0$ limit.
Another relatively popular functional  is the PBEsol\cite{PhysRevLett.100.136406-PBEsol}, which uses the value for $B_x$
reported in Refs.~\cite{Geldart-Taylor1970,Kleinman-Antoniewicz,Kleinman-Lee1988,Engel-Vosko1990}, and includes fourth-order derivative terms found within the GEA.

As already discussed, however, in this work he have demonstrated that in order to capture the correct $s \to 0$ and $r_s \to 0$ limits, the constraint to use in any GGA functional is
the value of $B_{xc}$ given by Eq.~\ref{Bxc_final}. In addition, since the separate exchange and correlation contributions do not have a
regulator-independent value, only the combination  $B_{xc}=B_x+B_c$
 should be considered.
Therefore, in the above-mentioned popular GGA functionals, as well
as in other functionals, incorrect constraints have been imposed.

We are currently working in the direction of
developing a GGA functional to be added to the LDA functional presented in
Ref.~\cite{benites2024} that obeys the constraints in the above-mentioned limits discussed. The work presented here should be of more general use, as any future GGA functional should be designed by
taking into account the constraints outlined in the present paper.

\section{Acknowledgments}
  This work was supported by the U.S. National Science Foundation under Grant No. NSF-EPM-2110814.

\appendix
\section{Useful identities involving the non-interacting Green's function $G^0(p^{\mu})$}
\label{Identities Green's function}
The calculation of the coefficients of $q^2$ of the proper-polarization functions requires using certain mathematical properties of the non-interacting Green's function $G^0(p^{\mu})$; in the rest of the appendix we use the four-vector notation to make our expressions more compact. The non-interacting Green's function can be written in compact form as:
\begin{equation}
G^0(p^{\mu}) = \frac{1}{p^0-\epsilon^0_p + \mu_0 +i\eta\text{sgn}(\epsilon^0_p-\mu_0)},
\label{G0}    
\end{equation}
where $\eta \to 0^{+}$ and $\epsilon^0_p=p^2/2m$ is the energy dispersion corresponding to the freely-interacting part of the Hamiltonian and $\mu_0=k_F^2/2m$ is the non-interacting chemical potential of the system. We can separate the real and imaginary parts of the non-interacting Green's function $G^0(p^{\mu})$ as follows:
\begin{equation}
G^0(p^{\mu}) = PV\left(\frac{1}{p^0-\epsilon^0_p+\mu_0}\right)-i\pi\text{sgn}(\epsilon^0_p-\mu_0)\delta(l^0),
\label{G0_2}    
\end{equation}
where $PV$ stands for the principal value and $l^0=p^0-\epsilon^0_p+\mu_0$ and we can evaluate $p^0=\epsilon^0_p-\mu_0$ inside of the sign function by using one of the identities of Dirac's delta. From this point, the non-interacting Green's function can be rewritten as follows:
\begin{equation}
G^0(p^{\mu}) = \frac{1}{p^0-\epsilon^0_p+\mu_0+i\eta\text{sgn}(p^0)},
\label{G^0_3}    
\end{equation}
where this expression is a convenient form of $G^0(p^{\mu})$ if regarded as a function of $\epsilon^0_p$ or $\mu$. The infinitesimal imaginary part of the Green's function from $i\eta$, has a coefficient that depends on specific variables as seen as in Eq.~\ref{G0} or in Eq.~\ref{G^0_3}, which implies that more careful treatment of $G^0(p^{\mu})$ is required in the calculation of the proper-polarization function $\Pi^{xc}(q,0)$. We can see this when calculating the partial derivatives of the Green's function, listed below:
\begin{eqnarray}
\frac{\partial G^0(p^{\mu})}{\partial \mu_0} = - \frac{1}{(p^0-\epsilon^0_p+\mu_0+i\eta \text{sgn}(p^0))^2},\label{partial_mu_G0}\\    
\frac{\partial G^0(p^{\mu})}{\partial p^0} = -\frac{1}{(p^0-\epsilon^0_p+\mu_0+i\eta\text{sgn}(\epsilon^0_p-\mu_0))^2}.\label{partial_p0_G0}    
\end{eqnarray}
The right-hand side expressions given in Eqs.~\ref{partial_mu_G0}--\ref{partial_p0_G0}, even though they look as if they are the square of $G^0(p^{\mu})$, these expressions are in fact not the same as explained by Ma-Brueckner\cite{Ma-Brueckner}. The Green's function is a distribution, but the square of this function cannot be defined without additional specifications. For instance, in Eq.~\ref{partial_mu_G0} there is a pole with degeneracy $2$ only if we regard it as a function of $\mu_0$ or $\epsilon^0_p$. While in Eq.~\ref{partial_p0_G0}, we can consider it to be the square of the Green's function if we regard it as a function of $p^0$. In fact, by taking the partial derivative with respect to $\mu_0$ on Eq.~\ref{G0_2}, one finds the following relationship:
\begin{equation}
\frac{\partial G^0(p^{\mu})}{\partial \mu_0} = \frac{\partial G^0(p^{\mu})}{\partial p^0}+2\pi i\delta(p^0)\delta(\mu_0-\epsilon^0_p),
\label{partial_mu_p0_connection}    
\end{equation}
where by using this expression, we can find the relationship between higher order partial derivatives of $G^0(p^{\mu})$ by taking a partial derivative with respect to $\mu_0$ in Eq.~\ref{partial_mu_p0_connection}. By doing this, we obtain the following expressions for the higher-order partial derivatives of $G^0(p^{\mu})$:
\begin{eqnarray}
&&\frac{\partial^2G^0(p^{\mu})}{\partial \mu^2_0} = \frac{\partial^2 G^0(p^{\mu})}{\partial {p^0}^{2}}+2\pi i \delta^{'}(p^0)\delta(\mu_0-\epsilon^0_p)\nonumber\\
 &&+2\pi i\delta(p^0) \delta^{'}(\mu_0-\epsilon^0_p),\label{partial_mu_mu_G0}\\   
&&\frac{\partial^3G^0(p^{\mu})}{\partial\mu^3_0} = \frac{\partial^3 G^0(p^{\mu})}{\partial {p^0}^{3}}+2\pi i [\delta^{''}(p^0)\delta(\mu_0 -\epsilon^0_p)\nonumber\\
  && +\delta^{'}(p^0) \delta^{'}(\mu_0-\epsilon^0_p)+\delta(p^0)\delta^{''}(\mu_0-\epsilon^0_p)].\label{partial_triple_mu_G0}    
\end{eqnarray}
We recall that the proper-polarization-functions illustrated in Fig.\ref{GA_Fock} and Fig.\ref{Pi_c}.
In the small wavevector limit, some of the proper-polarization functions can be written in terms of vertex functions within the static limit, such as: $\Lambda^{GW}_2(p^{\mu})$ or $\Lambda^3(p^{\mu})$, which technically have three external legs, where two legs correspond to a frequency-momentum $p^{\mu}$ and one leg is related with an infinitesimally-small momentum $\vec{\delta}$ as we explain later in the appendix section. In fact, we can find more simplified expressions for these vertex functions by using other identities related to $G^0(p^{\mu})$, in the $\vec{\delta} \to 0$ limit. One of such identities is:
\begin{equation}
\lim_{\vec{\delta}\rightarrow 0} G^0(\vec{p}+\vec{\delta},p^0)G^0(\vec{p},p^0) = -\frac{\partial G^0(\vec{p},p^0)}{\partial \mu_0},
\label{G0_square_delta}    
\end{equation}
where in this expression, the treatment of the pole is in the $\mu_0$ plane, which means that the partial derivative on the right-hand side of Eq.~\ref{G0_square_delta} is applied to $G^0(p^{\mu})$ written as in Eq.~\ref{G^0_3}. Another useful identity that even generalizes the identity from Eq.~\ref{G0_square_delta}, is given by:
\begin{equation}
\lim_{\vec{\delta} \rightarrow 0} G^0_{m}(p^{\mu}_{\vec{\delta}})G^0_{n}(p^{\mu})=-B(m,n)G^0_{m+n+1}(p^{\mu}),
\label{G0_partials_product_for_vertex}
\end{equation}
where we have used the shorthand notation for the momentum-frequency variables $p^{\mu}_{\vec{\delta}}=(\vec{p}+\vec{\delta},p^0)$, while $B(m.n)$ is the ``well-known" beta function for $m,n \in \mathbb{N}$. We have also defined the high-order partial derivative of the Green's function with respect to $\mu_0$:
\begin{equation}
G^0_{m}(p^{\mu}) = \frac{\partial^{m-1}G^0(p,p^0)}{\partial \mu^{m-1}_0},
\label{high_order_mu_derivative_G0}
\end{equation}
where in this notation, when the subindex $m=1$, it means that a partial derivative have not been taken with respect to $\mu_0$ over $G^0(p^{\mu})$. The interpretation of Eq.~\ref{G0_partials_product_for_vertex} is that the $m$th order pole from $G^0(\vec{p}+\vec{\delta},p^0)$ approaches the $n$th order pole from $G^0(p^{\mu})$ results in a $m+n$th order pole, where this relationship is found by writing $G^0(p^{\mu})$ as in Eq.~\ref{G^0_3}.

The last identity we give in this section gives the relationship between $(G^0(p^{\mu}))^2$, Dirac-delta functions, and a partial derivative of the non-interacting Green's function with respect to $\mu_0$. This last identity is very useful for keeping track of the integral expression of $\Pi^{xc}_2(q,0)$ given by Eq.~\ref{pixc_2} and is given by:
\begin{equation}
\left[G^0(p^{\mu}) \right]^2 = -\frac{\partial G^0(p^{\mu})}{\partial \mu_0} + 2\pi i\delta(p^0)\delta(p^0-\epsilon^0_p+\mu_0).
 \label{Green_function_square}   
\end{equation}

\section{Deriving the general expression of the $q^2$ coefficients of $\Pi^{xc}(q,0)$}
\label{Pi_xc_calculations}
In this section, we explain in more detail how we obtain the integral expressions for the $q^2$ coefficients $b_x$ and $b_c$-primed coefficients from Eqs.~\ref{b'_x}--\ref{b'''_xc}. These coefficients has been obtained in the same method of keeping track of the integrals that Ma-Brueckner did in Ref.~\cite{Ma-Brueckner}. 
\subsection{Obtaining the general expression of $b^{1}_{xc}$}
In section \ref{correct_calculation}, we have explained that the calculation of $\Pi^{xc}(q,0)$ relies on doing a Taylor expansion on a small wavevector $q$ to keep track of the high-dimensional integrals. We focus first on the calculation of $\Pi^{xc}(q,0)$, which corresponds to the sum of the vertex bubble on the left side of Fig.\ref{GA_Fock} with the last term on the right side from Fig.\ref{Pi_c}. By using the expression given by Eq.~\ref{pixc_1}, the first step is to do a Taylor expansion of the non-interacting Green's function $G^0(p^{\mu})$ in the $\vec{q} \to 0$ limit, we have:
\begin{eqnarray}
G^0(\vec{p}+\vec{q},p^0) &=& G^0(\vec{p}_{\vec{\delta}},p^0)+\vec{q}\cdot \nabla_p G^0(\vec{p}_{\vec{\delta}},p^0)\nonumber\\
 &+&\frac{1}{2}(\vec{q}\cdot \nabla_{\vec{p}})^2G^0(\vec{p}_{\vec{\delta}},p^0)+...,
\label{G0_taylor}    
\end{eqnarray}
where $\vec{p}_{\vec{\delta}}=\vec{p}+\vec{\delta}$ with $\vec{\delta} \rightarrow 0$. We carry over the vector $\vec{\delta}$ in every Taylor expanded form given in this appendix to indicate the treatment of the non-interacting Green's function, where in this case we are using the expression given by Eq.~\ref{G^0_3}. By doing this, Eq.~\ref{G0_taylor} is re-expressed as follows:
\begin{eqnarray}
G^0(p^{\mu}_{\vec{q}}) = G^0(p^{\mu}_{\vec{\delta}}) &-&\left(\frac{\vec{q}\cdot \vec{p}}{m}+\frac{q^2}{2m} \right) G^0_2(p^{\mu}_{\vec{\delta}})\nonumber\\ 
 &+& \frac{(\vec{q}\cdot \vec{p})^2}{2m^2} G^0_3(p^{\mu}_{\vec{\delta}}),
\label{G0_Taylor_2}    
\end{eqnarray}
where $p^{\mu}_{\vec{q}}=p^{\mu}+\vec{q}$ in our notation.

After using this expression in Eq.~\ref{pixc_1}, the proper-polarization function $\Pi^{xc}_1(q,0)$ yields three main terms that contribute up to the $q^2$ order given by:
\begin{equation}
\Pi^{xc}_1(q,0) = \sum_{i=1}^{3}\Pi^{xc}_{1i}(q,0) + O(q^4),
\label{pixc_1_decomposed}    
\end{equation}
then one proceeds to use the identity given by Eq.~\ref{G0_partials_product_for_vertex}. This yields the following three terms:
\begin{equation}
\Pi^{xc}_{11}(q,0) = 2\int d[{\bf p}] d[{\bf p'}]\frac{\tilde{V}(\vec{\kappa})}{\epsilon(k^{\mu})}G^0_2(p^{\mu}) G^0_2(p'^{\mu}),
\label{pixc_11}    
\end{equation}
\begin{equation}
\Pi^{xc}_{12}(q,0) = -\frac{q^2}{2m}\int d[{\bf p}] d[{\bf p'}] \frac{\tilde{V}(\vec{\kappa})}{\epsilon(\kappa^{\mu})} G^0_2(p'^{\mu}) G^0_3(p^{\mu}),
\label{pixc_12}    
\end{equation}
\begin{eqnarray}
\Pi^{xc}_{13}(q,0) = \frac{1}{6 m^2}\int d[{\bf p}] &d[{\bf p'}]& \frac{\tilde{V}(\vec{\kappa})(\vec{q}\cdot \vec{p})^2}{\epsilon(\kappa^{\mu})}\nonumber\\
      &\times& G^0_2(p'^{\mu})G^0_4(p^{\mu}),
\label{pixc_13}    
\end{eqnarray}
where $\kappa^{\mu}=p'^{\mu}-p^{\mu}$, where $\vec{\kappa}$ is the spatial part of this four-vector, and $\Pi^{xc}_{11}(q,0)$ is the only term that contributes to a constant, and is not the coefficient of interest in this section. Only $\Pi^{xc}_{12}(q,0)$ and $\Pi^{xc}_{13}(q,0)$ are the functions that contribute to the coefficient of $q^2$. We can rewrite both relevant terms as an integral expression which involves a specific vertex function $\Lambda^{GW}_2(p^{\mu})$, defined as:
\begin{equation}
\Lambda^{GW}_{2}(p^{\mu}) = \lim_{\vec{\delta'}\rightarrow 0} i\int d[{\bf p'}] \frac{\tilde{V}(\vec{\kappa}) }{\epsilon(\kappa^{\mu})}G^0(p'^{\mu}_{\vec{\delta'}})G^0(p'^{\mu}),
\label{lambda2_GW_definition}    
\end{equation}
which can be expressed in compact form by using Eq.~\ref{G0_square_delta}, which yields:
\begin{equation}
\Lambda^{GW}_2(p^{\mu}) = -i\int d[{\bf p'}] \frac{\tilde{V}(\vec{\kappa})}{\epsilon({\kappa}^{\mu})}\frac{\partial G^0(p'^{\mu})}{\partial \mu_0},
\label{Lambda2_frpa}    
\end{equation}
After using this expression of the vertex function $\Lambda^{GW}_2(p^{\mu})$, we exploit the symmetry at the integration level over the azimuthal and polar angles in Eq.~\ref{pixc_13}. This implies that we can do the replacement in this integral expression: $\vec{q}\cdot \vec{p} \rightarrow (qp)^2/3$. By doing this and expressing $\Pi^{xc}_{12}(q,0)$ and $\Pi^{xc}_{13}(q,0)$ in terms of $\Lambda^{GW}_2(p^{\mu})$, we have that the coefficient $b^{'}_{xc}$ of the $q^2$ term from the proper-polarization function $\Pi^{xc}_1(q,0)$ is given by the following expression:
\begin{eqnarray}
b^1_{xc} = -\frac{i}{m} \int d[{\bf p}] \Lambda^{GW}_2(p^{\mu}) (&\frac{1}{2}& G^0_3(p^{\mu})\nonumber\\
   &-&\frac{\epsilon^0_p}{9} G^0_4(p^{\mu}) ),
\label{b1_xc}    
\end{eqnarray}
where $\epsilon^0_p=p^2/2m$ is the energy dispersion of the freely interacting part of the Hamiltonian. 

\subsection{Obtaining the general expression of $b^{2}_{xc}$}
We now proceed with the calculation of the coefficient $b^{2}_{xc}$ of the $q^2$ term of $\Pi^{xc}_2(q,0)$, where we use Eq.~\ref{pixc_2} as the starting point. By using the identity given by Eq.~\ref{Green_function_square}, the proper-polarization function $\Pi^{xc}_2(q,0)$ is now separated into two terms as follows:
\begin{equation}
\Pi^{xc}_2(q,0) = \sum^{3}_{i=1} \Pi^{xc}_{2i}(q),
\label{pixc_2_decomposed}    
\end{equation}
where each of the $\Pi^{xc}_{2i}(q)$ terms (for $i=1,2,3$), we have:
\begin{eqnarray}
\Pi^{xc}_{21}(q) = &2i&\int d[{\bf p}] \frac{\partial G^0(p^{\mu})}{\partial \mu_0} \Sigma_{GW}(p^{\mu})\nonumber\\
 &\times&\left(G^0(p^{\mu}+q^{\mu}) + G^0(p^{\mu}-q^{\mu}) \right), 
\label{pixc_21}    
\end{eqnarray}
where $\Sigma_{GW}(p^{\mu})$ is known as the GW self-energy, which is defined as follows:
\begin{equation}
\Sigma_{GW}(p^{\mu}) \equiv i \int \frac{d^4p^{'}}{(2\pi)^4}\tilde{V}_e(p'^{\mu}-p^{\mu})G^0(p'^{\mu}),
\label{GW_self_energy_true_definition}
\end{equation}
which can be also expressed as the sum of the Fock self-energy $\Sigma_F(p,0)$ and the ring-like self-energy term $\Sigma_r(p^{\mu})$, as explained in appendix~\ref{derivation_general_expressions_b_xc_primed}. The second term that contributes to $\Pi^{xc}_2(q,0)$ can be found after calculating an easy frequency-momentum integral. It yields:
\begin{equation}
\Pi^{xc}_{22}(q,0) = -\frac{m^2 \Sigma_{GW}(k_F,0)}{\pi^2 q}\ln \left|\frac{k_F+q}{k_F-q} \right|.
\label{pixc_22}    
\end{equation}

At this point, we can do a Taylor expansion in small $\vec{q}$ to the non-interacting Green's function terms inside the integral expression of $\Pi^{xc}_{21}(q,0)$ by using the expression given by Eq.~\ref{G0_Taylor_2}. When this Taylor expansion is done, the term $\Pi^{xc}(q,0)$ yields three terms up to the $q^2$ order, given by the expressions below:
\begin{eqnarray}
&\Pi^{xc}_{21}(q,0)& = \sum^3_{i=1} \Pi^{xc}_{21i}(q,0) +O(q^4),\label{pixc21_decomposition}\\    
&\Pi^{xc}_{211}(q,0)& = \lim_{\vec{\delta} \rightarrow 0} 4i \int d[{\bf p}] A_{12}(p^{\mu}_{\vec{\delta}},p^{\mu}),\label{pixc_211}\\    
&\Pi^{xc}_{212}(q,0)& = \lim_{\vec{\delta} \rightarrow 0}-\frac{2q^2i}{m} \int d[{\bf p}] A_{22}(p^{\mu}_{\vec{\delta}},p^{\mu}), \label{pixc_212}\\    
&\Pi^{xc}_{212}(q,0)& = \lim_{\vec{\delta} \rightarrow 0} \frac{2i}{m^2} \int d[{\bf p}] (\vec{p}\cdot\vec{q})^2  A_{32}(p^{\mu}_{\vec{\delta}},p^{\mu}),
\label{pixc_213}
\end{eqnarray}
where the tensor $A_{mn}(p^{\mu}_1,p^{\mu}_2)$ is defined as follows:
\begin{equation}
A_{mn}(p^{\mu}_1,p^{\mu}_2) = \Sigma_{GW}(p^{\mu}) G^0_m(p^{\mu}_1)G^0_n(p^{\mu}_2),
\label{A_tensor}
\end{equation}
and $\Pi^{xc}_{211}(q,0)$ turns out to be a constant term. The next step is to exploit the symmetry when integrating over the azimuthal and polar angles, which allows to replace a term in the integral expression of $\Pi^{xc}_{213}(q,0)$ as follows: $(\vec{p}\cdot \vec{q})^2 = (p^2q^2)/3$. After such step, we apply the identity given by the Eq.~\ref{G0_Taylor_2} in the expressions from Eqs.~\ref{pixc_212}--\ref{pixc_213} to finally obtain the $q^2$ term of $\Pi^{xc}_{21}(q,0)$. When such a term is combined along with the $q^2$ contribution of $\Pi^{xc}_{22}(q,0)$, which is obtained by doing a Taylor expansion in the logarithm term in Eq.~\ref{pixc_22}, we finally obtain the total $q^2$ coefficient $b^2_{xc}$ of $\Pi^{xc}_2(q,0)$. Such coefficient is given by:
\begin{eqnarray}
b^2_{xc} = \frac{i}{m} \int d[{\bf p}] \Sigma_{GW}&(p^{\mu})& \left[\frac{G^0_4(p^{\mu})}{3}-\frac{\epsilon^0_p G^0_5(p^{\mu})}{9} \right]\nonumber\\
  &-&\frac{m^2 \Sigma_{GW}(k_F,0)}{12 \pi^3 k_F^3}.
\label{b2_xc}    
\end{eqnarray}

\subsection{Obtaining the general expression of $b^{3}_{xc}$}
Now we finally show the calculation of the proper-polarization term $\Pi^{xc}_{3}(q,0)$ in the $\vec{q} \to 0$ limit. In order to find this contribution, we only need to work with one of the diagrams that contributes to $\Pi^3_{xc}(q,0)$ given that these terms are related with each other by doing the momentum transformation $\vec{q}\rightarrow -\vec{q}$. In this case we focus on the first diagram on the left from Fig.\ref{Pi_c}, which we labeled as $\Pi^{xc}_{31}(q,0)$. The integral expression of $\Pi^{xc}_{31}(q,0)$ is given by Eq.~\ref{pixc_31}. For this term we proceed to do a Taylor expansion in small wavevector $\vec{q}$ on the effective potential terms $\tilde{V}_e(\vec{k}\pm\vec{q}/2,k^0)$. The Taylor expansion up to $q^2$ order is given by:
\begin{eqnarray}
\tilde{V}_{e}\left(\vec{k}+\frac{\vec{q}}{2},k^0\right) &=& \tilde{V}_e(k,k^0)\pm\frac{\vec{q}\cdot\nabla_k}{2}\tilde{V}_e(k,k^0)\nonumber\\
 &+&\frac{(\vec{q}\cdot\nabla_k)^2}{8} \tilde{V}_e(k,k^0)+ O(k^3). 
\label{Taylor_effective_potential}
\end{eqnarray}
After expanding the Taylor series of the effective potential terms and the non-interacting Green functions contained in the expression of $M(p^{\mu},p'^{\mu},q^{\mu})$ from Eq.~\ref{M}, we apply the identity given by Eq.~\ref{G0_Taylor_2} to simplify the expressions. We obtain three main contributions to $\Pi^{xc}_{31}(q,0)$ up to the $q^2$ order:
\begin{equation}
\Pi^{xc}_{31}(q,0)=\sum^3_{i=1} \Pi^{xc}_{31i}(q,0) + O(q^4),
\label{pixc_31_decomposition}
\end{equation}
where $\Pi^{xc}_{311}(q,0)$ is the only term that is independent of wavevector $\vec{q}$ given by the following expression:
\begin{eqnarray}
\Pi^{xc}_{311}(q,0) &=& -4i \int d[{\bf p}] d[{\bf p'}] d[{\bf k}] A(p^{\mu},p'^{\mu},k^{\mu})\nonumber\\
&\times& (\tilde{V}_e(k^{\mu}))^2 G^0_2(p^{\mu}) G^0_2(p'^{\mu}),
\label{pixc_311}    
\end{eqnarray}
while the rest of the terms are very extensive, so it is convenient to express their respective integral expressions in terms of the vertex function $\Lambda^3(p^{\mu})$, which is defined by the following expression:
\begin{eqnarray}
\Lambda^3(p^{\mu}) &\equiv& \lim_{\vec{\delta} \rightarrow 0} 2 \int d[{\bf p'}] d[{\bf k}] A(p^{\mu},p'^{\mu},k^{\mu})\nonumber\\
&\times& G^0(\vec{p'}_{\vec{\delta'}},{p'}^0)G^0(p',{p'}^0)(\tilde{V}_e(k^{\mu}))^2,
\label{lambda3_definition}    
\end{eqnarray}
where in the term $\vec{p'}_{\vec{\delta'}}=\vec{p'}+\vec{\delta'}$ and $\vec{\delta'} \to 0$. This infinitesimally-small vector $\vec{\delta'}$ is used to indicate the expression that we have used for $G^0(p,^{}p^0)$, given by Eq.~\ref{G0_Taylor_2}, when obtaining the Taylor expansion in small wavevector $\vec{q}$. By exploiting the identity given by Eq.~\ref{G0_square_delta}, we find a compact expression for the vertex function $\Lambda^3(p^{\mu})$ given by:
\begin{eqnarray}
\Lambda^3(p^{\mu}) = &-2& \int d[{\bf p'}] d[{\bf k}] A(p^{\mu},p'^{\mu},k^{\mu})\nonumber\\
&\times& \left(\tilde{V}_e(k^{\mu}) \right)^2 G^0_2(p'^{\mu}), 
\label{lambda3}    
\end{eqnarray}
In terms of the vertex function $\Lambda^3(p^{\mu})$, we write the expressions for $\Pi^{xc}_{312}(q,0)$ as follows:
\begin{equation}
\Pi^{xc}_{312}(q,0) = \sum^3_{i=1} \Pi_{312i}(q,0),
\label{pixc_312_decomposition}    
\end{equation}
where each individual term is given below by the following integral expressions:
\begin{equation}
\Pi^{xc}_{3121}(q,0) = -\frac{iq^2}{2m}\int d[{\bf p}] \Lambda^{3}(p^{\mu})\frac{\partial^2 G^0(p^{\mu})}{\partial\mu^2_0},
\label{pixc_3121}    
\end{equation}
\begin{eqnarray}
\Pi^{xc}_{3122}(q,0) &=& \frac{iq^2}{3m^2}\int d[{\bf p}]p^2\Lambda^3(p^{\mu})\nonumber\\
         &\times&\left(-\frac{1}{6}\frac{\partial^3G^0(p^{\mu})}{\partial\mu^3_0}+\frac{1}{3}\frac{\partial^3G^0(p^{\mu})}{\partial \mu^3_0}\right),
\label{pixc_3122}    
\end{eqnarray}
where in the latter expression we have already exploited a symmetry in the integration over the azimuthal and polar angles, where we have replaced in the integrand: ($\vec{k}\cdot\vec{q})^2 \rightarrow k^2 q^2/3 $.

 The expression of $\Pi^{xc}_{313}(q,0)$ from Eq.~\ref{pixc_31_decomposition}, can be written as only one term as follows:
\begin{eqnarray}
\Pi^{xc}_{313}(q,0) = -i\int d[{\bf p}]&d&[{\bf p'}]d[{\bf k}] R_2(\vec{q},k^{\mu})\nonumber\\
&\times&R_1(p^{\mu},p'^{\mu},k^{\mu}), 
\label{pixc_313}    
\end{eqnarray}
where the function $R_1(p^{\mu},p'^{\mu},k^{\mu})$ is expressed in terms of the non-interacting Green's function and also in terms of the partial derivative with respect to $\mu_0$ of this term. The expression for this function is given by:
\begin{eqnarray}
R_1(p^{\mu},p'^{\mu},k^{\mu}) &=& G^0(p^{\mu}-k^{\mu})\frac{\partial G^0(p^{\mu})}{\partial \mu_0}\nonumber\\
&\times& G^0(p'^{\mu}+k^{\mu})\frac{\partial G^0(p'^{\mu})}{\partial \mu_0},
\label{R1}    
\end{eqnarray}
while the function $R_2(\vec{q},k^{\mu})$ depends on the effective potential and its gradient terms, given by:
\begin{equation}
R_2(\vec{q},k^{\mu}) = \tilde{V}_e(k^{\mu})(\vec{q}\cdot\nabla_{\vec{k}})^2\tilde{V}_e(k^{\mu})-(\vec{q}\cdot\nabla_{\vec{k}}\tilde{V}_e(k^{\mu}))^2.
\label{R2}    
\end{equation}
One can find a more compact expression for $\Pi^{xc}_{313}(q,0)$, by recalling that the leading proper-polarization function $\Pi_0(k^{\mu})$ is given by the following expression:
\begin{equation}
\Pi_0(k^{\mu}) =-2i\int d[{\bf p'}]G^0(p'^{\mu}+k^{\mu})G^0(p'^{\mu}),
\label{pi0}    
\end{equation}
where we can now apply a partial derivative with respect to $\mu_0$ to this expression. After applying the product rule of partial derivatives, we do a change of variable given by: $p'^{\mu}\rightarrow p'^{\mu}-k^{\mu}$ on the integral that contains the partial derivative of $G^0(p'^{\mu}+k^{\mu})$ with respect to $\mu_0$. By exploiting that $\Pi^0(k^{\mu})$ is a function that depends on the magnitude of the wavevector $k$, then we finally obtain a compact expression:
\begin{equation}
\frac{\partial \Pi_0(k^{\mu})}{\partial \mu_0}=-4i\int d[{\bf p'}]G^0(p'^{\mu}+k^{\mu})\frac{\partial G^0(p'^{\mu})}{\partial \mu_0}.
\label{partial_mu_pi0}    
\end{equation}
By using this expression, we find that $\Pi^{xc}_{313}(q,0)$ is written in the following simplified expression:
\begin{equation}
\Pi^{xc}_{313}(q,0) = \frac{i}{16}\int d[{\bf k}]\left(\frac{\partial \Pi_0(k^{\mu})}{\partial \mu_0} \right)^2R_2(\vec{q},k^{\mu}).
\label{pixc_313_compact}    
\end{equation}

We finally extract the $q^2$ coefficients from Eqs.~\ref{pixc_3121},~\ref{pixc_3122},~and~\ref{pixc_313_compact}. We label the sum of the coefficients that comes from these three terms as $b^{xc}_{31}$. The associated $q^2$ coefficient $b^{32}_{xc}$, that comes from the second diagram (from the left side) of Fig.\ref{Pi_c}, should exactly be the same as the coefficient $b^{31}_{xc}$ by symmetry. For this reason, we obtain the total $q^2$ coefficient of the proper-polarization function $\Pi^{xc}_3(q,0)$ by multiplying $b^{31}_{xc}$ by a factor of two. Such coefficient we label it as $b^3_{xc}$, and it is given by the sum of two main terms:
\begin{equation}
b^3_{xc} = b^{3}_{xc,a}+b^3_{xc,b},
\label{bxc_3_combination}    
\end{equation}
\begin{eqnarray}
b^{3}_{xc,a} = -&2\frac{i}{m} \int& d[{\bf p}] \Lambda^3(p^{\mu})\nonumber\\
&\times& \left[\frac{G^0_3(p^{\mu})}{2} -\frac{\epsilon_p G^0_4(p^{\mu}) }{9} \right],\label{bxc_3a}\\    
b^{3}_{xc,b} = \frac{i}{24} &\int d[{\bf k}]&\left(\frac{\partial \Pi_0(k^{\mu})}{\partial \mu_0} \right)^2R_3(\vec{q},k^{\mu}),\label{bxc_3b} \\
R_3(\vec{q},k^{\mu})= &\tilde{V}_e(k^{\mu})&\nabla^2_q \tilde{V}_e(k^{\mu})-\left(\frac{\partial \tilde{V}_e(k^{\mu})}{\partial k} \right)^2,\label{R3}   
\end{eqnarray}
where we have exploited symmetries in the integration over the azimuthal and polar angles from the expression we had in Eq.~\ref{pixc_313_compact}. After this, we can re-express the terms in the integrand as follows: $(\vec{q}\cdot \nabla_{\vec{k}})^2\tilde{V}_e(k^{\mu}) \rightarrow q^2 \nabla^2_{\vec{k}}\tilde{V}_e(k^{\mu})/3$ and $(\vec{q}\cdot \nabla_{\vec{k}}\tilde{V}_e(k^{\mu}))^2 \rightarrow \frac{q^2}{3}(\frac{\partial}{\partial q} \tilde{V}_e(k^{\mu}))^2$.

\section{Deriving the general expressions of $b'_{xc}$,$b^{''}_{xc}$ and $b^{'''}_{xc}$}
\label{derivation_general_expressions_b_xc_primed}
We proceed to find the general expressions of all $b_{xc}$-primed coefficients by doing a combination of the final results we have previously found for $b^i_{xc}$ (for $i=1,2,3$). As we have mentioned before, the $b_{xc}$-primed coefficients emerge by reorganizing the sum of the three $q^2$ coefficients of $\Pi^{xc}(q,0)$ obtained from the diagrams illustrated in Fig.\ref{GA_Fock} and Fig.\ref{Pi_c}, as we have in Eq.~\ref{b_xc_sum}.

The expressions we have found for $b^i_{xc}$ (for $i=1,2,3$) are too complicated to calculate analytically per term individually due to the way they are written in terms of the vertex functions $\Lambda^{GW}_2(p^{\mu})$ and $\Lambda^3(p^{\mu})$. We can get around this difficulty by separating into two separate terms the GW self-energy $\Sigma_{GW}(p^{\mu})$ and the vertex functions $\Lambda^{GW}_2(p^{\mu})$ and $\Lambda^3(p^{\mu})$. We are forced, at the calculation level, to do a separation between the terms that have the instantaneous part of the interaction associated with the potential energy term $\tilde{V}(k)$ from the frequency-dependent part of the renormalized interaction. This is done because we can keep track of the integration along the frequency variables. For the GW self-energy, we have:
\begin{equation}
\Sigma_{GW}(p^{\mu}) = \Sigma_{F}(p,0)+\Sigma_r(p^{\mu}),
\label{Gw_self_energy_separation}    
\end{equation}
where $\Sigma_{F}(p^{\mu})$ is the ``well-known" Fock self-energy, which is usually defined by using $V_0(k)$. In our case, since we don't want to specify an expression of the Fourier transform of the interacting potential that we are using in our expressions yet, we define this self-energy as follows:
\begin{equation}
\Sigma_{F}(p^{\mu}) \equiv i\int \frac{d^4p'}{(2\pi)^4} e^{ip'^0 \eta} \tilde{V}(\vec{p'}-\vec{p}) G^0(p',p'^0),
\label{self-energy_Fock}    
\end{equation}
where $\eta \rightarrow 0^{+}$ is a regulator that emerges from using an instantaneous interaction term within time-ordering of the fermion fields and allows the frequency integral of Eq.~\ref{self-energy_Fock} to exist. The second term, $\Sigma_r(p^{\mu})$ from Eq.~\ref{Gw_self_energy_separation}, is the rest of the contribution to the GW self-energy, where the topology of such associated terms involve doing a summation of polarization bubbles $\Pi_0(p'^{\mu})$, which is why we attribute its subscript as for ``ring-like series". Such a self-energy term is defined as:
\begin{equation}
\Sigma_r(p^{\mu}) = i\int\frac{d^4p'}{(2\pi)^4}(\tilde{V}_e(\kappa^{\mu})-\tilde{V}(\vec{\kappa}))G^0(p'^{\mu}).
\label{self-energy_ring}    
\end{equation}
where we have used $\kappa^{\mu}=p'^{\mu}-p^{\mu}$.

In a similar way, as we separated the GW self-energy into two terms, the vertex function $\Lambda^{GW}_2(p^{\mu})$ can also be re-expressed as follows:
\begin{equation}
\Lambda^{GW}_2(p^{\mu}) = \Lambda_2(p^{\mu}) + \Lambda^r_2(p^{\mu}), 
\label{lambda2_GW_separation}    
\end{equation}
where $\Lambda_2(p^{\mu})$ is obtained by only using the instantaneous part of the interaction given by $\tilde{V}(k)$, while $\Lambda^r_2(p^{\mu})$ is obtained from substracting the $\tilde{V}(k)$ from the effective potential $\tilde{V}_e(k^{\mu})$. We have:
\begin{equation}
\Lambda_2(p^{\mu}) = -i\int\frac{d^4p'}{(2\pi)^4} \tilde{V}(\vec{\kappa})\frac{\partial G^0(p'^{\mu})}{\partial\mu_0}, 
\label{lambda_2}    
\end{equation}
\begin{equation}
\Lambda^r_2(p^{\mu}) = -i\int \frac{d^4p'}{(2\pi)^4}\tilde{V}_r(\kappa^{\mu})\frac{\partial G^0(p'^{\mu})}{\partial \mu_0}.
\label{lambda_2_r}    
\end{equation}
where $\tilde{V}_r(\kappa^{\mu})$ is given by:
\begin{equation}
\tilde{V}_r(\kappa^{\mu}) = \tilde{V}_e(\kappa^{\mu})-\tilde{V}(\vec{\kappa}),
\label{V_r}
\end{equation}
where $\vec{\kappa}$ corresponds to the spatial part of $\kappa^{\mu}$ we used in this appendix section.

After reorganizing the sum of the $b^i_{xc}$ coefficients (for $i=1,2,3$), we have that the $b_{xc}$-primed coefficients are given by:
\begin{equation}
b'_{xc} = b'_{xc,1}+b'_{xc,2},
\label{bxc_'}    
\end{equation}
where each of these coefficients are given by the following expressions:
\begin{equation}
b'_{xc,1} = \frac{i}{m}\int d[{\bf p}] \Sigma_{GW}(p^{\mu})\left(\frac{G^0_4(p^{\mu})}{2}-\frac{\epsilon^0_p G^0_5(p^{\mu})}{9} \right),
\label{b'_xc_1}    
\end{equation}
\begin{eqnarray}
b'_{xc,2} = -\frac{i}{m}&\int& d[{\bf p}](\Lambda^{GW}_2(p^{\mu})+2\Lambda^3(p^{\mu}))\nonumber\\
&\times&\left(\frac{1}{2}G^0_3(p^{\mu})-\frac{\epsilon^0_p}{9} G^0_4(p^{\mu}) \right),
\label{b'_xc_2}    
\end{eqnarray}
while the coefficient $b^{''}_{xc}$, however, is obtained by the following expression:
\begin{equation}
b^{''}_{xc} = -\frac{i}{m}\int d[{\bf p}]\Sigma_{GW}(p^{\mu})\frac{1}{6}\frac{\partial^3G^0(p^{\mu})}{\partial\mu^3_0}.
\label{b''_xc}    
\end{equation}
and finally $b^{'''}_{xc}$ is precisely the coefficient $b^3_{xc,b}$ we found in Eq.~\ref{bxc_3b}:
\begin{equation}
b^{'''}_{xc} = b^3_{xc,b}.
\label{b'''_xc_b}    
\end{equation}

Doing the separation of the terms of $\Lambda_{GW}(p^{\mu})$ and $\Sigma_{GW}(p^{\mu})$ forces the separation of the coefficients $b_{xc}$-primed coefficients as follows:
\begin{equation}
b^{'}_{xc} = b^{'}_x+b^{'}_c,
\label{b'_xc_separation}    
\end{equation}
\begin{equation}
b^{''}_{xc} = b^{''}_x+b^{''}_c,
\label{b''_xc_separation}    
\end{equation}
Another identity can be exploited within this separation of terms, which is straightforward to prove by using the expressions of every vertex function and self-energy we found in Eqs.~\ref{lambda3},~\ref{Gw_self_energy_separation}--\ref{lambda_2_r}. Firstly, there is a relationship between the vertex function $\Lambda_2(p^{\mu})$ and the partial derivative of the Fock self-energy with respect to $\mu_0$:
\begin{equation}
-\frac{\partial \Sigma_F(p^{\mu})}{\partial \mu_0} = \Lambda_2(p^{\mu})+\frac{{\lambda_c}^2\beta(k_F)}{2 \pi e^2}\frac{\partial \beta(k_{F})}{\partial \mu_0}\Delta_2(p^{\mu}). 
\label{Fock_lambda2_relation}    
\end{equation}
where the second term does not contribute to the $b_x$ coefficient at the end of the calculation since it yields zero after taking the $\lambda_c \to 0$ limit, and has been ignored in the rest of the calculation. The function $\Delta_2(p^{\mu})$ is given by the following expression:
\begin{equation}
\Delta_2(p^{\mu}) = i \int d[{\bf p'}] e^{ip'^0 \eta}\left[\tilde{V}\left(\vec{\kappa} \right)\right]^2G^0(p'^{\mu}).
\label{Delta_2_correction_partial}
\end{equation} 

Similarly, there is a relationship between the vertex functions $\Lambda^r_2(p^{\mu})$ and $\Lambda^3(p^{\mu})$ with the partial derivative of the self-energy (ring-like) $\Sigma_r(p^{\mu})$ with respect to $\mu_0$, given by:
\begin{eqnarray}
-\frac{\partial \Sigma_r(p^{\mu})]}{\partial \mu_0} &=& \Lambda^{r}_2(p^{\mu})+2\Lambda^3(p^{\mu}) \nonumber\\
         &+&\frac{{\lambda_c}^2}{2 \pi e^2}\frac{\partial [\beta(k_{F})]^2}{\partial \mu_0}\sum^{4}_{i=3}\Delta_i(p^{\mu}).
\label{lambda2_lambda3_ring_self}    
\end{eqnarray}
where the last term with the prefactor ${\lambda_c}^{2}$, does not contribute to the coefficient $b_c$ and this is why this term is ignored in the rest of the calculation. The functions $\Delta_i(p^{\mu})$ (for $i=3,4$) are given by the following expressions:
\begin{eqnarray}
\Delta_3(p^{\mu}) &=& i \int d[{\bf p'}]\tilde{V}(\vec{\kappa})\tilde{V}_r(\kappa^{\mu})G^0(p'^{\mu}),\label{Delta_3_correction_partial}\nonumber\\
\Delta_4(p^{\mu}) &=& i \int d[{\bf p'}] \tilde{V}_e(\vec{\kappa})\tilde{V}_r(\kappa^{\mu})G^0(p'^{\mu}),\label{Delta_4_correction_partial}.
\end{eqnarray}

After using these two relationships while summing the coefficients $b'_{xc,1}$ with $b'_{xc,2}$ and ignoring the corrections with a ${\lambda_c}^2$ prefactor from Eq.~\ref{Fock_lambda2_relation} and Eq.~\ref{lambda2_lambda3_ring_self}, we finally obtain the compact expressions for $b'_x$ and $b'_c$, given in Eq.~\ref{b'_x} and Eq.~\ref{b'_c}. The expressions of $b^{''}_x$ and $b^{''}_c$ given by Eq.~\ref{b''_x} and Eq.~\ref{b''_c} respectively, come directly from doing the separation of terms we did in Eq.~\ref{Gw_self_energy_separation} and Eq.~\ref{lambda2_GW_separation}.

\section{Explicit calculation of $b_x$}
\label{explicit_calculation_bx}
We proceed to calculate the total $q^2$ coefficient given by the sum of the $b_x$-primed coefficients and the self-energy term by using Eq.~\ref{b_x_sum}.
From the starting point of the calculation of these coefficients, it suffices to calculate the frequency integral from Eq.~\ref{self-energy_Fock}, which is calculated by using the residue theorem. Due to the exponential with the regulator in the integrand, the contour complex path used is a semi-circle that covers entirely the first and second quadrants in the complex plane. Then, the three-dimensional integral expression of the Fock self-energy (without specifying the potential $\tilde{V}(k)$) is given by the following expression:
\begin{equation}
\Sigma_F(p) = -\int \frac{d^3p'}{(2 \pi)^3} \tilde{V}(\vec{p'}-\vec{p}) \Theta(k_F-p'), 
\label{Fock_selfenergy_3D}    
\end{equation}
where this expression is completely independent of the frequency variable $p^0$, and we can exploit this fact when calculating the frequency integrals from the expressions of $b'_x$ and $b''_x$.

Calculating the frequency integrals from Eq.~\ref{b'_x} and Eq.~\ref{b''_x} involves using the expanded forms of the second and third order partial derivatives of the non-interacting Green's function with respect to $\mu_0$ given by the expressions Eqs.~\ref{partial_mu_mu_G0}--\ref{partial_triple_mu_G0}. After doing such an expansion, only the term of the form $\delta(p^0)\delta(\mu_0-\epsilon^0_p)$ yields non-zero values after doing the integration over the variable $p^0$. The reason for this is that the other terms from the expanded forms given in Eqs.~\ref{partial_mu_mu_G0}--\ref{partial_triple_mu_G0} contain derivative terms of the Dirac delta function $\delta(p^0)$ and high-order partial derivatives of $G^0(p^{\mu})$ with respect to the frequency variable $p^0$ that multiplies $\Sigma_F(p)$, which is a frequency-independent function. In the frequency integral of the term involving the second (third) order partial derivatives of $G^0(p^{\mu})$ with respect to $p^0$ can be justified to be zero due to the residue theorem applied to their corresponding second(third) degree complex poles. After performing the integral over the $p^0$ variable of the term with the form of $\delta(p^0)\delta(\mu_0-\epsilon^0_p)$, we obtain the three-dimensional integral expressions of $b'_x$ and $b^{''}_x$ given by Eqs.~\ref{b'_x_3d} and \ref{b''_x_3d}, respectively. 

It is straightforward to keep track of the momentum integrals given in the expressions given by Eqs.~\ref{b'_x_3d}--\ref{b''_x_3d} since there are partial derivatives acting on the Dirac delta function $\delta(\mu_0-\epsilon^{0}_p)$ at the level of the integrands. One expresses these integral expressions of $b^{'}_{x}$ and $b^{''}_x$ in a more convenient form by exploiting the following Dirac delta function identity:
\begin{equation}
\frac{\partial \delta(\mu_0 -\epsilon^0_p)}{\partial \mu_0} = -\frac{\partial \delta(\epsilon^0_p-\mu_0)}{\partial \epsilon^0_{p}}.
\label{Dirac_identity_derivative}    
\end{equation}
We can find more convenient forms of the momentum integral expressions for $b^{'}_x$ and $b^{''}_x$ by exploiting the properties of the Dirac delta functions and
by using the integration by parts technique. The convenient expressions
are given by:
\begin{eqnarray}
  &b^{'}_x&= -\int^{\infty}_0 \frac{dp}{6 \pi^2} \delta(\mu_0-\epsilon^0_p) \frac{\partial P(p)}{\partial p} + \nonumber \\
  &&\int^{\infty}_0\frac{dp}{36\pi^2} \delta(\mu_0-\epsilon^0_p)\left[2\frac{\partial^2(p \Sigma_F(p))}{\partial p\partial\epsilon^0_p } +p\frac{\partial^2P(p)}{\partial p^2} \right],
\label{b'_x_3d_2}  \\  
&b^{''}_{x}&= \int^{\infty}_0 \frac{dp}{12 \pi^2} \delta(\mu_0-\epsilon^0_p)\frac{\partial^2 (p \Sigma_F(p))}{\partial p \partial\epsilon^0_p},
\label{b''_x_3d_2}    
\end{eqnarray}
where $P(p)$ is obtained by applying a differential operator
on $p\Sigma_F(p)$ as follows: 
\begin{equation}
P(p) \equiv \left(\frac{\partial}{\partial \mu_0}+\frac{\partial}{\partial \epsilon^0_p} \right)(p\Sigma_F(p)).
\label{P}    
\end{equation}
From this point, we can start using a specific expression for the potential $\tilde{V}(k)$.

\subsection{Using the $k_F$-dependent regulator $\lambda$ (explicit steps)}
In this part of the appendix, we set $\tilde{V}(k)=V_{\lambda}(k)$ and proceed to explain the calculations for $b'_x$ and $b^{''}_x$. In most of our expressions, we introduce the superscript $\lambda$ to indicate that we regularize the Coulomb potential by using the $k_F$-dependent regulator, by using Eq.~\ref{V_generic}. 

The momentum integral given in Eq.~\ref{Fock_selfenergy_3D} is straightforward to calculate by using the generic potential $V_{\lambda}(k)$. We obtain:
\begin{equation}
\Sigma^{\lambda}_{F}(p) = -\frac{e^2 k_F}{\pi}F(\frac{p}{k_F},\frac{\lambda}{k_F}),
\label{self_energy_lambda_fock}    
\end{equation}
where $\lambda(k_F)=\lambda_c \beta(k_F)$, and $F(x,y)$ is given by:
\begin{equation}
F(x,y) = 1+\left( \frac{1+y^2-x^2}{4x} \right)g(x,y)-\frac{g_2(x,y)}{2},
\label{Fxy}    
\end{equation}
where we have $x=p/k_F$ and $y=\lambda(k_F)/k_F$. By using this expression of the Fock self-energy $\Sigma^{\lambda}_F(p)$, we can use Eq.~\ref{P} to find the specific expression of $P^{\lambda}(p)$ by treating $\lambda_c$ as a constant. We obtain the following:
\begin{equation}
P^{\lambda}(p) = -\frac{e^2m}{\pi}\left(H_1(\frac{p}{k_F},\frac{\lambda}{k_F})+H_2(\frac{p}{k_F},\frac{\lambda}{k_F})\right),
\label{P_lambda_2}    
\end{equation}
where $H_1(x,y)$ and $H_2(x,y)$ are the functions given below:
\begin{equation}
H_1(x,y) = \frac{2}{x}-\frac{g_2(x,y)}{2x},
\label{H_function}    
\end{equation}
\begin{equation}
H_2(x,y) = \lambda_c\frac{\partial \beta(k_F)}{\partial k_F}\left[\frac{y g(x,y)}{2}-\frac{xg_2(x,y)}{2y}\right].
\label{H2_function}    
\end{equation}

We can now use the expression given in Eq.~\ref{P_lambda_2} to find the necessary expressions to calculate the integrals given by Eqs.~\ref{b'_x_3d_2}--\ref{b''_x_3d_2}. The necessary expressions are given by:
\begin{equation}
\frac{\partial P^{\lambda}(p)}{\partial p} = -\frac{e^2 m}{\pi k_F} \left(P^{\lambda}_1(p)+P^{\lambda}_2(p) \right),
\label{partial_p_P_lambda}    
\end{equation}
\begin{equation}
 P^{\lambda}_1(p) = -\frac{2}{x^2}+\frac{2y(2y+(1+x^2+y^2)\lambda_c\frac{\partial \beta(k_F)}{\partial k_F})}{D(x,y) },
\label{P1_lambda}    
\end{equation}
\begin{equation}
 P^{\lambda}_2(p) = \frac{(y-x^2\lambda_c\frac{\partial \beta(k_F)}{\partial k_F})}{2yx^2}g_2(x,y),
\label{P2_lambda}    
\end{equation}
\begin{equation}
\frac{\partial^2 P^{\lambda}(p)}{\partial p^2} = -\frac{e^2m}{\pi k_F^2}\left(P^{\lambda}_3(p)+P^{\lambda}_4(p) \right),
\label{partial_p_partial_p_P_lambda}    
\end{equation}
\begin{equation}
P^{\lambda}_3(p) = \frac{4}{x^3}-\frac{4y^2}{x D(x,y)},
\label{P3_lambda}    
\end{equation}
\begin{eqnarray}
P^{\lambda}_4(p) &=& \frac{8xy(1-x^2-y^2)}{(D(x,y))^2}\nonumber\\
&\times&(2y+(1+x^2+y^2)\lambda_c\frac{\partial \beta(k_F)}{\partial k_F}),
\label{P4_lambda}
\end{eqnarray}
\begin{equation}
\frac{\partial^2(p \Sigma^{\lambda}_{F}(p))}{\partial p\partial \epsilon^0_p} = -\frac{me^2}{\pi k_F}\left(P^{\lambda}_5(p) + P^{\lambda}_6(p) \right),
\label{cross_partial_p_epsilonp_pselfenergy}    
\end{equation}
\begin{equation}
P^{\lambda}_5(p) = -\frac{2}{x^2} -\frac{2(1-x^2-y^2)}{D(x,y)},
\label{P5_lambda}    
\end{equation}
\begin{equation}
P^{\lambda}_6(p) = \frac{g_2(x,y)}{2x^2},
\label{P6_lambda}    
\end{equation}
\begin{equation}
D(x,y) = ((1+x)^2+y^2)((1-x)^2+y^2).
\label{D}    
\end{equation}
Where these expressions are integrated along with $\delta(\mu_0-\epsilon^0_p)$, yielding the results we have obtained in Eqs.~\ref{b'_x_final}--\ref{b''_x_final}, which are the necessary expressions for the coefficients to obtain $b^{\lambda \rightarrow 0}_x$ reported in Eq.~\ref{b_x_final}.  

\subsection{Setting $\lambda=0$ before calculating the integrals (explicit steps).}
In this part of the appendix, we set $\lambda=0$ before calculating the momentum integrals given by Eqs.~\ref{b'_x_3d_2}--\ref{b''_x_3d_2}. This means that we set $\tilde{V}(k)=\tilde{V}_0(k)$, and by using the momentum integral of the Fock self-energy given by Eq.~\ref{Fock_selfenergy_3D}, we obtain the same result we obtained in Eq.~\ref{self_energy_lambda_fock} at $\lambda=0$. In this appendix, we write $\lambda=0$ as a superscript in every relevant term in the integral expressions corresponding to the $b_x$-primed coefficients. We have:
\begin{equation}
\Sigma^{(\lambda=0)}_F(p) = -\frac{e^2 k_F}{\pi} \left[1+\left(\frac{k_F^2-p^2}{2pk_F} \right) \ln{\left|\frac{k_F+p}{k_F-p} \right|} \right].
\label{self_energy_fock_lambdazero}    
\end{equation}
The expression for $P^{(\lambda=0)}(p)$ and their respective partial derivatives are trivial to calculate for the $\lambda=0$ case:
\begin{equation}
P^{(\lambda=0)}(p) = -\frac{2me^2k_F}{\pi p},
\label{P_lambda=0}    
\end{equation}
\begin{equation}
\frac{\partial^2 P^{(\lambda=0}(p)}{\partial p^2} = -\frac{4me^2k_F}{\pi p^3},
\label{P_partial_partial_p_lambda=0}    
\end{equation}
while the last crossed partial derivative of the function $p\Sigma^{(\lambda=0)}_F(p)$ with respect to the energy dispersion $\epsilon^0_p$ and wavevector $p$, turns out to be a problematic term at $p=k_F$. Its corresponding expression is given by:
\begin{equation}
\frac{\partial^2 P^{(\lambda=0)}(p)}{\partial p\partial \epsilon^0_p} = \frac{2me^2 k_F^3}{\pi p^2 (k_F^2-p^2)}.
\label{cross_partial_p_epsilonp_pselfenergy_lambdazero}
\end{equation}
By substituting this term in the three-dimensional integral expressions of $b_x$-primed coefficients given by Eqs.~\ref{b'_x_3d_2}--\ref{b''_x_3d_2}, the Dirac delta function $\delta(\mu^0-\epsilon^0_p)$ in the integrand evaluates this crossed partial derivative term at $p=k_F$, making the final result of these two integrals diverge. This proves analytically that the non-existence of the two coefficients: $b^{'(\lambda=0)}_x$ and $b^{''(\lambda=0)}_x$, which is a strong indication that the exchange part of the functional within the gradient expansion does not exist either.     

\section{Explicit calculation of $b'_c$ and $b^{''}_c$}
\label{explicit_calculation_bc}
In this section, we explain the main mathematical steps of the calculation of $b_c$, which is given by the sum of the three $b_c$ primed terms. We can do this by keeping track of the integrals that defines the functions $I_i(k^{\mu})$ (for $i=1,2,3$), given by Eqs.~\ref{I1}--\ref{I3}. To do this, we first point out that these functions are being multiplied at the integrand level, by an even function in the frequency variable in the expressions of the two $b_c$-primed coefficients given by Eqs.~\ref{b'_c}--\ref{b''_c}. Then, we can express any of these $I_i(k^{\mu})$ functions (for $i=1,2,3$) in terms of a symmetrized and anti-symmetrized version of these functions under the flip of the sign of the frequency variable $k^0$, which we label them as $I^{S}(k^{\mu})$ and $I^{(A)}(k^{\mu})$ respectively. We have:
\begin{equation}
I_i(k^{\mu}) = I^{(S)}_i(k^{\mu})+I^{(A)}_i(k^{\mu}), \quad (\text{for} \ i=1,2,3),
\label{I_i_sym_antisym}   
\end{equation}
where the symmetrized and anti-symmetrized versions of the functions are given by the following expressions:
\begin{equation}
I^{(S)}_i(k^{\mu})=\frac{I_i(\vec{k},k^0)+I_i(\vec{k},-k^0)}{2}, \quad (\text{for} \ i=1,2,3),
\label{I_i_sym}    
\end{equation}
\begin{equation}
I^{(A)}_i(k^{\mu})=\frac{I_i(\vec{k},k^0)-I_i(\vec{k},-k^0)}{2}, \quad (\text{for} \ i=1,2,3).
\label{I_i_asym}    
\end{equation}
Since the integrals given by the Eqs.\ref{b'_c}--\ref{b''_c} are done over the whole frequency-momentum space, the only terms that will give a non-zero value when integrating over the frequency variable, are given by the symmetrized functions $I^{(S)}_i(k^{\mu})$ (for $i=1,2,3$). This means we can substitute $I_i(k^{\mu}) \rightarrow I^{(S)}(k^{\mu})$ in Eqs.~\ref{b'_c}--\ref{b''_c}.

Then, we find these ``symmetrized" functions $I^{(S)}(k^{\mu})$ under the change of the sign of the frequency variables by using Eqs.~\ref{I1}--\ref{I3}, and Eq.~\ref{I_i_sym}. Such calculation involves expanding the second and third order of the partial derivative terms of $G^0(p^{\mu})$ with respect to $\mu_0$ by using the expressions given by Eqs.~\ref{partial_mu_mu_G0}--\ref{partial_triple_mu_G0}. It also involves a tedious algebraic manipulation and reorganization of the sum of the terms within the expansion by also exploiting the integration by parts technique. After these steps, we obtain the following expressions:
\begin{eqnarray}
I^{(S)}_1(k^{\mu}) &=& \frac{1}{2} \frac{\partial^2 \Pi_0(k^{\mu})}{{\partial k^0}^2} -\frac{1}{2} \frac{\partial^2 J(k^{\mu})}{\partial k^0 \partial \mu_0}\nonumber\\
 &+& \frac{1}{4} \frac{\partial^2 \Pi_0(k^{\mu})}{\partial \mu^2_0}-2 \pi i \delta(k^0)S(k),
\label{Is_1}    
\end{eqnarray}
\begin{eqnarray}
I^{(S)}_2(k^{\mu}) &=& \frac{\partial}{\partial \mu_0} \left[\frac{3}{4}\frac{\partial^2 \Pi_0(k^{\mu})}{\partial {k^0}^2} -\frac{3}{4}\frac{\partial^2 J(k^{\mu})}{\partial k^0 \partial \mu_0}\right]\nonumber\\
&+&\frac{\partial}{\partial \mu_0}\left[\frac{1}{4}\frac{\partial^2 \Pi_0(k^{\mu})}{\partial \mu^2_0} -3 \pi i \delta(k^0) S(k)\right],
\label{Is_2}    
\end{eqnarray}
\begin{equation}
I^{(S)}_3(k^{\mu}) = I^{(S)}_{31}(k^{\mu})+I^{(S)}_{32}(k^{\mu}),
\label{Is_3}    
\end{equation}
\begin{eqnarray}
I^{(S)}_{31}(k^{\mu}) &=& \frac{k^0}{4} \frac{\partial^3 \Pi_0(k^{\mu})}{\partial {k^0}^3}+ \frac{1}{2}\frac{\partial^2 \Pi_0(k^{\mu})}{\partial \mu^2_0}-S_2(k^{\mu})\nonumber\\
&+&\frac{3}{4}\left(\frac{\partial^2 \Pi_0(k^{\mu})}{\partial {k^0}^2}-\frac{\partial^2 J(k^{\mu})}{\partial k^0 \partial \mu_0} \right),
\label{Is_31}    
\end{eqnarray}
\begin{equation}
I^{(S)}_{32}(k^{\mu}) = \mu_0 I^{(S)}_2(k^{\mu}),
\label{Is_32}    
\end{equation}
where the function $S_2(k^{\mu})=4\pi i\delta(k^0)S(k)$, while $S(k)$ and $J(k^{\mu})$ are defined through the following integral expressions:
\begin{eqnarray}
S(k) &=& \int \frac{d^3 p}{(2 \pi)^3} \delta(\mu_0-\epsilon_p)\delta(\mu_0-\epsilon_{\vec{p}+\vec{q}})\nonumber\\
 &=& \frac{m^2 \Theta(2 k_F-k)}{(2 \pi)^2 k},
\label{S(k)}    
\end{eqnarray}
\begin{equation}
J(k^{\mu}) = \int \frac{d^4p}{(2 \pi)^4} \frac{\Theta(\mu_0-\epsilon_p)+\Theta(\mu_0-\epsilon_{\vec{p}+\vec{k}})}{k^0-\Delta(\vec{p},\vec{k})+i\eta \text{sgn}(\Delta(\vec{p},\vec{k}))}, 
\label{J}    
\end{equation}
where $\Delta(\vec{p},\vec{k})= \epsilon^0_{\vec{p}+\vec{k}}-\epsilon^0_{\vec{p}}$.

The next important step for the calculation of the integrals given in Eqs.~\ref{b'_c}--\ref{b''_c} is to find the analytical expressions of $\Pi_0(k^{\mu})$, and also, the partial derivative of the function $J(k^{\mu})$ with respect to $\mu_0$ to check if there are any discontinuities at some value of frequency $k^{0}$. By calculating the four-dimensional integral given by Eq.~\ref{pi0}, we obtain the expression for $\Pi_0(k^{\mu})$ given by:
\begin{equation}
\Pi_0(k^{\mu}) = \Theta(2k_F-k)\Pi^1_0(k^{\mu}) + \Theta(k-2k_F)\Pi^2_0(k^{\mu}),
\label{Pi_0_decomposition}    
\end{equation}
where $\Pi^i_0(k^{\mu})$ (for $i=1,2$), decomposes in terms of specific functions listed below:
\begin{equation}
\Pi^1_0(k^{\mu}) = \sum^3_{i=1} \Pi^{1i}_{0}(k^{\mu})+ k_Fk,
\label{pi_1_0_decomposition}    
\end{equation}
\begin{equation}
\Pi^2_0(k^{\mu}) = \sum^{2}_{i=1} \Pi^{2i}_0(k^{\mu})+k_F k,
\label{pi_2_0}
\end{equation}
\begin{equation}
\Pi^{11}_0(k^{\mu}) = \frac{m}{\pi^2}\left(\frac{m^2\beta^2_{-}-k_F^2k^2}{4k^3} \right)\ln\left(\frac{\gamma^{-}_{-}+i\eta}{\gamma^{-}_{+}-i\eta} \right),
\label{pi_11_0}
\end{equation}
\begin{equation}
\Pi^{12}_0(k^{\mu}) = \frac{m}{\pi^2}\left(\frac{m^2\beta^2_+-k_F^2k^2}{4k^3} \right)\ln\left(\frac{\gamma^{+}_{+}-i\eta}{\gamma^{+}_{-}+i\eta} \right),
\label{pi_12_0}
\end{equation}
\begin{equation}
\Pi^{13}_0(k^{^\mu}) = -\frac{m^2 k^0}{2 \pi^2 k}\ln \left(\frac{k^0-i\eta}{k^0+i\eta} \right),
\label{pi_13_0}    
\end{equation}
\begin{equation}
\Pi^{21}_0(k^{\mu}) = \frac{m}{\pi^2}\left(\frac{m^2\beta^2_{-}-k_F^2k^2}{4k^3} \right)\ln\left(\frac{\gamma^{-}_{-}+i\eta}{\gamma^{-}_{+}+i\eta} \right),
\label{pi_21_0}
\end{equation}
\begin{equation}
\Pi^{22}_0(k^{\mu}) = \frac{m}{\pi^2}\left(\frac{m^2\beta^2_+-k_F^2k^2}{4k^3} \right)\ln\left(\frac{\gamma^{+}_{+}-i\eta}{\gamma^{+}_{-}-i\eta} \right),
\label{pi_22_0}
\end{equation}
where $\beta_{\pm}=k^0\pm k^2/2m$, $\gamma^{+}_{\pm} = \beta_{+}\pm k_Fk/m$ and $\gamma^{-}_{\pm} = \beta_{-}\pm k_F k/m$. These expressions are important since we can extract, from the second and third-order partial derivatives of $\Pi_0(k^{\mu})$ with respect to the frequency, Dirac delta functions $\delta(k^0)$. We can do this by separating the principal part values of the following terms:
\begin{equation}
\lim_{\eta \rightarrow0}\frac{1}{x\pm i\eta} = PV\left(\frac{1}{x}\right)\mp i\pi \delta(x).
\label{identity_denominator_principalvalue_separation}    
\end{equation}
The Dirac delta functions that arise from the derivative terms of $\Pi_0(k^{\mu})$, come specifically from the term $\Pi^{13}_0(k^{\mu})$. We have that the partial derivatives of this term are given by:
\begin{equation}
\frac{\partial^2 \Pi^{13}_0(k^{\mu})}{\partial {k^0}^2} = -4\pi i\delta(k^0)S(k),
\label{k0_partial_double_k0_pi0}    
\end{equation}
\begin{equation}
k^0\frac{\partial^3 \Pi^{13}_0(k^{\mu})}{\partial {k^0}^3} = 4\pi i \delta(k^0)S(k).
\label{k0_partial_triple_k0_pi0}    
\end{equation}
We reorganize the sum of the terms given in Eq.~\ref{Pi_0_decomposition} as follows:
\begin{equation}
\Pi_0(k^{\mu}) = \Pi_1(k^{\mu})+\Pi^{13}_0(k^{\mu}),
\label{Pi0_extract_delta}    
\end{equation}
\begin{eqnarray}
\Pi_1(k^{\mu}) &=& \Theta(2k_F-k)\sum^2_{i=1}\Pi^{1i}_0(k^{\mu})\nonumber\\ 
&+& \Theta(k-2k_F)\sum^2_{i=1}\Pi^{2i}_{0}(k^{\mu})+k_Fk.
\label{Pi1}    
\end{eqnarray}

Similarly as done for the polarization function $\Pi_0(k^{\mu})$, we obtain the expression for the partial derivative of $J(k^{\mu})$ with respect to $\mu_0$. We have:
\begin{eqnarray}
\frac{\partial J(k^{\mu})}{\partial \mu_0} &=& \Theta(2k_F-k)J_1(k^{\mu})\nonumber\\
&+&\Theta(k-2k_F)J_2(k^{\mu})+J_3(k^{\mu}),
\label{partial_mu_J_decomposition}    
\end{eqnarray}
where these functions are given by the following expressions:
\begin{equation}
J_{1}(k^{\mu}) = \frac{m^2}{2\pi^2k}\ln\left(\frac{(\gamma^{+}_{+}-i\eta)(\gamma^{-}_{+}-i\eta)}{(\gamma^{+}_{-}+i\eta)(\gamma^{-}_{-}+i\eta)} \right),
\label{J1}    
\end{equation}
\begin{equation}
J_{2}(k^{\mu}) = \frac{m^2}{2\pi^2k} \ln\left(\frac{(\gamma^{+}_{+}-i\eta)(\gamma^{-}_{+}+i\eta)}{(\gamma^{+}_{-}-i\eta)(\gamma^{-}_{-}+i\eta)} \right),
\label{J2}    
\end{equation}
\begin{equation}
J_3(k^{\mu}) = \frac{m^2}{\pi^2k}\Theta(2k_F-k)\ln\left(\frac{k^0+i\eta}{k^0-i\eta} \right),
\label{J3}    
\end{equation}
where the partial derivative terms of $J_3(k^{\mu})$ yield a Dirac delta function $\delta(k^0)$. The partial derivative terms of this function are given by:
\begin{equation}
\frac{\partial J_3(k^{\mu})}{\partial k^0} = -8 \pi i\delta(k^0)S(k),
\label{partial_k0_J3}    
\end{equation}
The extracted Dirac delta functions from the partial derivative terms of the polarization function $\Pi_0(k^{\mu})$ and $J_3(k^{\mu})$ cancels out the other Dirac delta functions that we have in Eqs.~\ref{Is_1}--\ref{Is_32}. After this cancellation, we can keep track the frequency integrals expressions of the $b_c$-primed coefficients, by mapping the integral along the imaginary frequency $i\nu$. This is done by using a specific complex contour as we did in our previous work when calculating the correlation energy functional in the LDA in the large density limit\cite{benites2024}. After doing this, we write the momentum and imaginary frequency into dimensionless variables as follows: $k=k_F k'$ and $\nu={k_F}^2\nu'/m$. After this, we use the change of variable $\nu'=k'y$ and finally we set $k'=2x$. These steps yields the following expressions for $I^{(S)}_i(k^{\mu})$ (for $i=1,2,3$):
\begin{eqnarray}
I^{(S)}_1(k^{\mu}) &=& \frac{m^3}{\pi^2k_F^3}\left(\frac{g(x,y)}{32x^3}+\frac{N_1(x,y)}{8x^2D(x,y)} \right),\label{Is_1_final}\\
N_1(x,y) &=& 3(x^2-y^2)-1-2x^2(x^2+y^2),\label{N1}\\  
I^{(S)}_2(k^{\mu}) &=& \frac{m^4}{2\pi^2k_F^5x^2}I'_2(x,y),\label{Is_2_final}\\    
I^{S}_3(k^{\mu}) &=& \frac{m^3}{4\pi^2k_F^3x^2}I'_3(x,y)+\mu_0 I_2(x,y),\label{Is_3_final}    
\end{eqnarray}
where $g(x,y)$ and $D(x,y)$ are given by Eqs.~\ref{g(x,y)} and \ref{D}, respectively. The functions $I'_2(x,y)$ and $I'_3(x,y)$ are given by the following expressions:
\begin{equation}
I'_2(x,y) = \frac{M_1(x,y)}{D(x,y)}+\frac{M_2(x,y)}{(D(x,y))^2},
\label{I'_2}    
\end{equation}
\begin{equation}
I'_3(x,y) = \frac{3g(x,y)}{16x}+\frac{M_3(x,y)+M_4(x,y)}{(D(x,y))^2},
\label{I'_3}    
\end{equation}
where the functions $M_i(x,y)$ (for $i=1,2,3,4$) are given by:
\begin{eqnarray}
M_1(x,y) &=& x^2y^2-\frac{1}{4}(y^2+1-x^2)(3+2x^2),\label{M1}\\    
M_2(x,y) &=& \left(\frac{3}{4}-x^2\right)N_2(x,y)+N_3(x,y),\label{M2}\\
N_2(x,y) &=& (y^2+1-x^2)^2-4x^2y^2,\label{N2}\\
N_3(x,y) &=& 4y^2(y^2+1-x^2)\left(\frac{3}{4}+x^2 \right),\label{N3}\\    
M_3(x,y) &=& N_4(x,y)D(x,y),\label{M3}\\
N_4(x,y) &=& \frac{-8x^4-x^2(8y^2-11)-13y^2-3}{4},\label{N4}\\    
M_4(x,y) &=& y^2(x^4+2x^2y^2+y^4-1),\label{M4}    
\end{eqnarray}
These are the required expressions to obtain $I^{(s)}_{i}(k^{\mu})$ (for $i=1,2,3$) by using Eqs.~\ref{Is_1_final}--\ref{Is_3_final}, that can now be substituted in the expressions of both $b_c$-primed coefficients given by Eqs.~\ref{b'_c}--\ref{b''_c}. The next step after this substitution of terms, we must specify the explicit expression of the potential $\tilde{V}(k)$.

\subsection{Calculation of $b'_c$ by using the $k_F$-dependent regulator $\lambda$ (explicit steps)}
In this part of the appendix we calculate $b'_c$ by using a regularized-Coulomb potential given by Eq.~\ref{V_generic}. We also carry over the constant part of the regulator $\lambda_c$ in the rest of the calculation of the $b^{'}_c$ coefficient. 

At this point of our calculation, we have done the substitution of the functions $I^{(S)}_i(k^{\mu})$ in the expressions of $b'_c$ given by Eq.~\ref{b'_c}, which results in a four-dimensional integral that can be reduced into a two-dimensional expression. This is obtained after integrating over the two spherical angle variables. The two-dimensional integral is obtained from the expression given in Eq.~\ref{b'_c_2d}, which depends on the function $C(x,y)$. This is given by:
\begin{eqnarray}
C(x,y) = \frac{M_5(x,y)}{D(x,y)}&+&\frac{M_6(x,y)}{9(D(x,y))^2}\nonumber\\
&-&\frac{g(x,y)}{24 x} + \frac{I'_2(x,y)}{9},
\label{C_xy}    
\end{eqnarray}
where the functions $M_5(x,y)$ and $M_6(x,y)$ are given below:
\begin{equation}
M_5(x,y) = \left(\frac{1}{6}-\frac{5x^2}{18}\right)(1+y^2-x^2)+\frac{y^2}{9}(2+5x^2), 
\label{M5}    
\end{equation}
\begin{equation}
M_{6}(x,y) = y^2(x^4+2x^2y^2+y^4-1).
\label{M6}    
\end{equation}

From this point, we explain first on how to obtain the leading contribution in $r_s$ for the $b'_c$ coefficient. The expression for $b'_c$ is expressed as a sum of three terms because of the partial derivative with respect to $\mu_0$ operating over the integral given in Eq.~\ref{b'_c_2d}. We have:
\begin{equation}
b^{'\lambda}_c = \sum^{3}_{i=1} b^{'}_i,
\label{b'_c_lambda_decomposition}    
\end{equation}
\begin{equation}
b'_{1} = -\frac{e^4 m^3}{2 \pi^5 k_F^3} \int^{\infty}_0 dx \int^{\infty}_0 dy \frac{xC(x,y)Q(x,y)}{(\overline{\epsilon_{\lambda}}(x,y))^2},
\label{b'_lambda_1}    
\end{equation}
\begin{eqnarray}
b'_{2} &=& \frac{e^4 m^3}{4\pi^5 k_F^3} \int^{\infty}_0 dx \int^{\infty}_0 dy \frac{\lambda'^2 C(x,y)Q(x,y)}{(\overline{\epsilon_{\lambda}}(x,y))^2}\nonumber\\
 &\times& \frac{x}{x^2+\frac{\lambda'^2}{4}}(1-k_F \frac{\partial \ln(\beta(k_F)}{k_F}),\label{b'_lambda_2}\\    
b'_{3} &=& \frac{\lambda'^2 m}{2k_F^2} \int^{\infty}_0 dx \int^{\infty}_0 dy \frac{C(x,y)Z_{\lambda}(x,y)}{x^2+\frac{\lambda'^2}{4}}\nonumber\\
&\times&(1-k_F\frac{\partial \ln(\beta(k_F))}{\partial k_F}),
\label{b'_lambda_3}    
\end{eqnarray}
where in all of these expressions we have $\lambda'=\lambda/k_F$. In $b'_1$, we can ignore the $\lambda'$ term since it is an infinitesimally small parameter and is only contained in $\overline{\epsilon_{\lambda}}(x,y)$. This term is precisely Geldart-Rasolt's second term in the expression of $b'$ reported in \cite{Geldart-Rasolt}. The coefficient $b'_2$ is proven to be zero by doing the change of variable $x=\lambda' x'$, which ends up being an expression that is quadratic in $\lambda$, where at this stage we take the limit of $\lambda_c \to 0$. Similarly for $b'_3$, we apply again the change of variable $x=\lambda'x'$. After doing this, we keep up the dominant terms of the integrand and we keep the zeroth order of the Taylor expansion in the first argument of the functions $C(\lambda'x,y)$ and $Q(\lambda'x,y)$ since we are interested in taking the $\lambda_c \to 0$ limit, obtaining the following expression:
\begin{eqnarray}
b'_3 &=& \frac{e^2m^2}{\pi^4k_F^2}\int^{\infty}_0dx' \int^{\infty}_0 dy \frac{C(0,y)x'}{({x'}^2+\frac{1}{4})^2}\nonumber\\
   &\times& \left(1-k_F\frac{\partial \ln(\beta(k_F))}{\partial k_F}\right)     ,
\label{b'_lambda_3_2nd}
\end{eqnarray}
where this integral can be calculated analytically. In fact, when we integrate over $x'$, we obtain Geldart-Rasolt's one-dimensional integral expression of $b'$ given in Ref.~\cite{Geldart-Rasolt}. This one-dimensional integral is straightforward to calculate with common integration techniques, so we obtain that the leading term of $b'_3$ in $r_s$ is given by:
\begin{equation}
b'_3 = \frac{5e^2m^2}{36\pi^3k_F^2}\left(1-k_F\frac{\partial \ln(\beta(k_F))}{\partial k_F}\right).
\label{b'_lambda_3_final}    
\end{equation}

Finding the leading contribution in $r_s$ from $b'_1$ can be extracted by only focusing in the region of integration $x\in(0,1/2)$, which is equivalent of calculating the integral in the region $k \in (0,k_F)$ due to the change of variables done in this work. In this region, we can keep the zeroth-order of the Taylor expansion in the $x$ variable for both functions: $C(x,y)$ and $Q(x,y)$. By doing this we obtain the following two-dimensional integral:
\begin{equation}
b^{'}_1 = -\frac{e^4 m^3}{2 \pi^5 k_F^3} \int^{\frac{1}{2}}_0 dx\int^{\infty}_0dy\frac{xQ(0,y)C(0,y)}{(x^2+\frac{\alpha r_s Q(0,y)}{4 \pi})^2},
\label{b'_c_lambda_correction_2}    
\end{equation}
where it is straightforward to do the integration over variable $x$, where we obtain the following one-dimensional integral:
\begin{equation}
b'_1 \approx -\frac{e^2m^2}{36\pi^4k_F^2} \int^{\infty}_0 dy \frac{y^2(13+9y^2)}{(1+y^2)(1+\frac{\alpha r_s}{\pi}Q(0,y))},  
\label{b'_lambda_1_2nd}
\end{equation}
where we can now perform a Taylor expansion in small $r_s$ to the denominator term. Then the integration over the $y$ variable can be performed, where we obtain the final result of $b'_1$:
\begin{equation}
b'_1 = -\frac{5e^2m^2}{72\pi^3k_F^2}.
\label{b'_lambda_1_final}    
\end{equation}
Then, the final expression of $b^{'\lambda}_c$ is given precisely by using the sum in Eq.~\ref{b'_c_lambda_decomposition}, which reproduces the result reported in Eq.~\ref{b'_c_final_result}.

\subsection{Explicit calculation of $b^{'}_c$ without using a regulator}
In this part of the appendix we prove that the integral given in the expression of $b^{'}_c$, given by Eq.~\ref{b'_c}, does not converge. The same steps of reducing this expression into a two-dimensional integral, as we have done by using the $k_F$-dependent $\lambda$, also applies for the $\lambda=0$ case. By doing this, we obtain the following two-dimensional integral for $b'_c$:
\begin{equation}
b^{'(\lambda=0)}_c = \frac{\partial}{\partial \mu_0}\int^{\infty}_0 dx \int^{\infty}_0 dy C(x,y)Z(x,y),
\label{b'_c_lambda_0}    
\end{equation}
where $Z(x,y)$ is obtained from the expression of $Z_{\lambda}(x,y)$ given in Eq.~\ref{Z_function}, by setting $\lambda=0$. At this point of the calculation, to obtain the leading contribution to the coefficient $b'_c$, we focus on the region of integration $x \in (0,1/2)$. In this region of integration, the non-converging part of the integral comes from the second term of the integral given by Eq.~\ref{b'_c_lambda_0}. By doing this, we can do a Taylor expansion  of the functions $C(x,y)$ and $Q(x,y)$ and keep their respective zeroth-order terms. We obtain:
\begin{eqnarray}
b^{'(\lambda=0)}_c &=& \frac{\partial}{\partial \mu_0} \int^{\frac{1}{2}}_0 dx \int^{\infty}_0 dy \frac{C(0,y)Q(0,y)}{xD_2(x,y)},\label{b'_c_lambda_0_2}\\
D^{\lambda}_2(x,y) &=& \frac{2 \pi^5 k_F}{e^4 m^2} \left(x^2+\frac{\lambda'^2}{4} + \frac{\alpha r_s Q(0,y)}{4\pi}\right),\label{D2_lambda_function}    
\end{eqnarray}
where $D_2(x,y)$ is obtained by setting $\lambda=0$ in the expression of $D^{\lambda}_2(x,y)$. We notice that the integral from Eq.~\ref{b'_c_lambda_0_2} will diverge when integrating over the $x$ variable. This implies that the value of the coefficient $b^{'}_c$ do not exist, given that the partial derivative with respect to $\mu_0$ is acting over a divergent integral. 

\subsection{Calculation of $b^{''}_c$ by using the $k_F$-dependent regulator $\lambda$ (Explicit Steps)}
In this part of the appendix, the calculation of the coefficient $b^{''}_c$ is independent on the form of $\beta(k_F)$ that we used in this work. At this level we have already substituted the functions $I^{S}_i(x,y)$ (for $i=1,2,3$) in the integral expression of $b^{''}_c$ given by Eq.~\ref{b''_c}, and that the mapping of the integral in the frequency variable $k^0$ into the imaginary frequency $i\nu$. After we have set either $\tilde{V}(k)=V_{\lambda}(k)$, we can calculate the integrals over the two spherical angles, given that the integrands only depend on the magnitude of the momentum and the imaginary frequency. These steps have been done by being consistent with the change of variables that we have explained so far in this appendix section and we obtain the two-dimensional integral expression for $b^{''}_{c}$ given by Eq.~\ref{b''_c_2d}.

We can obtain the leading contribution to the coefficient $b^{''}_c$ by focusing the integral over the $x$ variable in the region of integration $x \in (0,1/2)$, as we have done in the calculation of $b^{'}_c$. Within this region of integration we can keep the zeroth order of the Taylor expansion of the functions $I'_2(x,y)$  and $Q(x,y)$ in the expression $\overline{\epsilon_{\lambda}}(x,y)$, where we obtain the two integrals:
\begin{equation}
b^{''}_c = b^{''}_1+b^{''}_2,
\label{b''_c_decomposition_appendix}    
\end{equation}
where these coefficients are given by the following terms:
\begin{equation}
b^{''}_1 = -\frac{4\pi}{3 e^2 k_F}\int^{\frac{1}{2}}_0 dx\int^{\infty}_0 dy\frac{x I'_2(0,y)}{D^{\lambda}_2(x,y)},
\label{b''_1}    
\end{equation}
\begin{equation}
b^{''}_2 = \frac{2m^2 e^2}{3\pi^4k_F^2}\int^{\frac{1}{2}}_0 dx \int^{\infty}_0 dy \frac{x I'_2(0,y)}{x^2+\frac{{\lambda'}^2}{4}},
\label{b''_2}    
\end{equation}
where we have used again $\lambda'=\lambda(k_F)/k_F$ as a shorthand expression.

We proceed to do the calculation of the integrals associated to the expression $b^{''}_2$. After integrating over the variable $x$ we obtain the following one-dimensional integral:
\begin{equation}
b^{''}_2 = \frac{m^2 e^2}{4 \pi^4 k_F^2}\ln\left|\frac{1+\lambda'^2}{{\lambda'}^2} \right| \int^{\infty}_0 dy \frac{y^2(3-y^2)}{(1+y^2)^3},
\label{b''_2_2}
\end{equation}
where the integral over the $y$ variable yields zero, therefore $b^{''}_2=0$.

The integral over the $x$ variable, given by Eq.~\ref{b''_1}, is straightforward to calculate. We obtain the following expression for $b^{''}_1$:
\begin{equation}
b^{''}_1 = -\frac{m^2e^2}{3 \pi^4 k_F^2}\int^{\infty}_0 dy I'_2(0,y) \ln\left|\frac{D^{\lambda}_2(\frac{1}{2},y)}{D^{\lambda}_2(0,y)} \right|,
\label{b''_1_3}    
\end{equation}
where we can ignore $\lambda'$ since it is infinitesimally small. After this, a Taylor expansion of the logarithm terms is performed in small $r_s$. By keeping only the leading term in $r_s$ and exploiting the fact that the integral of $I'_2(0,y)$ in the region $y \in (0,\infty)$ yields zero, we obtain:
\begin{equation}
b^{''}_1 = \frac{m^2 e^2}{4 \pi^4 k_F^2} \int^{\infty}_0 dy \frac{y^2(3-y^2)}{(1+y^2)^3} \ln\left|Q(0,y)\right|,
\label{b''_1_4}    
\end{equation}
where $Q(0,y)=4(1-y\tan^{-1}(1/y))$, and by calculating the integral over the $y$ variable, we obtain the expression given by Eq.~\ref{b''_c_final_result}. This proves that if we use a $k_F$-dependent regulator in $\tilde{V}(q)$, this coefficient always converges to a unique value. 

\subsection{Explicit calculation of $b^{''}_c$ without using a regulator}
In this part of the appendix we set $\tilde{V}(k)=\tilde{V}_0(k)$, where we prove that the integral from the expression of $b^{''}_c$ does not converge. The same set of steps used to obtain the two-dimensional integral expressions of $b^{''}_c$ by carrying over a regulator is, also applied in this case where we set $\lambda=0$ in Eq.~\ref{b''_c_2d} at the integrand level. In the region of integration $x \in (0,1/2)$, by keeping the zeroth-order of the Taylor expansion of the functions $I'_2(x,y)$ and $Q(x,y)$, where we obtain the same expressions given by Eqs.~\ref{b''_c_decomposition_appendix}--\ref{b''_2} and setting $\lambda=0$. The problematic part of the obtained integral expression comes from $b^{''(\lambda=0)}_2$, given by:
\begin{equation}
b^{''(\lambda=0)}_2 = \frac{2m^2e^2}{3\pi^4 k_F^2} \int^{\frac{1}{2}}_0 dx \int^{\infty}_0 dy \frac{I'_2(0,y)}{x},
\label{b''_lambda_0_2}    
\end{equation}
where the superscript indicates that we have set the regulator $\lambda$ to zero at the integrand level. This specific integral depends on the order of which variables are integrated. In this case, if the integral over the $y$ variable is calculated first, we obtain that the integral is zero. If we integrate over the $x$ variable first, however, the integral diverges, which proves that the integral does not exist, which was also previously pointed out by \cite{Kleinman-Tamura}.

\section{Explicit calculation of $b^{'''}_{xc}$}
\label{b'''_xc_calculations}
In this section we give more details of the calculation of the leading term of $b^{'''}_{xc}$ in the $r_s \to 0$ limit. We point out that in these steps, it is not required to specify whether a regulator was used in the expression of the potential $\tilde{V}(k)$. This is because $\lambda$ can be neglected in the expressions of $\overline{\epsilon_{\lambda}}(x,y)$ since the regulator is infinitesimally small. In the expression of $b^{'''}_{xc}$ given by Eq.~\ref{b'''_xc}, we can handle the frequency $k^0$ integral by using the contour complex path we used for the calculation of both $b_c$-primed coefficients. This maps the frequency integral into the imaginary frequency line, which means $k^0=i \nu$, where the expression of $b^{'''}_{xc}$ can be re-expressed as follows:
\begin{equation}
b^{'''}_{xc} = b^{'''}_1+b^{'''}_2,
\label{b'''_xc_decomposition}    
\end{equation}
\begin{eqnarray}
b'''_{1} = -\frac{1}{384 \pi^4} \int d^3k \int^{\infty}_{-\infty}&d\nu& \left(\frac{\partial \Pi_0(k^{\mu})}{\partial \mu} \right)^2\nonumber \\
&\times& \tilde{V}_e(k,i\nu)\nabla^2_{k}\tilde{V}_e(k,i\nu),
\label{b'''_1}    
\end{eqnarray}
\begin{equation}
b'''_{2} = \int\frac{d^3k}{384\pi^4} \int^{\infty}_{-\infty}d\nu \left(\frac{\partial \Pi_0(k^{\mu}_{c})}{\partial \mu} \right)^2 \left(\partial_k \tilde{V}_e(k^{\mu}_{c})\right)^2.
\label{b'''_2}    
\end{equation}
where we have used the shorthand notation $k^{\mu}_{c}=(\vec{k},i\nu)$. At this point of the calculation, we use the same sequence of transformation of variables used for the calculation of $b'_c$ and $b^{''}_c$. After such changes of variables, we end up having $k=2k_F x$ and $k^0=i2k_F^2xy/m$. At this point, we work out each term that we had in Eq.~\ref{b'''_xc}. One term involves a divergence term (in spherical coordinates) acting over the renormalized-potential $\tilde{V}_e(k^{\mu})$ within the RPA. The other term involves the square of the derivative of this renormalized potential with respect to magnitude $k$. We list the relevant terms for the calculation of the $b^{'''}_{xc}$ coefficient, within this transformation of variables:
\begin{equation}
\frac{\partial V_{e}(k^{\mu}_c)}{\partial k} = -\frac{\pi e^2}{k_F^3} \frac{M_7}{(\overline{\epsilon})^2},
\label{V_eff_partial_q}    
\end{equation}
\begin{equation}
\frac{\partial^2 \tilde{V}_e(k^{\mu}_c)}{\partial k^2} = -\frac{\pi e^2}{4 k_F^4} \left(\frac{M_8}{(\overline{\epsilon})^2}-\frac{8(M_7)^2}{(\overline{\epsilon})^3} \right),
\label{V_eff_double_partial_q}    
\end{equation}
\begin{equation}
M_7 = x+\frac{\alpha r_s}{4\pi}\left(\frac{2}{x}-\frac{(y^2+x^2)}{2x^2}g(x,y)-\frac{Q(x,y)}{2x}\right),
\label{M7}    
\end{equation}
\begin{equation}
M_8 = 2+\frac{\alpha r_s}{4\pi}(M_{8,1}+M_{8,2}),
\label{M8}    
\end{equation}
\begin{equation}
M_{8,1} = -\frac{2}{x^2}+\left(\frac{5y^2+x^2}{4x^3}\right)g(x,y)+\frac{Q(x,y)}{2x^2},
\label{M81}
\end{equation}
\begin{equation}
M_{8,2} = \frac{(x^2+y^2)(x^2-3y^2-1)}{x^2 D(x,y)},
\label{M82}
\end{equation}
where we have used $\overline{\epsilon}$ as a shorthand notation for $\overline{\epsilon}(x,y)$. This function is obtained from the expression of $\overline{\epsilon_{\lambda}}(x,y)$ given by Eq.~\ref{epsilon_bar} by setting $\lambda=0$. By using these expressions in Eqs.\ref{b'''_xc_decomposition}--\ref{b'''_2}, we finally obtain the two-dimensional integral expression of $b^{'''}_{xc}$ we reported in Eq.~\ref{b'''_c_2d}, which are expressed in terms of the functions $M_9(x,y)$ and $M_{10}(x,y)$, given below:
\begin{eqnarray}
M_9(x,y) = 4&-&\left(\frac{7y^2+3x^2}{4x}\right)g(x,y)\nonumber\\
     &-&\frac{3}{2}Q(x,y)-x^2M_{8,2},
\label{M9}    
\end{eqnarray}
\begin{equation}
M_{10}(x,y) = \frac{4-3 Q(x,y)}{2}-\left(\frac{y^2+x^2}{2x}\right)g(x,y).
\label{M10}    
\end{equation}
We extract the leading $r_s$ contribution of the coefficient $b^{'''}_{xc}$  by focusing in the region of integration $x\in(0,1/2)$ and carefully keeping track of the dominant contributions to the integral. In this region, we treat $x\ll1$ and we keep the zeroth order term from the Taylor expanded forms of $Q(x,y)$. By doing this, we obtain the following expressions:
\begin{equation}
b^{'''}_{xc} = -\frac{e^4 m^3}{3 \pi^5 {k_F}^3}(M_{11}(r_s)+M_{12}(r_s)),
\label{proof_3_prime_step1}
\end{equation}
where $M_{11}(r_s)$ and $M_{12}(r_s)$ are a more compact one-dimensional integral expressions, obtained after integrating over the $x$ variable, which are given by:
\begin{eqnarray}
M_{11}(r_s) &=& \frac{\pi}{2 \alpha r_s} \int^{\infty}_{0} dy \frac{1}{Q(0,y)(1+y^2)^2}\nonumber\\
              &\times&\left(1+\frac{16}{3(1+y^2)^2 (Q(0,y))^2} \right), \label{M11}\\
M_{12}(r_s) &=& -\frac{\pi}{2 \alpha r_s} \int^{\infty}_0 dy \frac{1}{(1+y^2)Q(0,y)}\nonumber\\
            &\times&\left(3+\frac{4y^2}{(1+y^2)^2 Q(0,y)} \right),              
\end{eqnarray}
where $\alpha=(4/9\pi)^{1/3}$ and the sum of these two integrals can be reduced into the following simplified integral expression:
\begin{equation}
b^{'''}_{xc} = \frac{e^2m^2}{9\pi^4 k_F^2}\int^{\infty}_0 dy \frac{(3+2y^2)}{(1+y^2)^3}\frac{1}{Q(0,y)},
\label{b'''_{xc}_1D}    
\end{equation}
where this integral expression was obtained by Ma-Brueckner and can be calculated numerically. By calculating this integral, we finally obtain the leading term  in $r_s$  of $b^{'''}_{xc}$ which we reported in Eq.~\ref{b'''_c_final_result}.


\begin{thebibliography}{33}%
\makeatletter
\providecommand \@ifxundefined [1]{%
 \@ifx{#1\undefined}
}%
\providecommand \@ifnum [1]{%
 \ifnum #1\expandafter \@firstoftwo
 \else \expandafter \@secondoftwo
 \fi
}%
\providecommand \@ifx [1]{%
 \ifx #1\expandafter \@firstoftwo
 \else \expandafter \@secondoftwo
 \fi
}%
\providecommand \natexlab [1]{#1}%
\providecommand \enquote  [1]{``#1''}%
\providecommand \bibnamefont  [1]{#1}%
\providecommand \bibfnamefont [1]{#1}%
\providecommand \citenamefont [1]{#1}%
\providecommand \href@noop [0]{\@secondoftwo}%
\providecommand \href [0]{\begingroup \@sanitize@url \@href}%
\providecommand \@href[1]{\@@startlink{#1}\@@href}%
\providecommand \@@href[1]{\endgroup#1\@@endlink}%
\providecommand \@sanitize@url [0]{\catcode `\\12\catcode `\$12\catcode
  `\&12\catcode `\#12\catcode `\^12\catcode `\_12\catcode `\%12\relax}%
\providecommand \@@startlink[1]{}%
\providecommand \@@endlink[0]{}%
\providecommand \url  [0]{\begingroup\@sanitize@url \@url }%
\providecommand \@url [1]{\endgroup\@href {#1}{\urlprefix }}%
\providecommand \urlprefix  [0]{URL }%
\providecommand \Eprint [0]{\href }%
\providecommand \doibase [0]{https://doi.org/}%
\providecommand \selectlanguage [0]{\@gobble}%
\providecommand \bibinfo  [0]{\@secondoftwo}%
\providecommand \bibfield  [0]{\@secondoftwo}%
\providecommand \translation [1]{[#1]}%
\providecommand \BibitemOpen [0]{}%
\providecommand \bibitemStop [0]{}%
\providecommand \bibitemNoStop [0]{.\EOS\space}%
\providecommand \EOS [0]{\spacefactor3000\relax}%
\providecommand \BibitemShut  [1]{\csname bibitem#1\endcsname}%
\let\auto@bib@innerbib\@empty
\bibitem [{\citenamefont {Wigner}(1934)}]{PhysRev.46.1002}%
  \BibitemOpen
  \bibfield  {author} {\bibinfo {author} {\bibfnamefont {E.}~\bibnamefont
  {Wigner}},\ }\bibfield  {title} {\bibinfo {title} {On the interaction of
  electrons in metals},\ }\href {https://doi.org/10.1103/PhysRev.46.1002}
  {\bibfield  {journal} {\bibinfo  {journal} {Phys. Rev.}\ }\textbf {\bibinfo
  {volume} {46}},\ \bibinfo {pages} {1002} (\bibinfo {year}
  {1934})}\BibitemShut {NoStop}%
\bibitem [{\citenamefont {Bohm}\ and\ \citenamefont
  {Pines}(1951)}]{PhysRev.82.625}%
  \BibitemOpen
  \bibfield  {author} {\bibinfo {author} {\bibfnamefont {D.}~\bibnamefont
  {Bohm}}\ and\ \bibinfo {author} {\bibfnamefont {D.}~\bibnamefont {Pines}},\
  }\bibfield  {title} {\bibinfo {title} {A collective description of electron
  interactions. i. magnetic interactions},\ }\href
  {https://doi.org/10.1103/PhysRev.82.625} {\bibfield  {journal} {\bibinfo
  {journal} {Phys. Rev.}\ }\textbf {\bibinfo {volume} {82}},\ \bibinfo {pages}
  {625} (\bibinfo {year} {1951})}\BibitemShut {NoStop}%
\bibitem [{\citenamefont {Pines}\ and\ \citenamefont
  {Bohm}(1952)}]{PhysRev.85.338}%
  \BibitemOpen
  \bibfield  {author} {\bibinfo {author} {\bibfnamefont {D.}~\bibnamefont
  {Pines}}\ and\ \bibinfo {author} {\bibfnamefont {D.}~\bibnamefont {Bohm}},\
  }\bibfield  {title} {\bibinfo {title} {A collective description of electron
  interactions: Ii. collective $\mathrm{vs}$ individual particle aspects of the
  interactions},\ }\href {https://doi.org/10.1103/PhysRev.85.338} {\bibfield
  {journal} {\bibinfo  {journal} {Phys. Rev.}\ }\textbf {\bibinfo {volume}
  {85}},\ \bibinfo {pages} {338} (\bibinfo {year} {1952})}\BibitemShut
  {NoStop}%
\bibitem [{\citenamefont {Bohm}\ and\ \citenamefont
  {Pines}(1953)}]{PhysRev.92.609}%
  \BibitemOpen
  \bibfield  {author} {\bibinfo {author} {\bibfnamefont {D.}~\bibnamefont
  {Bohm}}\ and\ \bibinfo {author} {\bibfnamefont {D.}~\bibnamefont {Pines}},\
  }\bibfield  {title} {\bibinfo {title} {A collective description of electron
  interactions: Iii. coulomb interactions in a degenerate electron gas},\
  }\href {https://doi.org/10.1103/PhysRev.92.609} {\bibfield  {journal}
  {\bibinfo  {journal} {Phys. Rev.}\ }\textbf {\bibinfo {volume} {92}},\
  \bibinfo {pages} {609} (\bibinfo {year} {1953})}\BibitemShut {NoStop}%
\bibitem [{\citenamefont {Pines}(1953)}]{PhysRev.92.626}%
  \BibitemOpen
  \bibfield  {author} {\bibinfo {author} {\bibfnamefont {D.}~\bibnamefont
  {Pines}},\ }\bibfield  {title} {\bibinfo {title} {A collective description of
  electron interactions: Iv. electron interaction in metals},\ }\href
  {https://doi.org/10.1103/PhysRev.92.626} {\bibfield  {journal} {\bibinfo
  {journal} {Phys. Rev.}\ }\textbf {\bibinfo {volume} {92}},\ \bibinfo {pages}
  {626} (\bibinfo {year} {1953})}\BibitemShut {NoStop}%
\bibitem [{\citenamefont {Gell-Mann}\ and\ \citenamefont
  {Brueckner}(1957)}]{PhysRev.106.364}%
  \BibitemOpen
  \bibfield  {author} {\bibinfo {author} {\bibfnamefont {M.}~\bibnamefont
  {Gell-Mann}}\ and\ \bibinfo {author} {\bibfnamefont {K.~A.}\ \bibnamefont
  {Brueckner}},\ }\bibfield  {title} {\bibinfo {title} {Correlation energy of
  an electron gas at high density},\ }\href
  {https://doi.org/10.1103/PhysRev.106.364} {\bibfield  {journal} {\bibinfo
  {journal} {Phys. Rev.}\ }\textbf {\bibinfo {volume} {106}},\ \bibinfo {pages}
  {364} (\bibinfo {year} {1957})}\BibitemShut {NoStop}%
\bibitem [{\citenamefont {Nozi\`eres}\ and\ \citenamefont
  {Pines}(1958)}]{PhysRev.109.741}%
  \BibitemOpen
  \bibfield  {author} {\bibinfo {author} {\bibfnamefont {P.}~\bibnamefont
  {Nozi\`eres}}\ and\ \bibinfo {author} {\bibfnamefont {D.}~\bibnamefont
  {Pines}},\ }\bibfield  {title} {\bibinfo {title} {Electron interaction in
  solids. general formulation},\ }\href
  {https://doi.org/10.1103/PhysRev.109.741} {\bibfield  {journal} {\bibinfo
  {journal} {Phys. Rev.}\ }\textbf {\bibinfo {volume} {109}},\ \bibinfo {pages}
  {741} (\bibinfo {year} {1958})}\BibitemShut {NoStop}%
\bibitem [{\citenamefont {Hedin}(1965)}]{PhysRev.139.A796}%
  \BibitemOpen
  \bibfield  {author} {\bibinfo {author} {\bibfnamefont {L.}~\bibnamefont
  {Hedin}},\ }\bibfield  {title} {\bibinfo {title} {New method for calculating
  the one-particle green's function with application to the electron-gas
  problem},\ }\href {https://doi.org/10.1103/PhysRev.139.A796} {\bibfield
  {journal} {\bibinfo  {journal} {Phys. Rev.}\ }\textbf {\bibinfo {volume}
  {139}},\ \bibinfo {pages} {A796} (\bibinfo {year} {1965})}\BibitemShut
  {NoStop}%
\bibitem [{\citenamefont {Hohenberg}\ and\ \citenamefont
  {Kohn}(1964)}]{PhysRev.136.B864}%
  \BibitemOpen
  \bibfield  {author} {\bibinfo {author} {\bibfnamefont {P.}~\bibnamefont
  {Hohenberg}}\ and\ \bibinfo {author} {\bibfnamefont {W.}~\bibnamefont
  {Kohn}},\ }\bibfield  {title} {\bibinfo {title} {Inhomogeneous electron
  gas},\ }\href {https://doi.org/10.1103/PhysRev.136.B864} {\bibfield
  {journal} {\bibinfo  {journal} {Phys. Rev.}\ }\textbf {\bibinfo {volume}
  {136}},\ \bibinfo {pages} {B864} (\bibinfo {year} {1964})}\BibitemShut
  {NoStop}%
\bibitem [{\citenamefont {Ma}\ and\ \citenamefont
  {Brueckner}(1968)}]{Ma-Brueckner}%
  \BibitemOpen
  \bibfield  {author} {\bibinfo {author} {\bibfnamefont {S.-K.}\ \bibnamefont
  {Ma}}\ and\ \bibinfo {author} {\bibfnamefont {K.~A.}\ \bibnamefont
  {Brueckner}},\ }\bibfield  {title} {\bibinfo {title} {Correlation energy of
  an electron gas with a slowly varying high density},\ }\href
  {https://doi.org/10.1103/PhysRev.165.18} {\bibfield  {journal} {\bibinfo
  {journal} {Phys. Rev.}\ }\textbf {\bibinfo {volume} {165}},\ \bibinfo {pages}
  {18} (\bibinfo {year} {1968})}\BibitemShut {NoStop}%
\bibitem [{\citenamefont {Geldart}\ and\ \citenamefont
  {Taylor}(1970)}]{Geldart-Taylor1970}%
  \BibitemOpen
  \bibfield  {author} {\bibinfo {author} {\bibfnamefont {D.~J.~W.}\
  \bibnamefont {Geldart}}\ and\ \bibinfo {author} {\bibfnamefont
  {R.}~\bibnamefont {Taylor}},\ }\bibfield  {title} {\bibinfo {title}
  {Wave-number dependence of the static screening function of an interacting
  electron gas. i. lowest-order hartree--fock corrections},\ }\href
  {https://doi.org/10.1139/p70-022} {\bibfield  {journal} {\bibinfo  {journal}
  {Canadian Journal of Physics}\ }\textbf {\bibinfo {volume} {48}},\ \bibinfo
  {pages} {155} (\bibinfo {year} {1970})}\BibitemShut {NoStop}%
\bibitem [{\citenamefont {Geldart}\ and\ \citenamefont
  {Rasolt}(1976)}]{Geldart-Rasolt}%
  \BibitemOpen
  \bibfield  {author} {\bibinfo {author} {\bibfnamefont {D.~J.~W.}\
  \bibnamefont {Geldart}}\ and\ \bibinfo {author} {\bibfnamefont
  {M.}~\bibnamefont {Rasolt}},\ }\bibfield  {title} {\bibinfo {title} {Exchange
  and correlation energy of an inhomogeneous electron gas at metallic
  densities},\ }\href {https://doi.org/10.1103/PhysRevB.13.1477} {\bibfield
  {journal} {\bibinfo  {journal} {Phys. Rev. B}\ }\textbf {\bibinfo {volume}
  {13}},\ \bibinfo {pages} {1477} (\bibinfo {year} {1976})}\BibitemShut
  {NoStop}%
\bibitem [{\citenamefont {Sham}(1971)}]{Sham1971}%
  \BibitemOpen
  \bibfield  {author} {\bibinfo {author} {\bibfnamefont {L.~J.}\ \bibnamefont
  {Sham}},\ }\bibinfo {title} {Approximations of the exchange and correlation
  potentials},\ in\ \href {https://doi.org/10.1007/978-1-4684-1890-3_36} {\emph
  {\bibinfo {booktitle} {Computational Methods in Band Theory}}},\ \bibinfo
  {editor} {edited by\ \bibinfo {editor} {\bibfnamefont {P.~M.}\ \bibnamefont
  {Marcus}}, \bibinfo {editor} {\bibfnamefont {J.~F.}\ \bibnamefont {Janak}},\
  and\ \bibinfo {editor} {\bibfnamefont {A.~R.}\ \bibnamefont {Williams}}}\
  (\bibinfo  {publisher} {Springer US},\ \bibinfo {address} {Boston, MA},\
  \bibinfo {year} {1971})\ pp.\ \bibinfo {pages} {458--468}\BibitemShut
  {NoStop}%
\bibitem [{\citenamefont {Langreth}\ and\ \citenamefont
  {Perdew}(1980)}]{PhysRevB.21.5469_Langreth_Perdew}%
  \BibitemOpen
  \bibfield  {author} {\bibinfo {author} {\bibfnamefont {D.~C.}\ \bibnamefont
  {Langreth}}\ and\ \bibinfo {author} {\bibfnamefont {J.~P.}\ \bibnamefont
  {Perdew}},\ }\bibfield  {title} {\bibinfo {title} {Theory of nonuniform
  electronic systems. i. analysis of the gradient approximation and a
  generalization that works},\ }\href
  {https://doi.org/10.1103/PhysRevB.21.5469} {\bibfield  {journal} {\bibinfo
  {journal} {Physical Review B}\ }\textbf {\bibinfo {volume} {21}},\ \bibinfo
  {pages} {5469} (\bibinfo {year} {1980})}\BibitemShut {NoStop}%
\bibitem [{\citenamefont {Ceperley}\ and\ \citenamefont
  {Alder}(1980)}]{PhysRevLett.45.566}%
  \BibitemOpen
  \bibfield  {author} {\bibinfo {author} {\bibfnamefont {D.~M.}\ \bibnamefont
  {Ceperley}}\ and\ \bibinfo {author} {\bibfnamefont {B.~J.}\ \bibnamefont
  {Alder}},\ }\bibfield  {title} {\bibinfo {title} {Ground state of the
  electron gas by a stochastic method},\ }\href
  {https://doi.org/10.1103/PhysRevLett.45.566} {\bibfield  {journal} {\bibinfo
  {journal} {Phys. Rev. Lett.}\ }\textbf {\bibinfo {volume} {45}},\ \bibinfo
  {pages} {566} (\bibinfo {year} {1980})}\BibitemShut {NoStop}%
\bibitem [{\citenamefont {Gross}\ and\ \citenamefont
  {Dreizler}(1981)}]{Gross1981}%
  \BibitemOpen
  \bibfield  {author} {\bibinfo {author} {\bibfnamefont {E.~K.~U.}\
  \bibnamefont {Gross}}\ and\ \bibinfo {author} {\bibfnamefont {R.~M.}\
  \bibnamefont {Dreizler}},\ }\bibfield  {title} {\bibinfo {title} {Gradient
  expansion of the coulomb exchange energy},\ }\href
  {https://doi.org/10.1007/BF01413038} {\bibfield  {journal} {\bibinfo
  {journal} {Zeitschrift f{\"u}r Physik A Atoms and Nuclei}\ }\textbf {\bibinfo
  {volume} {302}},\ \bibinfo {pages} {103} (\bibinfo {year}
  {1981})}\BibitemShut {NoStop}%
\bibitem [{\citenamefont {Kleinman}(1984)}]{Kleinman1984}%
  \BibitemOpen
  \bibfield  {author} {\bibinfo {author} {\bibfnamefont {L.}~\bibnamefont
  {Kleinman}},\ }\bibfield  {title} {\bibinfo {title} {Exchange
  density-functional gradient expansion},\ }\href
  {https://doi.org/10.1103/PhysRevB.30.2223} {\bibfield  {journal} {\bibinfo
  {journal} {Physical Review B}\ }\textbf {\bibinfo {volume} {30}},\ \bibinfo
  {pages} {2223} (\bibinfo {year} {1984})}\BibitemShut {NoStop}%
\bibitem [{\citenamefont {Antoniewicz}\ and\ \citenamefont
  {Kleinman}(1985)}]{Kleinman-Antoniewicz}%
  \BibitemOpen
  \bibfield  {author} {\bibinfo {author} {\bibfnamefont {P.~R.}\ \bibnamefont
  {Antoniewicz}}\ and\ \bibinfo {author} {\bibfnamefont {L.}~\bibnamefont
  {Kleinman}},\ }\bibfield  {title} {\bibinfo {title} {Kohn-sham exchange
  potential exact to first order in
  \ensuremath{\rho}(k\ensuremath{\rightarrow})/${\ensuremath{\rho}}_{0}$},\
  }\href {https://doi.org/10.1103/PhysRevB.31.6779} {\bibfield  {journal}
  {\bibinfo  {journal} {Phys. Rev. B}\ }\textbf {\bibinfo {volume} {31}},\
  \bibinfo {pages} {6779} (\bibinfo {year} {1985})}\BibitemShut {NoStop}%
\bibitem [{\citenamefont {Kleinman}\ and\ \citenamefont
  {Lee}(1988)}]{Kleinman-Lee1988}%
  \BibitemOpen
  \bibfield  {author} {\bibinfo {author} {\bibfnamefont {L.}~\bibnamefont
  {Kleinman}}\ and\ \bibinfo {author} {\bibfnamefont {S.}~\bibnamefont {Lee}},\
  }\bibfield  {title} {\bibinfo {title} {Gradient expansion of the
  exchange-energy density functional: Effect of taking limits in the wrong
  order},\ }\href {https://doi.org/10.1103/PhysRevB.37.4634} {\bibfield
  {journal} {\bibinfo  {journal} {Physical Review B}\ }\textbf {\bibinfo
  {volume} {37}},\ \bibinfo {pages} {4634} (\bibinfo {year}
  {1988})}\BibitemShut {NoStop}%
\bibitem [{\citenamefont {Engel}\ and\ \citenamefont
  {Vosko}(1990)}]{Engel-Vosko1990}%
  \BibitemOpen
  \bibfield  {author} {\bibinfo {author} {\bibfnamefont {E.}~\bibnamefont
  {Engel}}\ and\ \bibinfo {author} {\bibfnamefont {S.~H.}\ \bibnamefont
  {Vosko}},\ }\bibfield  {title} {\bibinfo {title} {Wave-vector dependence of
  the exchange contribution to the electron-gas response functions: An analytic
  derivation},\ }\href {https://doi.org/10.1103/PhysRevB.42.4940} {\bibfield
  {journal} {\bibinfo  {journal} {Physical Review B}\ }\textbf {\bibinfo
  {volume} {42}},\ \bibinfo {pages} {4940} (\bibinfo {year}
  {1990})}\BibitemShut {NoStop}%
\bibitem [{\citenamefont {Svendsen}\ and\ \citenamefont {von
  Barth}(1996)}]{PhysRevB.54.17402}%
  \BibitemOpen
  \bibfield  {author} {\bibinfo {author} {\bibfnamefont {P.~S.}\ \bibnamefont
  {Svendsen}}\ and\ \bibinfo {author} {\bibfnamefont {U.}~\bibnamefont {von
  Barth}},\ }\bibfield  {title} {\bibinfo {title} {Gradient expansion of the
  exchange energy from second-order density response theory},\ }\href
  {https://doi.org/10.1103/PhysRevB.54.17402} {\bibfield  {journal} {\bibinfo
  {journal} {Physical Review B}\ }\textbf {\bibinfo {volume} {54}},\ \bibinfo
  {pages} {17402} (\bibinfo {year} {1996})}\BibitemShut {NoStop}%
\bibitem [{\citenamefont {Svendsen}\ and\ \citenamefont
  {Von~Barth}(1995)}]{Svendsen1995}%
  \BibitemOpen
  \bibfield  {author} {\bibinfo {author} {\bibfnamefont {P.~S.}\ \bibnamefont
  {Svendsen}}\ and\ \bibinfo {author} {\bibfnamefont {U.}~\bibnamefont
  {Von~Barth}},\ }\bibfield  {title} {\bibinfo {title} {On the gradient
  expansion of the exchange energy within linear response theory and beyond},\
  }\href {https://doi.org/10.1002/qua.560560421} {\bibfield  {journal}
  {\bibinfo  {journal} {International Journal of Quantum Chemistry}\ }\textbf
  {\bibinfo {volume} {56}},\ \bibinfo {pages} {351} (\bibinfo {year}
  {1995})}\BibitemShut {NoStop}%
\bibitem [{\citenamefont {Giuliani}\ and\ \citenamefont
  {Vignale}(2005)}]{Vignale}%
  \BibitemOpen
  \bibfield  {author} {\bibinfo {author} {\bibfnamefont {G.}~\bibnamefont
  {Giuliani}}\ and\ \bibinfo {author} {\bibfnamefont {G.}~\bibnamefont
  {Vignale}},\ }\href@noop {} {\emph {\bibinfo {title} {The quantum theory of
  the electron liquid}}}\ (\bibinfo  {publisher} {Cambridge University Press},\
  \bibinfo {address} {Cambridge},\ \bibinfo {year} {2005})\BibitemShut
  {NoStop}%
\bibitem [{\citenamefont {Benites}\ \emph {et~al.}(2024)\citenamefont
  {Benites}, \citenamefont {Rosado},\ and\ \citenamefont
  {Manousakis}}]{benites2024}%
  \BibitemOpen
  \bibfield  {author} {\bibinfo {author} {\bibfnamefont {M.}~\bibnamefont
  {Benites}}, \bibinfo {author} {\bibfnamefont {A.}~\bibnamefont {Rosado}},\
  and\ \bibinfo {author} {\bibfnamefont {E.}~\bibnamefont {Manousakis}},\
  }\bibfield  {title} {\bibinfo {title} {Accurate electron correlation energy
  functional: Expansion in the interaction renormalized by the random-phase
  approximation},\ }\href {https://doi.org/10.1103/PhysRevB.110.195151}
  {\bibfield  {journal} {\bibinfo  {journal} {Phys. Rev. B}\ }\textbf {\bibinfo
  {volume} {110}},\ \bibinfo {pages} {195151} (\bibinfo {year}
  {2024})}\BibitemShut {NoStop}%
\bibitem [{\citenamefont {Fetter}\ and\ \citenamefont
  {Walecka}(1971)}]{Fetter}%
  \BibitemOpen
  \bibfield  {author} {\bibinfo {author} {\bibfnamefont {A.~L.}\ \bibnamefont
  {Fetter}}\ and\ \bibinfo {author} {\bibfnamefont {J.~D.}\ \bibnamefont
  {Walecka}},\ }\href@noop {} {\emph {\bibinfo {title} {Quantum Theory of
  Many-Particle Systems}}}\ (\bibinfo  {publisher} {McGraw-Hill},\ \bibinfo
  {address} {New York},\ \bibinfo {year} {1971})\BibitemShut {NoStop}%
\bibitem [{\citenamefont {Mahan}(2000)}]{Mahan}%
  \BibitemOpen
  \bibfield  {author} {\bibinfo {author} {\bibfnamefont {G.}~\bibnamefont
  {Mahan}},\ }\href@noop {} {\emph {\bibinfo {title} {Many-Particle Physics}}}\
  (\bibinfo  {publisher} {Kluwer Academic/Plenum},\ \bibinfo {address} {New
  York},\ \bibinfo {year} {2000})\BibitemShut {NoStop}%
\bibitem [{\citenamefont {Pines}(1961)}]{Pines}%
  \BibitemOpen
  \bibfield  {author} {\bibinfo {author} {\bibfnamefont {D.}~\bibnamefont
  {Pines}},\ }\href@noop {} {\emph {\bibinfo {title} {The Many-Body Problem}}}\
  (\bibinfo  {publisher} {W. A. Benjamin},\ \bibinfo {address} {N.Y},\ \bibinfo
  {year} {1961})\BibitemShut {NoStop}%
\bibitem [{\citenamefont {Abrikosov}\ \emph {et~al.}(1963)\citenamefont
  {Abrikosov}, \citenamefont {Gor’kov},\ and\ \citenamefont
  {Dzyaloshinski}}]{Gorkov}%
  \BibitemOpen
  \bibfield  {author} {\bibinfo {author} {\bibfnamefont {A.~A.}\ \bibnamefont
  {Abrikosov}}, \bibinfo {author} {\bibfnamefont {L.~P.}\ \bibnamefont
  {Gor’kov}},\ and\ \bibinfo {author} {\bibfnamefont {I.~E.}\ \bibnamefont
  {Dzyaloshinski}},\ }\href@noop {} {\emph {\bibinfo {title} {Methods of
  quantum field theory in statistical physics}}}\ (\bibinfo  {publisher} {Dover
  Publications},\ \bibinfo {address} {New York},\ \bibinfo {year}
  {1963})\BibitemShut {NoStop}%
\bibitem [{\citenamefont {Kohn}\ and\ \citenamefont
  {Sham}(1965)}]{PhysRev.140.A1133}%
  \BibitemOpen
  \bibfield  {author} {\bibinfo {author} {\bibfnamefont {W.}~\bibnamefont
  {Kohn}}\ and\ \bibinfo {author} {\bibfnamefont {L.~J.}\ \bibnamefont
  {Sham}},\ }\bibfield  {title} {\bibinfo {title} {Self-consistent equations
  including exchange and correlation effects},\ }\href
  {https://doi.org/10.1103/PhysRev.140.A1133} {\bibfield  {journal} {\bibinfo
  {journal} {Phys. Rev.}\ }\textbf {\bibinfo {volume} {140}},\ \bibinfo {pages}
  {A1133} (\bibinfo {year} {1965})}\BibitemShut {NoStop}%
\bibitem [{\citenamefont {Perdew}\ \emph {et~al.}(1996)\citenamefont {Perdew},
  \citenamefont {Burke},\ and\ \citenamefont
  {Ernzerhof}}]{PhysRevLett.77.3865}%
  \BibitemOpen
  \bibfield  {author} {\bibinfo {author} {\bibfnamefont {J.~P.}\ \bibnamefont
  {Perdew}}, \bibinfo {author} {\bibfnamefont {K.}~\bibnamefont {Burke}},\ and\
  \bibinfo {author} {\bibfnamefont {M.}~\bibnamefont {Ernzerhof}},\ }\bibfield
  {title} {\bibinfo {title} {Generalized gradient approximation made simple},\
  }\href {https://doi.org/10.1103/PhysRevLett.77.3865} {\bibfield  {journal}
  {\bibinfo  {journal} {Phys. Rev. Lett.}\ }\textbf {\bibinfo {volume} {77}},\
  \bibinfo {pages} {3865} (\bibinfo {year} {1996})}\BibitemShut {NoStop}%
\bibitem [{\citenamefont {Perdew}\ \emph {et~al.}(2008)\citenamefont {Perdew},
  \citenamefont {Ruzsinszky}, \citenamefont {Csonka}, \citenamefont {Vydrov},
  \citenamefont {Scuseria}, \citenamefont {Constantin}, \citenamefont {Zhou},\
  and\ \citenamefont {Burke}}]{PhysRevLett.100.136406-PBEsol}%
  \BibitemOpen
  \bibfield  {author} {\bibinfo {author} {\bibfnamefont {J.~P.}\ \bibnamefont
  {Perdew}}, \bibinfo {author} {\bibfnamefont {A.}~\bibnamefont {Ruzsinszky}},
  \bibinfo {author} {\bibfnamefont {G.~I.}\ \bibnamefont {Csonka}}, \bibinfo
  {author} {\bibfnamefont {O.~A.}\ \bibnamefont {Vydrov}}, \bibinfo {author}
  {\bibfnamefont {G.~E.}\ \bibnamefont {Scuseria}}, \bibinfo {author}
  {\bibfnamefont {L.~A.}\ \bibnamefont {Constantin}}, \bibinfo {author}
  {\bibfnamefont {X.}~\bibnamefont {Zhou}},\ and\ \bibinfo {author}
  {\bibfnamefont {K.}~\bibnamefont {Burke}},\ }\bibfield  {title} {\bibinfo
  {title} {Restoring the density-gradient expansion for exchange in solids and
  surfaces},\ }\href {https://doi.org/10.1103/PhysRevLett.100.136406}
  {\bibfield  {journal} {\bibinfo  {journal} {Physical Review Letters}\
  }\textbf {\bibinfo {volume} {100}},\ \bibinfo {pages} {136406} (\bibinfo
  {year} {2008})}\BibitemShut {NoStop}%
\bibitem [{\citenamefont {Kleinman}\ and\ \citenamefont
  {Tamura}(1989)}]{Kleinman-Tamura}%
  \BibitemOpen
  \bibfield  {author} {\bibinfo {author} {\bibfnamefont {L.}~\bibnamefont
  {Kleinman}}\ and\ \bibinfo {author} {\bibfnamefont {T.}~\bibnamefont
  {Tamura}},\ }\bibfield  {title} {\bibinfo {title} {Ma-brueckner correlation
  energy},\ }\href {https://doi.org/10.1103/PhysRevB.40.4191} {\bibfield
  {journal} {\bibinfo  {journal} {Phys. Rev. B}\ }\textbf {\bibinfo {volume}
  {40}},\ \bibinfo {pages} {4191} (\bibinfo {year} {1989})}\BibitemShut
  {NoStop}%
\bibitem [{\citenamefont {Perdew}\ and\ \citenamefont
  {Wang}(1992)}]{PhysRevB.45.13244}%
  \BibitemOpen
  \bibfield  {author} {\bibinfo {author} {\bibfnamefont {J.~P.}\ \bibnamefont
  {Perdew}}\ and\ \bibinfo {author} {\bibfnamefont {Y.}~\bibnamefont {Wang}},\
  }\bibfield  {title} {\bibinfo {title} {Accurate and simple analytic
  representation of the electron-gas correlation energy},\ }\href
  {https://doi.org/10.1103/PhysRevB.45.13244} {\bibfield  {journal} {\bibinfo
  {journal} {Phys. Rev. B}\ }\textbf {\bibinfo {volume} {45}},\ \bibinfo
  {pages} {13244} (\bibinfo {year} {1992})}\BibitemShut {NoStop}%
\end{thebibliography}
\end{document}